\pdfoutput=1  

\documentclass[aps,twocolumn,superscriptaddress,floatfix,preprintnumbers,nofootinbib,eqsecnum]{revtex4-1}
\usepackage{amssymb,amsmath,graphicx,multirow,xcolor}
\usepackage{slashed}

\usepackage{hyperref}

\usepackage{epstopdf} 
\usepackage{placeins}

\usepackage{ulem}
\usepackage{comment}
\usepackage{soul}

\begin{document}

\title{Non-Minimal Dark Sectors:\\  Mediator-Induced Decay Chains and Multi-Jet Collider Signatures}

\author{Keith R. Dienes}
\email{dienes@email.arizona.edu}
\affiliation{Department of Physics, University of Arizona, Tucson, AZ 85721 USA}
\affiliation{Department of Physics, University of Maryland, College Park, MD 20742 USA}
\author{Doojin Kim}
\email{doojin.kim@tamu.edu}
\affiliation{Department of Physics, University of Arizona, Tucson, AZ 85721 USA}
\affiliation{Mitchell Institute for Fundamental Physics and Astronomy, Department of 
        Physics and Astronomy, Texas A\&M University, College Station, TX 77843 USA}
\author{Huayang Song}
\email{huayangs@email.arizona.edu}
\affiliation{Department of Physics, University of Arizona, Tucson, AZ 85721 USA}
\author{Shufang Su}
\email{shufang@email.arizona.edu}
\affiliation{Department of Physics, University of Arizona, Tucson, AZ 85721 USA}    
\author{\\ Brooks Thomas}  
\email{thomasbd@lafayette.edu}
\affiliation{Department of Physics, Lafayette College, Easton, PA 18042 USA}
\author{David Yaylali}
\email{yaylali@email.arizona.edu}
\affiliation{Department of Physics, University of Arizona, Tucson, AZ 85721 USA}

\preprint{MI-TH-1937}

\begin{abstract}
A preponderance of astrophysical and cosmological evidence indicates that the universe contains 
not only visible matter but also dark matter.  In order to suppress the couplings between the dark 
and visible sectors, a standard assumption is that these two sectors communicate only through a 
mediator.  In this paper we make a simple but important observation:  if the dark sector contains 
multiple components with similar quantum numbers, then this mediator also generically gives rise 
to dark-sector decays, with heavier dark components decaying to lighter components.  This in turn 
can even give rise to relatively long dark decay chains, with each step of the decay chain also 
producing visible matter.  The visible byproducts of such mediator-induced decay chains can 
therefore serve as a unique signature of such scenarios.  In order to investigate this possibility 
more concretely, we examine a scenario in which a multi-component dark sector is connected through 
a mediator to Standard-Model quarks.  We then demonstrate that such a scenario gives rise to 
multi-jet collider signatures, and we examine the properties of such jets at both the parton and 
detector levels.  Within relatively large regions of parameter space, we find that such multi-jet signatures are not excluded by existing monojet and multi-jet searches.  Such decay cascades 
therefore represent a potential discovery route for multi-component dark sectors at current and 
future colliders.
\end{abstract}

\maketitle

\newcommand{\newc}{\newcommand}
\newc{\gsim}{\lower.7ex\hbox{$\;\stackrel{\textstyle>}{\sim}\;$}}
\newc{\lsim}{\lower.7ex\hbox{$\;\stackrel{\textstyle<}{\sim}\;$}}
\makeatletter
\newcommand{\biggg}{\bBigg@{3}}
\newcommand{\Biggg}{\bBigg@{4}}
\makeatother

\newcommand{\cm}{\text{ cm}}
\newcommand{\kmpersec}{\text{ km}/\text{s}}
\newcommand{\ev}{\text{ eV}}
\newcommand{\kev}{\text{ keV}}
\newcommand{\mev}{\text{ MeV}}
\newcommand{\gev}{\text{ GeV}}
\newcommand{\tev}{\text{ TeV}}
\newcommand{\pb}{\text{ pb}}
\newcommand{\mb}{\text{ mb}}
\newcommand{\fb}{\text{ fb}}
\newcommand{\gweak}{g_{\text{weak}}}
\newcommand{\mweak}{m_{\text{weak}}}
\newcommand{\mplanck}{M_{\text{Pl}}}
\newcommand{\ipb}{\text{ pb}^{-1}}
\newcommand{\ifb}{\text{ fb}^{-1}}
\newcommand{\del}{\partial}
\newcommand{\Ecm}{E_\text{cm}}

\def\vac#1{{\bf \{{#1}\}}}

\def\beq{\begin{equation}}
\def\eeq{\end{equation}}
\def\beqn{\begin{eqnarray}}
\def\eeqn{\end{eqnarray}}
\def\bea{\begin{eqnarray}}
\def\eea{\end{eqnarray}}
\def\calM{{\cal M}}
\def\calV{{\cal V}}
\def\calF{{\cal F}}
\def\half{{\textstyle{1\over 2}}}
\def\quarter{{\textstyle{1\over 4}}}
\def\ie{{\it i.e.}\/}
\def\eg{{\it e.g.}\/}
\def\etc{{\it etc}.\/}


\def\inbar{\,\vrule height1.5ex width.4pt depth0pt}
\def\IR{\relax{\rm I\kern-.18em R}}
 \font\cmss=cmss10 \font\cmsss=cmss10 at 7pt
\def\IQ{\relax{\rm I\kern-.18em Q}}
\def\IZ{\relax\ifmmode\mathchoice
 {\hbox{\cmss Z\kern-.4em Z}}{\hbox{\cmss Z\kern-.4em Z}}
 {\lower.9pt\hbox{\cmsss Z\kern-.4em Z}}
 {\lower1.2pt\hbox{\cmsss Z\kern-.4em Z}}\else{\cmss Z\kern-.4em Z}\fi}
\def\TBBN{T_{\mathrm{BBN}}}
\def\OmegaCDM{\Omega_{\mathrm{CDM}}}
\def\OmegaDM{\Omega_{\mathrm{CDM}}}
\def\Omegatot{\Omega_{\mathrm{tot}}}
\def\rhocrit{\rho_{\mathrm{crit}}}
\def\tnow{t_{\mathrm{now}}}
\def\arcsinh{\mbox{arcsinh}}
\def\Omegatotnow{\Omega_{\mathrm{tot}}^\ast}
\def\mij{m_{jj}}
\def\mijmin{m_{jj}^{(\mathrm{min})}}
\def\mijmax{m_{jj}^{(\mathrm{max})}}
\def\mmax{m_{\mathrm{max}}}
\def\epsig{\epsilon_{\mathrm{sig}}}
\def\Lint{\mathcal{L}_{\mathrm{int}}}
\def\MT2{M_{T2}}
\def\MTtwomax{M_{T2}^{\mathrm{max}}}
\def\BR{\mathrm{BR}}
\def\dnarw{\downarrow}
\def\Njet{N_{\mathrm{jet}}}
\def\Nkin{N_{\mathrm{kin}}}
\def\Lint{\mathcal{L}_{\mathrm{int}}}
\newcommand{\Dsle}[1]{\hskip 0.09 cm \slash\hskip -0.23 cm #1}
\newcommand{\Dirsl}[1]{\hskip 0.09 cm \slash\hskip -0.20 cm #1}
\newcommand{\met}{{\Dsle E_T}}
\newcommand{\mpt}{{\Dsle \vec{p}_T}}

\newcommand{\dkc}[1]{\textbf{\color{purple}{(#1 -- DK)}}}
\newcommand{\dk}[1]{{\color{purple}{#1}}}

\newcommand{\Shufang}[1]{\textcolor{red} {SS: #1}}
\newcommand{\Huayang}[1]{{\color{blue} [HS: #1]}}
\newcommand{\hy}[1]{{\color{blue} #1}}

\newcommand{\changed}[2]{{\protect\color{red}\sout{#1}}{\protect\color{blue}\uwave{#2}}}


\input epsf





\FloatBarrier
\section{Introduction\label{sec:Introduction}}


One of the most exciting implications of the mounting observational 
evidence~\cite{DMReviews} for particle dark matter
is that particle species beyond those of the Standard Model (SM) likely exist in 
nature.  Nevertheless, despite an impressive array of experiments designed to probe 
the particle properties of these dark-sector species, the only conclusive evidence we 
currently have for the existence of dark matter is due to its gravitational influence on 
visible-sector particles.  The fact that no non-gravitational signals for dark matter have been 
definitively observed would suggest that interactions between the dark and visible sectors are 
highly suppressed.  While it is certainly possible that these two sectors communicate with each 
other only through gravity, it is also possible that they might communicate through some 
additional field or fields which serve as mediators between the two sectors as well.
These mediators play a crucial role in the phenomenology of any scenario
in which they appear, providing a portal linking the dark and visible sectors and giving 
rise to production, scattering, and annihilation processes involving dark-sector particles.   

Moreover, while we know very little about how the dark and visible sectors
interact, we know perhaps even less about the structure of the dark sector itself.  
While it is possible that the dark sector comprises merely a single particle 
species, it is also possible that the dark sector is non-minimal either in terms of the 
number of particle species it contains or the manner in which these 
species interact with each other.  For example, multi-component dark-matter 
scenarios have recently attracted a great deal of 
attention~\cite{BoehmFayetSilk,MaNeutrinoMulticomp,HurLeeNasriMulticomp,ShadowDM,Wimpless,
CheoKangLimMulticomp,HuhKimMulticomp,FairbairnMulticomp,ZurekMulticomp,BaerAxionAxino,
BatellPospelozRitz,ProfumoSigurdsonMulticomp,ChenClineMulticomp,ZhangLiMulticomp,MiXDM,
GaoKangLiMulticomp,dEramoThaler,Feldman:2010wy,WinslowMulticomp,DDM1,DDM2,Belanger:2011ww,
Cho:2012er,DDMLHC,Aoki:2012ub,Dienes:2012cf,Chialva:2012rq,AokiKuboMulticomp,DoubleDisk1,
DoubleDisk2,Medvedev:2013vsa,Dienes:2013xff,DoubleDiskExothermic,BhattacharyaDrozdMulticomp,
Geng:2013nda,DDMCutsAndCorrelations,Agashe:2014yua,Boddy:2016fds,Boddy:2016hbp,Arcadi:2016kmk,
Kim:2016zjx,Kim:2017qaw,AhmedDuchMulticomp,Giudice:2017zke,Chatterjee:2018mej,
Poulin:2018kap,DDMMATHUSLA,Heurtier:2019rkz} --- in large part because such 
scenarios can lead to novel signatures at colliders, direct-detection experiments, and 
indirect-detection experiments.
Moreover, the dark sector may also include additional particle species which are not
sufficiently long-lived to contribute to the dark-matter abundance at present time, but
nevertheless play an important role in the phenomenology of the dark sector.   
  
In this paper, we make a simple but important observation: in scenarios involving non-minimal 
dark sectors, any mediator which provides a portal linking the dark and visible sectors
generically also gives rise to processes through which the particles in the dark sector decay.
For example, in scenarios in which the dark-sector particles have similar quantum numbers and 
interact with the fields of the visible sector via a common mediator, processes generically arise 
in which heavier dark-sector species decay to final states including both lighter dark-sector 
species and SM particles.  Successive decays of this nature can then lead to extended decay 
cascades wherein both visible and dark-sector particles are produced at each step.
Depending on the masses and couplings of the particles involved, these decay cascades can 
have a variety of phenomenological consequences.  

In this paper, we shall consider the implications of such mediator-induced decay cascades at 
colliders.  In particular, we shall consider a scenario in which the dark sector comprises a 
large number of matter fields $\chi_n$, all of which couple directly to a common mediator 
particle which also couples to SM quarks.
Cascade decays in this scenario give rise to signatures at hadron colliders involving 
large numbers of hadronic jets in the final state, either with or without significant missing 
transverse energy $\met$.  Signatures of this sort can be somewhat challenging to resolve 
experimentally, since the jet multiplicities associated with such decay cascades can be quite 
large.  Indeed, the energy associated with any new particle produced at a collider is partitioned 
among the final-state objects that ultimately result from its decays.  Thus, as one searches 
for events with increasing numbers of such objects and adjusts the event-selection criteria 
accordingly, it becomes more likely that a would-be signal event would be rejected on the 
grounds that too few of these objects have sufficient transverse momentum $p_T$. 

Of particular interest within scenarios of this sort is the regime in which the number of particles 
within the ensemble is relatively large, in which the mass spacings between successively heavier 
$\chi_n$ are relatively small, and in which each $\chi_n$ preferentially decays in such a way that 
the resulting daughter $\chi_m$ is only slightly less massive than the parent $\chi_n$.
Within this regime, the decay of each of the heavier $\chi_n$ typically proceeds
through a long decay chain involving a significant number of steps.  Since each step in the decay 
chain produces one or more quarks or gluons at the parton level, such scenarios give rise to 
events with large jet multiplicities, distinctive kinematics, and a wealth of jet substructure.
The collider signatures which arise from these mediator-induced decay cascades are in
many ways qualitatively similar to those which have been shown to arise in scenarios 
involving large numbers of additional scalar degrees of freedom which couple directly to the 
SM Higgs field~\cite{Cohen:2018cnq,DAgnoloLow} and in superymmetric models in which 
a softly-broken conformal symmetry gives rise to a closely-spaced discretum of squark and 
gluino states~\cite{GluinoContinuum}.  Furthermore, we note that if the 
lifetimes of the lighter states in the dark sector are sufficiently long, these events could also 
involve displaced vertices or substantial missing transverse energy.  

A variety of search strategies relevant for the detection
of signals involving large jet multiplicities have already been implemented at the LHC.~  
Searches for events involving a large number $\Njet \geq 8$ of 
isolated, high-$p_T$ jets with or without $\met$~\cite{Sirunyan:2017cwe,Aaboud:2017hdf}
have been performed, motivated in part by the predictions of both 
$R$-parity-conserving~\cite{AlwallJetsPlusMet,StealthSUSY1,StealthSUSY2,Squadron,NBodyJetsPlusMet} 
and $R$-parity-violating~\cite{RPVSUSYSignals} supersymmetry and in part by 
the predictions of other scenarios, such as those involving colorons~\cite{ColoronSignal} 
or additional quark generations~\cite{FourthGenQuarks}.
Searches have also been performed for events involving significant 
numbers of high-$p_T$ final-state objects --- regardless of their identity --- in 
conjunction with a large scalar sum of $p_T$ over all such objects in the 
event~\cite{Aad:2015mzg,Sirunyan:2018xwt}.  Searches of this sort are motivated largely by
the prospect of observing signatures associated with extended objects such as miniature 
black holes~\cite{DimopoulosBlackHoles,GiddingsBlackHoles}, 
string balls~\cite{StringBalls,LowScaleStringStates}, and 
sphalerons~\cite{RingwaldSphaleron,HenryBlochWave,EllisSakurai}.  
Searches for events involving multiple soft jets originating from a displaced
vertex~\cite{ATLASDisplacedJet,CMSDisplacedJet} have been performed as well,
motivated by the predictions of hidden-valley 
models~\cite{HiddenValley1,HiddenValley2,HiddenValley3}, scenarios involving 
strongly-coupled dark sectors~\cite{BaiDarkQCD}, and certain realizations of 
supersymmetry~\cite{MiniSplitSUSY1,MiniSplitSUSY2,StealthSUSY1,StealthSUSY2}.

The bounds obtained from these searches impose non-trivial constraints on scenarios in which 
multiple dark-sector states couple to SM quarks via a common mediator as well.  Ultimately, 
however, we shall show that such scenarios can give rise to extended mediator-induced decay 
cascades while simultaneously remaining consistent with existing constraints from ATLAS and 
CMS searches in both the monojet and multi-jet channels.  Future colliders --- or potentially 
even alternative search strategies at the LHC --- could therefore potentially uncover evidence of 
such extended decay cascades and thereby shed light on the structure of the dark sector.
 
This paper is organized as follows.  In Sect.~\ref{sec:model}, we describe a simple, illustrative 
model involving an ensemble of unstable dark-sector particles with similar quantum numbers, along
with a mediator through which these particles couple to the fields of the visible sector.  We
also discuss the processes through which these dark-sector particles can be produced at
a hadron collider.  In Sect.~\ref{sec:pheno}, we investigate the decay phenomenology of the 
dark-sector particles within this framework 
and examine the underlying kinematics and combinatorics of the 
corresponding mediator-induced decay chains at the parton level.  We also discuss several preliminary 
parton-level constraints on our model.  In Sect.~\ref{sec:expobs}, we perform a detector-level 
analysis of the model and identify a number of kinematic collider variables which are particularly 
suited for resolving multi-jet signatures of these decay chains from the sizable 
SM background.  In Sect.~\ref{sec:LHCseaches}, we investigate the constraints from existing LHC 
monojet and multi-jet searches.  In Sect.~\ref{sec:scan}, we identify regions of 
model-parameter space which can potentially be probed by alternative search strategies at the 
forthcoming LHC run and beyond.  In Sect.~\ref{sec:noncolliderpheno}, we comment on additional 
considerations from flavor physics and cosmology which constrain our illustrative 
model.  
Finally, in Sect.~\ref{sec:conclusions}, we summarize our main 
results and discuss a number of interesting directions for future work.  We also briefly discuss 
search strategies which could improve the discovery reach for such theories at future colliders 
and comment on the phenomenological implications of mediator-induced decay cascades at the 
upcoming LHC run.


\FloatBarrier
\section{An Illustrative Framework \label{sec:model}}

 
Many scenarios for physics beyond the SM give rise to large ensembles of decaying states, 
including theories involving large extra spacetime dimensions, theories involving 
strongly-coupled hidden sectors, theories involving large spontaneously-broken
symmetry groups, and many classes of string theories.  Such ensembles also arise in the 
Dynamical Dark Matter framework~\cite{DDM1,DDM2}.  In order to incorporate all of these 
possibilities within our analysis, we shall adopt an illustrative and fairly model-independent 
approach towards describing our $\chi_n$ ensemble.  In particular, we shall adopt a set of 
rather generic parametrizations for the masses and decays of such states.

Toward this end, in this paper we consider an ensemble consisting of 
$N$ Dirac fermions $\chi_n$, with $n = 0,1,\ldots, N-1$, where these particles
are labeled in order of increasing mass, such that $m_{n+1}> m_{n}$ for all $n$.  
For concreteness, we shall further assume that the masses $m_n$ 
of these ensemble constituents scale across the ensemble according to a general relation
of the form
\begin{equation}
  m_n ~=~ m_0 + n^\delta \Delta m~,
\label{masses}
\end{equation}
with positive $m_0$,  $\Delta m$, and $\delta$.
Thus, the mass spectrum of our ensemble
is described by three parameters $\lbrace m_0,\Delta m, \delta\rbrace$:
$m_0$ is the mass of the lightest ensemble constituent,
$\Delta m$ controls the overall scale of the mass splittings within the ensemble,
and $\delta$ is a dimensionless scaling exponent.

The general relation in Eq.~(\ref{masses}) is capable of describing the masses of states 
$\chi_n$ in a number of different scenarios for physics beyond the SM.  For example, if the 
$\chi_n$ are the Kaluza-Klein excitations of a five-dimensional scalar field with 
four-dimensional mass $m$ compactified on a circle or line segment of radius/length $R$,
we have $\lbrace m_0,\Delta m,\delta\rbrace = \lbrace m, 1/R, 1\rbrace$ if $m R \ll1$ or 
$\lbrace m_0,\Delta m,\delta\rbrace =\lbrace m, 1/(2 m R^2), 2\rbrace$ if $mR\gg 1$.
Likewise, if the ensemble constituents are the bound states of a strongly-coupled
gauge theory, or even the gauge-neutral bulk (oscillator) states within many classes of string 
theories, we have $\delta = 1/2$, where $\Delta m$ and $m_0$ are related to the Regge slopes and 
intercepts of these theories, respectively.  Thus $\delta=1/2$, $\delta=1$, and $\delta=2$ serve 
as particularly compelling ``benchmark'' values.  We shall nevertheless take $m_0$, $\Delta m$, 
and $\delta$ to be free parameters in what follows.

\begin{figure*}[t!]
  \begin{center}
        \includegraphics[width=5.3in]{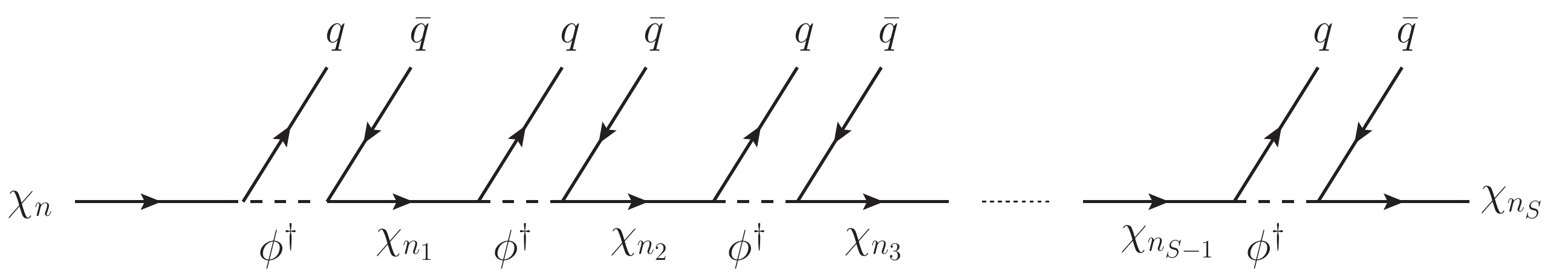}
  \end{center}
\caption{A decay chain in which an ensemble constituent $\chi_{n}$
  experiences $S$ successive decays into increasingly lighter constituents.
  Each individual decay occurs through a three-body process of the form
  $\chi_{n_k}\rightarrow \bar{q}q\chi_{n_{k+1}}$ involving an off-shell $\phi^\dagger$
   and resulting in the emission of two quarks (or parton-level ``jets'').
  Each decay chain effectively terminates once a collider-stable constituent 
   is reached.}
  \label{fig:DecayChain}
\end{figure*}

Having parametrized the masses of our dark ensemble states $\chi_n$, we now turn to consider 
the manner in which these states interact with the particles of the visible-sector through 
a mediator.  One possibility is that these interactions occur through an $s$-channel 
mediator $\phi$.  Assuming that the SM fields $\psi$ which couple directly to $\phi$ are  
fermions, the interaction Lagrangian then takes the schematic form
\begin{equation}
  \mathcal {L}_{\rm int} ~=~  \sum_\psi  c_\psi \phi \bar{\psi} \psi 
    +  \sum_{m,n = 0}^{N-1} c_{mn} \phi \bar{\chi}_{m} \chi_n~,
\label{eq:s_channel}
\end{equation}
where $c_\psi$ and $c_{mn}$ denote the couplings between the mediator and the fields of the 
visible and dark sectors, respectively.  An alternative possibility is that these interactions  
take place via a $t$-channel mediator.  The interaction Lagrangian in this case takes the 
schematic form
\begin{equation}
  \mathcal{L}_{\rm int} ~=~ 
    \sum_\psi\sum_{n=0}^{N-1} c_{\psi n} \phi^\dagger \bar{\chi}_n \psi + {\rm h.c.}
\label{eq:t_channel}
\end{equation}
While both possibilities allow our dark-sector constituents $\chi_n$ to be produced at 
colliders --- and also potentially allow these states to decay, with the simultaneous emission 
of visible-sector states~\cite{OffDiagDM1,OffDiagDM2} --- the mediator $\phi$ in the $t$-channel 
case can carry SM charges.  If these include color charge, mediator particles can be 
copiously pair-produced on shell at hadron colliders, and decay cascades precipitated by the 
subsequent decays of these mediators can therefore contribute significantly to 
the signal-event rate in the detection channels which are our main interest in 
this paper.  The interaction in Eq.~(\ref{eq:t_channel}) is also comparatively minimal, with the 
production and decay processes occurring through a single common interaction.

We shall therefore focus on the case of a $t$-channel mediator $\phi$ in this paper.  In 
particular, we shall assume that each of the $\chi_n$ couples to an additional heavy scalar 
mediator particle $\phi$ of mass $m_\phi$ which transforms as a fundamental triplet under the 
$SU(3)_c$ gauge group of the SM and has hypercharge $Y_\phi = -2/3$.  We shall then take the 
coupling between $\phi$ and each of the $\chi_n$ to be given by the interaction Lagrangian
\begin{equation}
  \mathcal{L}_{\mathrm{int}} ~=~ \sum_q \sum_{n=0}^{N-1} \big[ c_{n q} \phi^\dagger
     \bar{\chi}_n P_R q  + \mathrm{h.c.}\big]~,
\label{eq:Lint}
\end{equation}
where $q \in \{u,c,t\}$ denotes an up-type SM quark, where $P_R = \frac{1}{2}(1 + \gamma^5)$ is 
the usual right-handed projection operator, and where $c_{n q}$ is a dimensionless coupling 
constant which in principle depends both on the identity of the ensemble constituent 
and on the flavor of the quark.  For concreteness, we shall assume that the $c_{n q}$ 
scale according to the power-law relation
\begin{equation}
  c_{n q} ~=~ c_{0 q} \left(\frac{m_n}{m_0}\right)^\gamma~,
\label{eq:CouplingScaling}
\end{equation}
where the masses $m_n$ are given in Eq.~(\ref{masses}), where $c_{0 q}>0$ is an overall 
normalization for the couplings, and where $\gamma$ is a scaling exponent.  

Generally speaking, the interaction Lagrangian in Eq.~(\ref{eq:Lint}) can 
give rise to flavor-changing neutral currents (FCNCs), which are stringently constrained by 
data.  However, such constraints can easily be satisfied.   
These issues will be discussed in greater detail in Sect.~\ref{sec:noncolliderpheno}. 

The interaction Lagrangian in Eq.~(\ref{eq:Lint}) simultaneously describes two critical 
features of our model.  First, we see that our mediator field generically allows the heavier 
ensemble constituents to decay to successively lighter constituents, thereby 
forming a decay chain.  Indeed, according to our interaction Lagrangian, each step of the decay 
chain proceeds through an effective three-body decay process of the form
$\chi_k \rightarrow q\overline{q}'\chi_\ell$ 
involving an off-shell mediator $\phi$ particle,
where $m_\ell< m_k$.  Such a decay chain is illustrated in 
Fig.~\ref{fig:DecayChain}, with each step of the decay resulting in two parton-level
jets.  Indeed, such a decay chain effectively terminates only 
when a collider-stable constituent is reached.
If the parameters which govern our model are such that 
each ensemble constituent $\chi_k$ decays primarily to those daughters $\chi_\ell$ whose masses 
$m_\ell$ are only slightly less than $m_k$, relatively long decay chains involving multiple 
successive such decays can develop before a collider-stable constituent is reached, especially 
if the first constituent $\chi_n$ that is produced is relatively massive.  In such cases, 
relatively large numbers of parton-level ``jets'' --- \ie, quarks or gluons --- can be emitted.

\begin{figure}[t!]
  \begin{center}
	\includegraphics[width=1.5in]{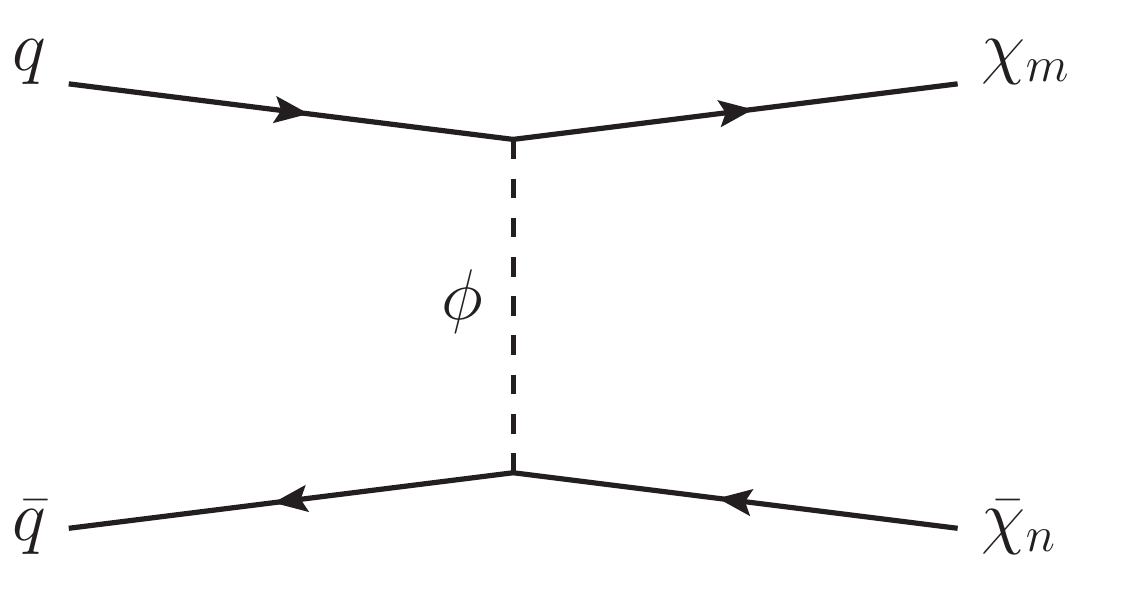}
  \end{center}
\caption{Feynman diagram for the process in which $\chi_m$ and $\overline{\chi}_n$ are 
  produced directly via the process $pp\rightarrow \chi_m\bar{\chi}_n$.}
  \label{fig:ChiChiDiagram}
  \begin{center}
  	\includegraphics[width=1.5in]{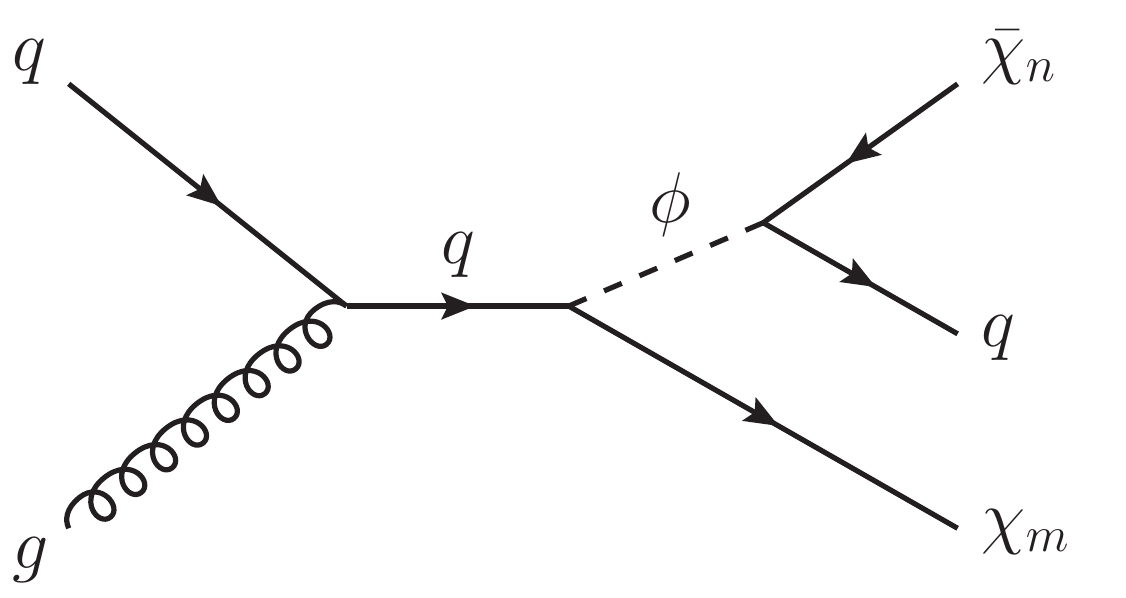}~~
  	\includegraphics[width=1.5in]{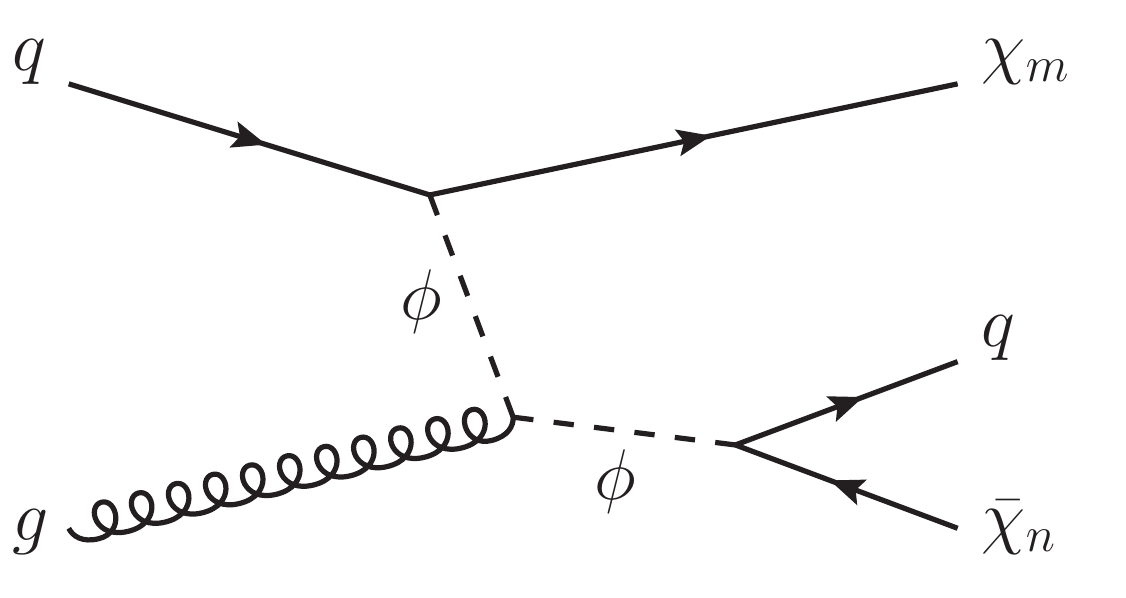}
  \end{center}
\caption{Representative Feynman diagrams for collider processes in 
  which the initial production process $pp\rightarrow \phi\chi_m $ is followed by a decay 
  of the form $\phi\rightarrow q \bar{\chi}_n$.  }
  \label{fig:ChiPhiDiagram}
  \begin{center}
	\includegraphics[width=1.5in]{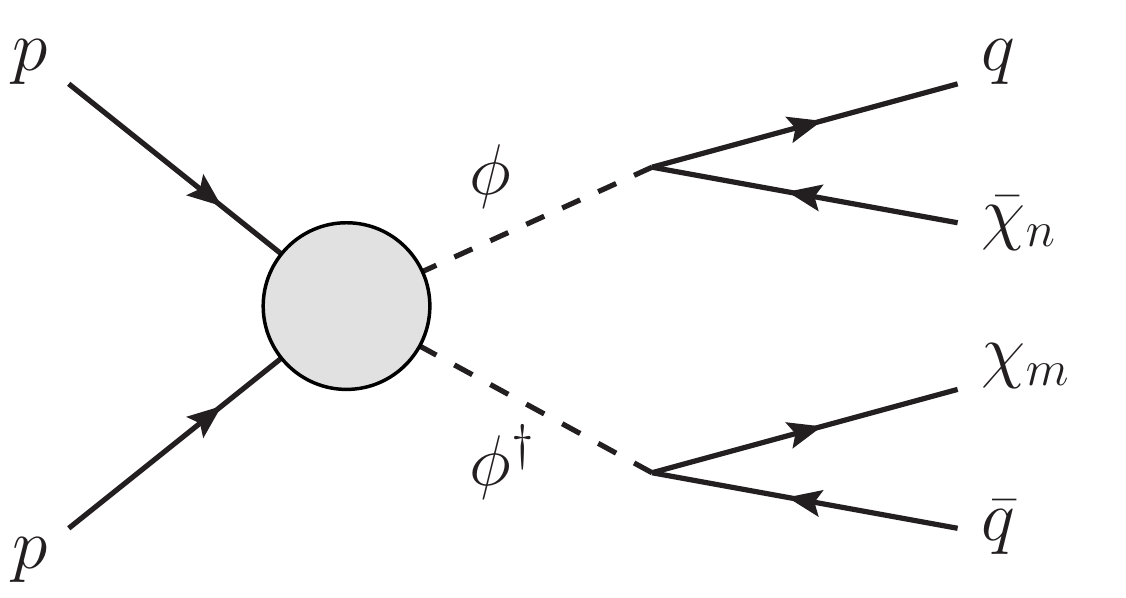}
  \end{center}
\caption{Representative Feynman diagram for collider processes in
   which $pp\rightarrow \phi^\dagger\phi$ production is followed by decays of the form 
   $\phi\rightarrow q \bar{\chi}_n$ and $\phi^\dagger \rightarrow \chi_m \bar{q}$.}
  \label{fig:PhiPhiDiagram}
\end{figure}

We see, then, that any $\chi_n$ that is produced --- unless it happens to be collider-stable ---  
will generate a subsequent decay chain.  The only remaining issue therefore concerns the manner 
in which such $\chi_n$ particles might be produced at a hadron collider such as the LHC.~
However, the relevant production processes are also described by our interaction Lagrangian in
Eq.~(\ref{eq:Lint}) in conjunction with our assumption that $\phi$ is an $SU(3)_c$ color triplet. 
Indeed, given this interaction Lagrangian, there are a number of distinct possibilities for how 
the production of the $\chi_n$ might take place:
\begin{itemize}
\item  The $\chi_n$ may be produced directly via the process $pp\rightarrow \chi_m\bar{\chi}_n$ 
   at leading order.  The Feynman diagram for this process is shown in 
   Fig.~\ref{fig:ChiChiDiagram}.   
\item The $\chi_n$ may be produced via the process $pp\rightarrow \phi\chi_m$, followed by a 
  decay of the form $\phi\rightarrow q \bar{\chi}_n$.  In such cases, one constituent $\chi_n$ 
  particle is produced directly while the other results from a subsequent $\phi$ decay.  Two 
  representative Feynman diagrams for such processes are shown in Fig.~\ref{fig:ChiPhiDiagram}.
\item Finally, because the $\phi$ particles are $SU(3)_c$ triplets, the $\chi_n$ may also be 
  produced via the process $pp\rightarrow \phi^\dagger\phi$ followed by decays of the form
  $\phi\rightarrow q \bar{\chi}_n$ and $\phi^\dagger \rightarrow \chi_m \bar{q}$.  In such cases, 
  both $\chi_m$ and $\bar{\chi}_n$ are produced via the decays of $\phi$ particles.  A 
  representative Feynman diagram for such a process is shown in Fig.~\ref{fig:PhiPhiDiagram}.
\end{itemize}

These different production processes have very different phenomenologies.
For example, since the amplitude for each contributing diagram in Fig.~\ref{fig:ChiChiDiagram}
is proportional to the product $c_m c_n$, the cross-section --- and therefore the
event rate --- for the overall process is proportional to $c_0^4$.  By contrast, the event
rates for the overall processes in Fig.~\ref{fig:ChiPhiDiagram} are proportional to $c_0^2$ 
when $\phi$ is on shell, since the factor $c_n$ from the decay vertex affects the decay width 
of $\phi$ but not the cross-section for  $pp\rightarrow \phi\chi_m $. 
Finally, the event rate for the process shown in Fig.~\ref{fig:PhiPhiDiagram} is
essentially independent of $c_0$, as $\phi$ is an $SU(3)$ color triplet and can
therefore be pair-produced through diagrams involving strong-interaction vertices alone. 

Another distinction between these processes is the manner in which their overall 
cross-sections scale with the number of kinematically accessible components $\chi_n$ within the 
ensemble.  For example, the event rate for the process shown in Fig.~\ref{fig:PhiPhiDiagram} is
essentially set by the cross-section for the initial process $pp\to \phi^\dagger \phi$
and is thus largely insensitive to the multiplicity of states within the ensemble. 
By contrast, processes such as those shown in Figs.~\ref{fig:ChiChiDiagram} and \ref{fig:ChiPhiDiagram}
scale with the multiplicity of the $\chi_n$ states that are kinematically accessible,
as the contributions from the production of each separate constituent $\chi_n$ must be added 
together.  For large ensembles, this can lead to a significant enhancement of the total 
cross-sections for such processes.

All of these processes are capable of giving rise to large numbers of parton-level jets,
particularly if the $\chi_n$ that are produced give rise to long subsequent decay chains.
Additional parton-level jets may also be produced as initial-state radiation 
or radiated off any internal lines associated with strongly-interacting particles.
However, these different processes differ in the {\it minimum}\/ numbers of parton-level jets 
which may be produced.  For example, the direct-production process in Fig.~\ref{fig:ChiChiDiagram}
can in principle be entirely jet-free as long as only collider-stable ensemble constituents are produced.
Likewise, the processes in Fig.~\ref{fig:ChiPhiDiagram} must give rise to at least one jet, and indeed
processes of this form involving an on-shell $\phi$ particle often turn out to provide the 
dominant contribution to the $pp\rightarrow \chi_m\bar{\chi}_n + j$ monojet production rate 
at the LHC within our model.  By contrast, the process in Fig.~\ref{fig:PhiPhiDiagram} must give 
rise to at least two jets.

In order to streamline the analysis of our model, we shall make two further assumptions in what 
follows.  First, we shall assume that the $\chi_n$ couple only to the up quark, 
taking $c_{0 q} = 0$ for $q = \{c,t\}$.  Thus only the $c_{nu}$ coefficients are non-zero, 
and we shall henceforth adopt the shorthand notation $c_n\equiv c_{n u}$ for all $n$.
Alternative coupling structures shall be discussed in Sect.~\ref{sec:noncolliderpheno}.  
Second, we shall assume that $N$, the total number of constituents in our ensemble,
is not only finite but also chosen so as to maximize the size of the ensemble while nevertheless
ensuring that all of the ensemble constituents $\lbrace \chi_0,\chi_1,...,\chi_{N-1}\rbrace$ are 
kinematically accessible via the decays of $\phi$.  In other words, we shall take $N$ to be 
the largest integer such that
\beq 
       N ~\leq~ 1 + \left( \frac{ m_\phi - m_0 - m_q}{\Delta m} \right)^{1/\delta}~,
\label{eq:Ndeff}
\eeq
where $m_q$ is the mass of the final-state (up) quark. 
While this last assumption is not required for the self-consistency of our model,
we shall see that it simplifies the resulting analysis and leads to an interesting phenomenology.

With these simplifications, our framework is characterized by six free parameters:  
$\lbrace m_0, \Delta m, \delta, m_\phi, c_0, \gamma\rbrace$.  These six parameters determine the 
masses of the ensemble constituents $\chi_n$, the probabilities for producing these different 
ensemble constituents from the decays of $\phi$, and the branching fractions that govern the 
possible subsequent decays of these constituents.  Indeed, depending on the values of these 
parameters, many intricate patterns of potential decay chains are possible which collectively 
contribute to jet production.  For example, in some regions of parameter space, the lifetimes 
of the heavier ensemble states are shorter than those of the lighter states, while in other 
regions the opposite is true.  Of course, the lightest state receives no contribution to its 
width from the interaction Lagrangian in Eq.~(\ref{eq:Lint}) and is therefore either absolutely 
stable or else decays only as a result of additional, highly suppressed interactions on a 
timescale that far exceeds collider timescales.
Likewise, in some regions of parameter space, each $\chi_n$ preferentially decays to daughters 
$\chi_\ell$ for which $m_\ell\ll m_n$, while in other regions the preferred daughters 
$\chi_\ell$ are only slightly lighter than $\chi_n$.  Finally, in some regions of parameter space, 
the contributions to jet production coming from the processes illustrated within 
Figs.~\ref{fig:ChiChiDiagram} and \ref{fig:ChiPhiDiagram} might dominate, while in other regions 
of parameter space the contributions from the process illustrated within 
Fig.~\ref{fig:PhiPhiDiagram} might dominate.  Thus, even though our framework is governed by 
only the single interaction in Eq.~(\ref{eq:Lint}), this framework is extremely rich and many 
different resulting phenomenologies are possible.

In our analysis of this framework, we shall be interested primarily in those regions of parameter
space which potentially give rise to extended jet cascades at colliders such 
as the LHC.~  We shall therefore be interested in those regions of parameter space that give
rise to a relatively large number of kinematically accessible ensemble constituents $\chi_n$ 
which decay promptly on collider timescales and for which the corresponding decays occur along
decay chains involving a relatively large number of steps.  Beyond this, however, we will not 
make any further assumptions concerning the values of these parameters.  
Of course, within our parameter-space regions of interest, there may exist subregions in which 
some of the other constituents will have very long lifetimes ---  lifetimes which potentially 
exceed the age of the universe.  In such cases, these long-lived constituents might serve as 
potential dark-matter candidates of the sort intrinsic to the Dynamical Dark Matter 
framework~\cite{DDM1,DDM2}, with the decay cascades arising from the decays of the shorter-lived 
ensemble constituents potentially serving as a signature of this framework.  Since our focus in 
this paper is on the collider phenomenology of the mediator-induced decay chains that arise in 
this model, we make no additional assumption as to whether the $\chi_n$ collectively 
contribute a non-negligible fraction of the overall present-day dark-matter abundance.  
However, we comment on the possibility that the $\chi_n$ might contribute 
significantly to this abundance in Sect.~\ref{sec:noncolliderpheno}.
  

\FloatBarrier
\section{Decay-Chain Phenomenology and the Generation of Extended Jet Cascades\label{sec:pheno}}


We shall now demonstrate that the model presented in Sect.~\ref{sec:model} is capable of 
giving rise to extended jet cascades at the LHC.~  In this section our analysis shall be purely at 
the parton level, while in Sect.~\ref{sec:expobs} we shall pass to the detector level.

In principle, mediator-induced decay cascades can arise from any of the processes illustrated in 
Figs.~\ref{fig:ChiChiDiagram}--\ref{fig:PhiPhiDiagram}.  Of course, our eventual goal in this 
paper is not merely to demonstrate that cascades of this sort with large jet multiplicities are 
possible, but that they might emerge while simultaneously satisfying existing LHC monojet and 
multi-jet constraints.  For this, of course, the contributions from {\it all}\/ of the processes 
discussed in Sect.~\ref{sec:model} will ultimately matter.  This will be discussed in 
Sect.~\ref{sec:scan}.

We shall begin by outlining the kinematics and combinatorics of the mediator-induced decay 
chains precipitated by the production processes illustrated in 
Figs.~\ref{fig:ChiChiDiagram}--\ref{fig:PhiPhiDiagram}.  We shall then discuss how the emergence 
of extended decay chains yielding large numbers of jets depends on the parameters which 
characterize  our model, and identify a region of parameter space within which such extended
decay chains emerge naturally while satisfying certain internal self-consistency constraints.

\subsection{The structure of the decay chain:  Kinematics and combinatorics\label{firstpart}}

Each of the processes illustrated in Figs.~\ref{fig:ChiChiDiagram}--\ref{fig:PhiPhiDiagram}
eventually results in decay chains of the sort illustrated in Fig.~\ref{fig:DecayChain}.
In cases such as that illustrated in Fig.~\ref{fig:ChiChiDiagram}, 
our ensemble constituents $\chi_m$ and $\overline{\chi}_n$ are produced directly.  
Each then becomes the heaviest component of a subsequent decay chain.
By contrast, in cases such as that illustrated in Fig.~\ref{fig:PhiPhiDiagram},
the particles that are produced directly are the mediator particles $\phi$ and $\phi^\dagger$.
It is the subsequent decays of these mediators which then trigger the unfolding of our decay 
chains.  Finally, cases such as those illustrated in Fig.~\ref{fig:ChiPhiDiagram}
exhibit what may be considered a ``mixture'' between these two production mechanisms.

In this section, rather than analyze each process separately, we shall instead treat them 
together by focusing on the two primary classes of decays which establish and sustain their decay 
chains.  These are 
\beqn
            \phi^\dagger ~&\rightarrow&~ \overline{q} \, \chi_n \nonumber\\
            \chi_n ~&\rightarrow&~ \overline{q}' q \, \chi_\ell ~.
\label{decaychainprocesses}
\eeqn  
Note that although we have written these decay processes in generality, we shall --- as discussed in 
Sect.~\ref{sec:model} --- restrict our attention to the case in which all quarks participating
in these processes are up-quarks ({\it i.e.}\/, $q=q'=u$) in what follows.  For cases involving 
the initial production of a mediator $\phi$, the first process in Eq.~(\ref{decaychainprocesses}) 
in some sense ``initializes'' the decay chain by producing the heaviest $\chi_n$ constituent within 
the chain.  This initialization process simultaneously produces one jet.  The second process 
then iteratively generates the subsequent decays --- each producing two jets ---  
which collectively give rise to the decay chain through which this heaviest constituent $\chi_n$
sequentially decays into lighter constituents.  By contrast, for cases involving the direct 
production of an ensemble constituent $\chi_n$, only the second process in 
Eq.~(\ref{decaychainprocesses}) is relevant for generating the subsequent decay chain.

Even with a fixed initial state, each of the decay processes in Eq.~(\ref{decaychainprocesses})
can result in a variety of different daughter particles.  Indeed, starting from a given mediator 
particle $\phi$, it is possible for any kinematically-allowed constituent $\chi_n$ to be produced 
via the first process, each with a different probability.  Likewise, a given $\chi_n$ can generally 
decay into {\it any}\/ lighter constituents via the second process, with each possible daughter 
state occurring with a different probability as well.  The sequential repetitions of this latter 
process thus lead to a proliferation of independent decay chains, with each decay chain terminating 
only when the lightest ensemble constituent is ultimately reached.  (For practical purposes we may 
also consider a given decay chain to have effectively terminated if the lifetimes for further decays 
exceed collider timescales.)  Thus, combining these effects, we see that each of the processes 
sketched in Figs.~\ref{fig:ChiChiDiagram}--\ref{fig:PhiPhiDiagram} actually spawns a large set of 
many different possible decay chains, each with its own relative probability for occurring
and each potentially producing a different number of jets.
 
It is not difficult to study these decay chains analytically.  Within any particular region of 
the model parameter space, the first step is to calculate 
the partial widths $\Gamma_{\phi n}\equiv \Gamma(\phi^\dagger \rightarrow \bar{q}\chi_n)$ and 
$\Gamma_{n\ell} \equiv \Gamma(\chi_n \rightarrow \bar{q}'q\chi_\ell)$ associated with the processes 
in Eq.~(\ref{decaychainprocesses}).  With $q=q'=u$ and with the up-quark treated as having a 
negligible mass, we find that $\Gamma_{\phi n}$ for any $n\leq N-1$ is to a very good approximation 
given by  
\begin{equation} 
  \Gamma_{\phi n} ~=~ \frac{c_{n}^2}{16 \pi}  
    \frac{( m_\phi^2 - m_n^2)^2}{m_{\phi}^3}~.
  \label{eq:Gammaphi}
\end{equation}
Likewise, we find that $\Gamma_{n\ell}$ 
takes the form  
\begin{eqnarray}
  \Gamma_{n \ell} ~&=&~ \frac{3c_{n}^2c_{\ell}^2}
    {256 \pi^2}\frac{m_\phi}{ r_{\phi n}^3} \Bigg[
    f_{\phi n \ell }^{(1)} -  f_{\phi n \ell }^{(2)} \ln (r_{n \ell}) \nonumber \\
    & &~~ + f_{\phi n \ell }^{(3)} \ln\left(\frac{1-r_{\phi n}^2}{1-r_{\phi n}^2r_{n \ell }^2}\right) 
    \Bigg]~,
  \label{eq:Gammanmeq}
\end{eqnarray}
where $r_{ij}\equiv m_j/m_i$, where $r_{\phi n}\equiv m_n/m_\phi$, and where 
\begin{eqnarray}
  f_{\phi n \ell }^{(1)} ~&\equiv&~ 6r_{\phi n}^2(1-r_{n \ell}^2) 
      - 5r_{\phi n}^4(1-r_{n \ell}^4) \nonumber \\
    ~& &~ + 2r_{\phi n}^6r_{n \ell}^2 (1-r_{n \ell}^2)\nonumber \\
  f_{\phi n \ell }^{(2)} ~&\equiv&~ 4r_{\phi n}^8r_{n \ell}^4 \nonumber \\
  f_{\phi n \ell}^{(3)} ~&\equiv&~ 6 - 8r_{\phi n }^2(1+r_{n \ell}^2) 
      - 2r_{\phi n}^8r_{n \ell}^4 \nonumber \\
    ~& &~ + 2r_{\phi n}^4(1+4r_{n \ell}^2+r_{n \ell}^4)~.
\end{eqnarray}

Under the assumption that no additional interactions beyond those in Eq.~(\ref{decaychainprocesses})
contribute non-negligibly to the total width of either $\phi$ or the $\chi_n$, the total decay 
width $\Gamma_\phi$ of $\phi$ is then simply
\begin{equation}
  \Gamma_\phi ~=~ \sum_{n=0}^{N-1} \Gamma_{\phi n}~,  
\end{equation} 
with a corresponding $\phi$ lifetime $\tau_\phi \equiv 1/\Gamma_\phi$.
Likewise, the total decay width $\Gamma_n$ for each ensemble constituent $\chi_n$ is simply 
\begin{equation}
  \Gamma_n ~=~ \sum_{\ell=0}^{n-1} \Gamma_{n \ell}~,
  \label{eq:TotalWidth}  
\end{equation} 
with a corresponding constituent lifetime $\tau_n\equiv 1/\Gamma_n$.  
As discussed in Sect.~\ref{sec:model}, the lightest ensemble constituent $\chi_0$ receives no 
contribution to its width from the interaction Lagrangian in Eq.~(\ref{eq:Lint}) and is 
therefore assumed to be stable on collider timescales.  Of course,
the results in Eqs.~(\ref{eq:Gammaphi}) and (\ref{eq:Gammanmeq}) assume that the $\phi$ and 
$\chi_n$ particles have total decay widths which are relatively small compared with their masses.  
This is a self-consistency constraint which will ultimately be found to hold across our eventual 
parameter-space regions of interest.

While $\Gamma_\phi$ and $\Gamma_n$ determine the overall timescales for particle decays within our 
model, it is the branching fractions $\mathrm{BR}_{\phi n} \equiv \Gamma_{\phi n}/\Gamma_\phi$ and 
$\mathrm{BR}_{n \ell} \equiv \Gamma_{n \ell}/\Gamma_n$ which effectively determine the 
probabilities associated with the various possible decay chains that can arise.  The behavior of 
these branching fractions is essentially determined by the interplay between two factors.  The first 
of these factors is purely kinematic in origin and arises due to phase-space considerations
which suppress the partial widths for decays involving heavier ensemble constituents 
in the final state.  Thus, this factor always decreases as the index which labels this final-state 
ensemble constituent increases.  The second factor arises as a result of the scaling 
of the individual coupling constants $c_{n}$ in Eq.~(\ref{eq:CouplingScaling}) across the ensemble. 
Depending on the value of the scaling exponent $\gamma$, this factor may either increase or 
decrease with the final-state index.

\begin{figure}[t]
  \begin{center}
	\includegraphics[width=3.2in]{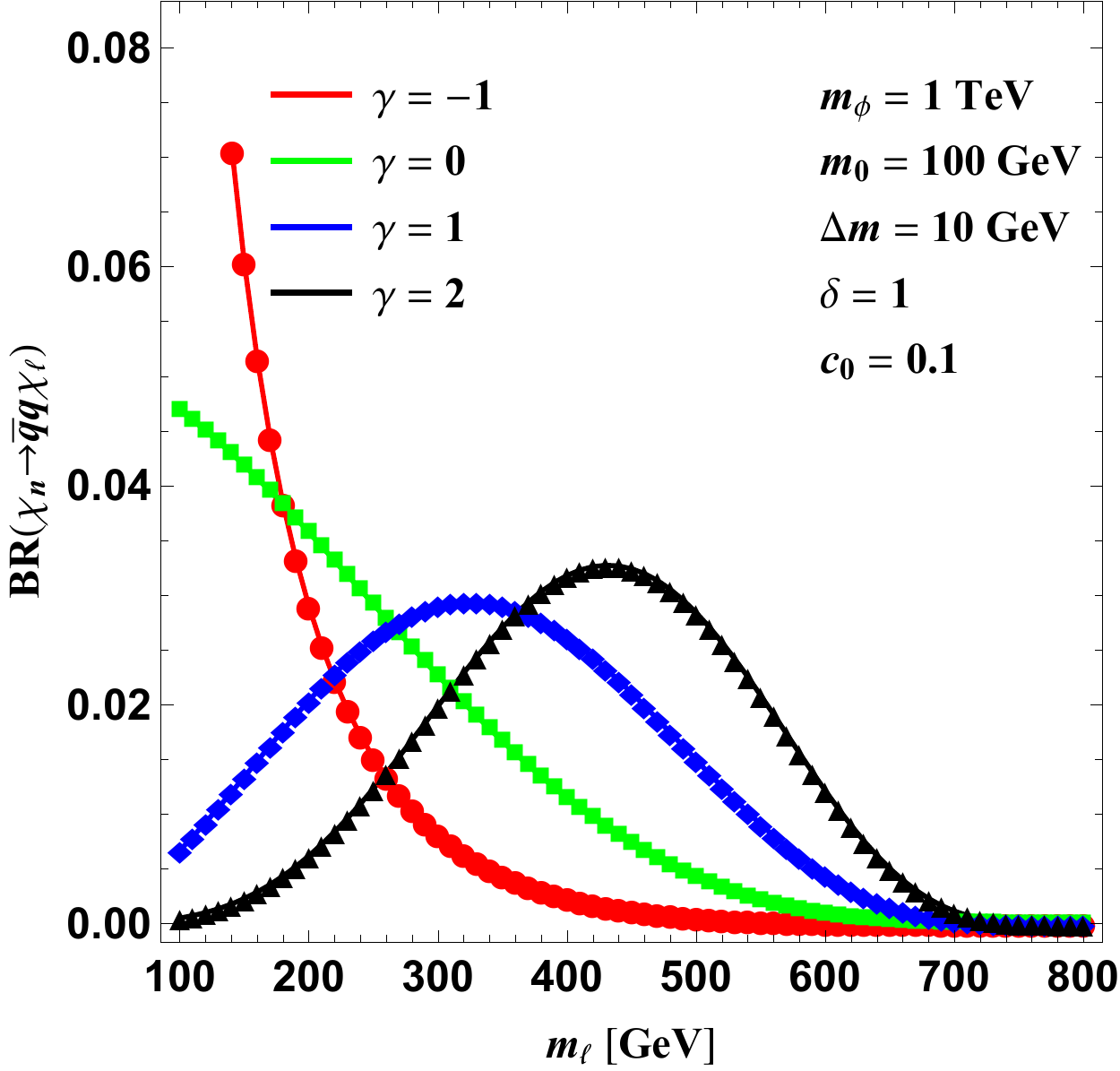}
  \end{center}
\caption{The branching fraction $\mathrm{BR}_{n\ell}$ for a decaying ensemble constituent 
  $\chi_n$, plotted as a function of the daughter mass $m_\ell$ for different values of the 
  scaling exponent $\gamma$.  The results shown here correspond to the parameter values $n=70$, 
  $m_\phi = 1$~TeV, $m_0 = 100$~GeV, $\Delta m = 10$~GeV, $\delta = 1$, and $c_{0}=0.1$ --- a 
  choice of parameters for which the mass of the parent is $m_n=800$~GeV.~  We see that when 
  $\gamma$ is large (even if only moderately so), the couplings $c_\ell$ which increase with 
  $\ell$ are able to partially overcome the increasingly severe phase-space suppressions that 
  also arise for larger $\ell$, allowing the parent $\chi_n$ to decay preferentially to daughters 
  $\chi_\ell$ with intermediate values of $\ell$.  This phenomenon underpins the existence of 
  decay chains with many intermediate steps, allowing such long decay chains to dominate
  amongst the set of all possible decay chains that emerge from a given parent $\chi_\ell$.   
\label{fig:BRScalingDiagram800GeV}}
\end{figure}

In the regime in which $\gamma \lesssim 0$, the mediator $\phi$ and all 
of the $\chi_n$ decay preferentially to $\chi_\ell$ with relatively small values of $\ell$.  Thus, 
for these parameters, the corresponding decay chains typically involve only one or a few steps and do 
not give rise to large multiplicities of jets.  By contrast, in the opposite regime in which $\gamma$ 
is positive and sufficiently large that the enhancement in $c_{\ell}^2$ with increasing $\ell$ 
overcomes the phase-space suppression, decays to $\chi_\ell$ with intermediate values of $\ell$ are 
preferred.  Within this regime, long decay chains can develop and events involving large
numbers of hadronic jets naturally arise.

In Fig.~\ref{fig:BRScalingDiagram800GeV}, we plot $\mathrm{BR}_{n \ell}$ as a function of the 
daughter-particle mass $m_\ell$ for several different choices of $\gamma$, holding $n$ fixed.
For these plots we have chosen the illustrative values $m_\phi = 1$~TeV, $m_0 = 100$~GeV, 
$\Delta m = 10$~GeV, $\delta = 1$, and $c_{0}=0.1$.  We have also chosen $n=70$ for the parent,
implying a parent mass $m_n = 800$~GeV.~  On the one hand, we observe from 
Fig.~\ref{fig:BRScalingDiagram800GeV} that $\mathrm{BR}_{n \ell}$ indeed decreases monotonically 
with $\ell$ for negative $\gamma$ --- and indeed even for $\gamma = 0$ --- as expected.  On the 
other hand, we also observe that decays to final states with $\ell > 0$ are strongly preferred even 
for $\gamma = 1$.  Thus, even a moderate positive value of $\gamma$ is sufficient to ensure that 
decay cascades with multiple steps will be commonplace.  Indeed, the shapes of the curves in 
Fig.~\ref{fig:BRScalingDiagram800GeV} do not depend sensitively on the chosen values of $\Delta m$ 
or $\delta$ as long as the  number of constituents $\chi_\ell$ lighter than $\chi_n$ is sufficiently 
large.  This is because for fixed $m_\phi$ and $m_n$, the branching fraction $\mathrm{BR}_{n \ell}$ 
can be viewed as a function of the single variable $r_{n\ell}$.  Thus, while changing $\Delta m$ 
and $\delta$ changes the values of $r_{n\ell}$ at which this function is evaluated, it has no 
effect on form of the function itself.

Given our results for the relevant branching fractions, we now have the ingredients
with which to calculate the probabilities associated with particular {\it sequences}\/ of 
decays --- \ie, particular decay chains --- in our model.  For simplicity, let us focus on the 
regime in which all $\chi_n$ with $n>0$ decay promptly within the detector.  Under this assumption, 
each decay chain precipitated by the production of a given ensemble constituent effectively 
terminates only when 
$\chi_0$ (the lightest element within the ensemble) is produced.  Within this regime, then, the 
probability $\mathcal{\hat P}(S)$ that such a decay chain will have precisely $S$ steps after the 
initial production of an ensemble constituent (\ie, the probability that our decay chain proceeds
according to a schematic of the form $\chi_{n_0} \to \chi_{n_1}\to ...\to \chi_{n_{S-1}} \to \chi_0$)
is given by
\beq
   {\cal \hat P}(S)~=~ \!\!\!\!\!
    \sum_{n_0,n_1,...,n_{S-1}=0}^{N-1} \!\!\!\!\!
     {\rm BR}_{n_0}^{\rm (prod)}
     {\rm BR}_{n_0,n_1} \cdots {\rm BR}_{n_{S-1},0}~
     \label{eq:prob}
\eeq
for $0\leq S\leq N-1$, where we of course understand that ${\rm BR}_{ij}=0$ for all $j\geq i$
and where the initial factor ${\rm BR}_{n_0}^{\rm (prod)}$ is the relative probability that the 
specific ensemble constituent $\chi_{n_0}$ is originally produced.  This last factor depends on 
the production process, with ${\rm BR}^{\rm (prod)}_{n_0} ={\rm BR}_{\phi n_0}$ in the case of 
indirect production through the mediator $\phi$ and ${\rm BR}^{\rm (prod)}_{n_0}=1$ for direct 
$\chi_{n_0}$ production.

This result then allows us to calculate the probabilities $P(\Njet)$ that each of the processes in 
Figs.~\ref{fig:ChiChiDiagram}--\ref{fig:PhiPhiDiagram} yields precisely $\Njet$ jets at the parton 
level.  First, we observe that each of these processes directly or indirectly gives rise to
two ensemble constituents $\chi_n$ and $\overline{\chi}_m$.  While producing these ensemble 
constituents, each process also produces a certain number $\zeta$ of parton-level jets;
indeed $\zeta=0,1,2$ for the processes sketched in Figs.~\ref{fig:ChiChiDiagram}, 
\ref{fig:ChiPhiDiagram}, and \ref{fig:PhiPhiDiagram}, respectively.  Each of these two constituents 
then spawns a set of decay chains, with each step producing exactly two parton-level jets.  Thus, 
for each process in Figs.~\ref{fig:ChiChiDiagram}--\ref{fig:PhiPhiDiagram}, the corresponding 
probability $P(\Njet)$ that a single event will yield a specified total number $\Njet$ of parton-level 
jets (from either quarks or anti-quarks) is therefore given by
\begin{equation}
  P(\Njet) ~=~ \sum_{S_1 = 0}^{(\Njet-\zeta)/2 } 
    \mathcal{\hat P}(S_1) \, \mathcal{\hat P}(\Njet/2 - \zeta/2 - S_1 )~.
\end{equation} 
Of course, for each process $\Njet$ is restricted to the values $\zeta+n$ where 
$0\leq n\leq 4N-4$, $n\in 2\IZ$. 
 
\begin{figure}[t!]
  \begin{center}
	\includegraphics[width=8.3cm]{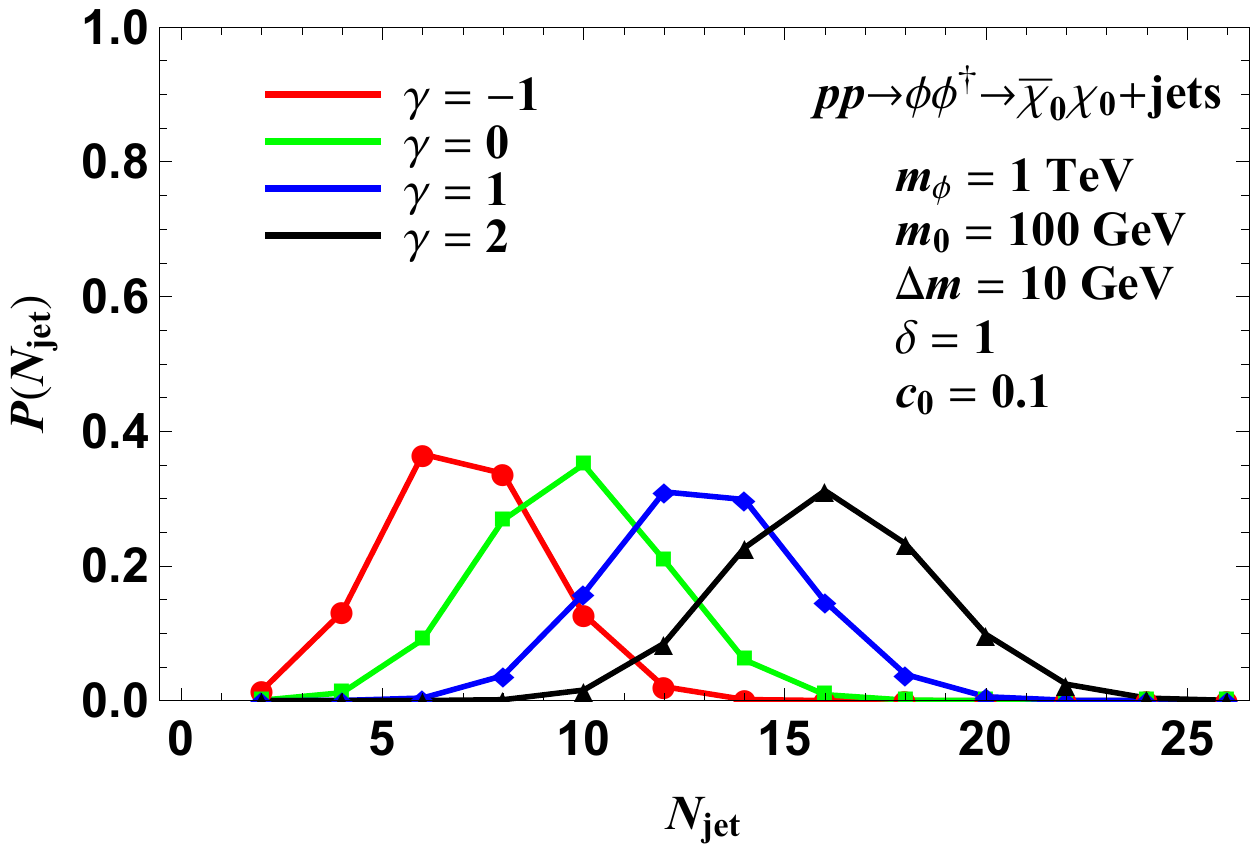}
  \end{center}
\caption{The probability $P(\Njet)$ for obtaining a total number $\Njet$ 
  of jets ({\it i.e.}\/, quarks or anti-quarks) at the parton level from the
  decay of a pair of mediator particles $\phi$ and $\phi^\dagger$ as in Fig.~\ref{fig:PhiPhiDiagram},
  plotted as a function of $\Njet$ for several different values of the scaling exponent $\gamma$.  
  The red, green, blue, and black curves correspond to the choices $\gamma =\{-1,0,1,2\}$ for 
  this parameter, respectively.  The remaining model parameters have been assigned the benchmark 
  values $m_\phi = 1$~TeV, $m_0 = 100$~GeV, $\Delta m = 10$~GeV, $\delta = 1$, and $c_0 = 0.1$. 
  The results shown in the figure indicate that large jet multiplicities can indeed arise within 
  our framework --- especially for large values of $\gamma$.    
  \label{fig:ProbJPlotgamma}}
\end{figure}

In Fig.~\ref{fig:ProbJPlotgamma}, we plot $P(\Njet)$ as a function of $\Njet$ for
several different choices of the scaling exponent $\gamma$.  For this figure we have again 
taken the illustrative values $m_\phi = 1$~TeV, $m_0 = 100$~GeV, $\Delta m = 10$~GeV, 
$\delta = 1$, and $c_0 = 0.1$, which together imply $N= 90$.  For concreteness we have also chosen 
$\zeta=2$, corresponding to the process sketched in Fig.~\ref{fig:PhiPhiDiagram} for which 
${\rm BR}^{\rm (prod)}_{n_0}={\rm Br}_{\phi,n_0}$.  For this choice of parameters, 
we see that the decay cascades initiated by parent-particle decays can indeed give rise to 
significant numbers of jets at the parton level.  Indeed, we observe from this figure that for 
$\gamma \gtrsim 1$, the majority of events in which a pair of mediator particles is produced have 
$\Njet \gtrsim 10$.  Similar results also emerge for the processes in Figs.~\ref{fig:ChiChiDiagram} 
and \ref{fig:ChiPhiDiagram}.
 
We conclude, then, that the example model described in Sect.~\ref{sec:model} is capable of giving 
rise to extended jet cascades at the parton level.  Indeed, the existence of this signature does not 
require any fine-tuning, and emerges as an intrinsic part of the phenomenology of the model.

\FloatBarrier
\subsection{Constraining the model parameter space \label{secondpart}}

Our analysis in Sect.~\ref{firstpart} focused on the general kinematic and combinatoric structure of
the decay chains that give rise to extended jet cascades in our model.  
However, there are a number of additional constraints which must also be addressed before we can 
claim that our model is actually capable of giving rise to signatures involving large jet 
multiplicities at a collider such as the LHC.~  Some of these additional constraints are fairly 
generic, and can be discussed even at the parton 
level.  Indeed, as we shall now demonstrate, satisfactorily addressing these concerns will enable us to 
place several important additional constraints on the parameter space of our model.  However, 
other constraints are more phenomenological and process-specific, having to do with existing LHC 
bounds on monojet and multi-jet signatures.  Discussion of these latter constraints will therefore 
be deferred to Sect.~\ref{sec:LHCseaches}.

As discussed in Sect.~\ref{sec:model}, our model is described by six parameters:  
$\lbrace m_0, \Delta m, \delta, m_\phi, c_0, \gamma\rbrace$.  The first three of these parameters 
together describe the entire mass spectrum $m_n$ of the ensemble constituents, and the fourth is 
nothing but the mass $m_\phi$ of the mediator $\phi$.  As we have seen, however, the all-important
branching fractions ${\rm BR}_{\phi n}$ and ${\rm BR}_{n\ell}$ depend on only the {\it ratios}\/ of 
these masses.  Likewise, the quantity $N$ which sets an upper limit on the number
of possible jets that can be produced (and which was defined in Sect.~\ref{sec:model} as 
the number of ensemble constituents which are kinematically accessible via the decays of $\phi$) 
also implicitly depends on these ratios.  Together, these considerations then govern the choices 
of mass  ratios in our system.

However, this still leaves an overall mass scale which we may take to be $m_\phi$ itself.
Likewise, we have not yet constrained the two parameters $c_0$ and $\gamma$ which together describe
the spectrum of couplings in our model through Eq.~(\ref{eq:CouplingScaling}).  Of course, we have 
already seen in Figs.~\ref{fig:BRScalingDiagram800GeV} and~\ref{fig:ProbJPlotgamma} that only when 
$\gamma$ is sufficiently positive and large do our decays preferentially proceed through sufficiently 
small steps that allow decay chains with sufficiently large numbers of steps to develop.  However, 
this still leaves $m_\phi$ and $c_0$ unconstrained.  Fortunately, there exist additional 
phenomenological constraints which will enable us to determine suitable ranges for these two 
remaining parameters as well.

\begin{figure}[t!]
\centering
  \includegraphics[width=8.3cm]{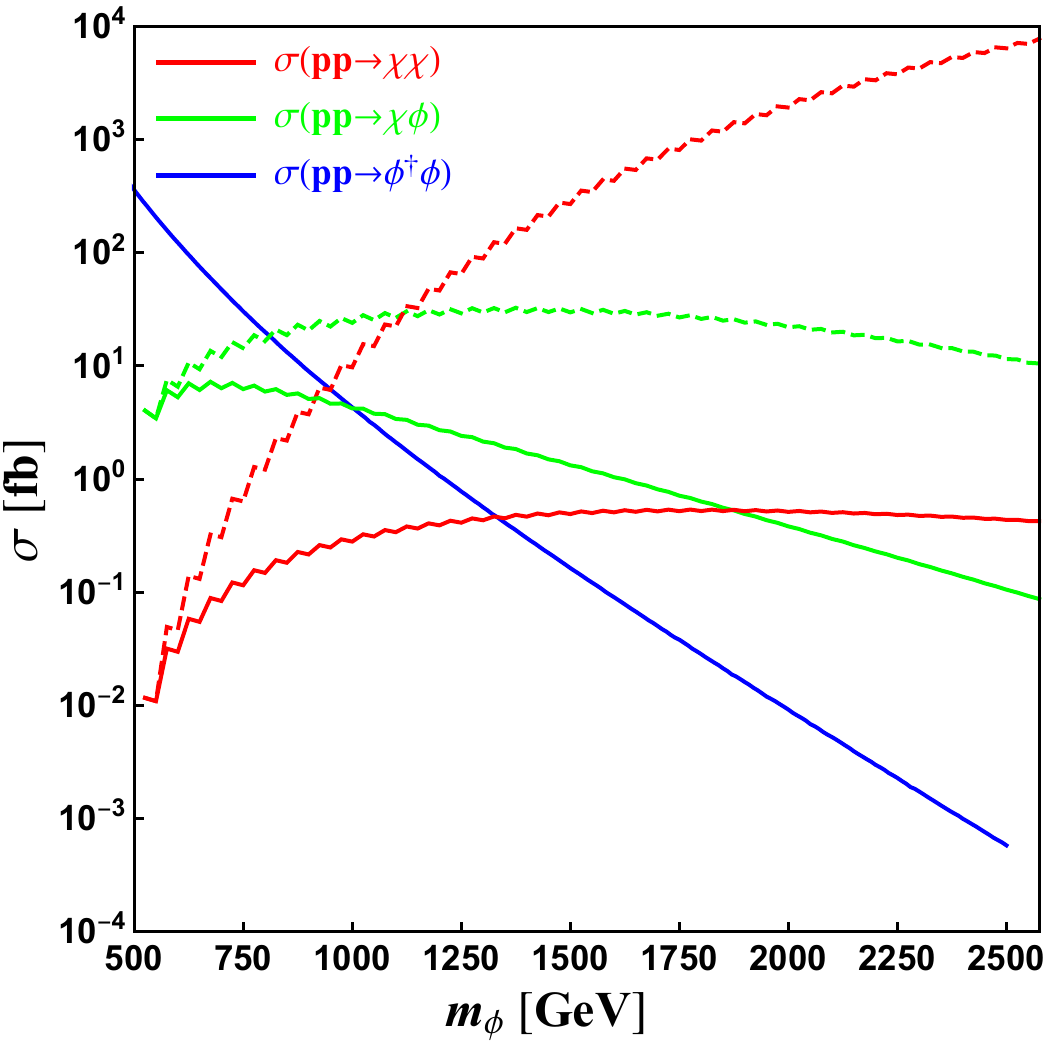}
\caption{The three production cross-sections $\sigma_{\chi\chi}$, $\sigma_{\phi\chi}$, and
   $\sigma_{\phi\phi}$ in Eq.~(\ref{eq:productionprocesses}), plotted as functions of $m_\phi$
   for  $\sqrt{s}=13$~TeV.~ The solid curves correspond to the parameter choices
   $m_0= 500$~GeV, $\Delta m= 50$~GeV, $c_0= 0.1$, and $\delta=\gamma=1$, while the dashed 
   curves correspond to the same parameter choices but with $\gamma=3$.  We see that 
   $\sigma_{\chi\chi}$ tends to dominate for large $m_\phi$, while $\sigma_{\phi\phi}$  
   tends to dominate for small $m_\phi$.
\label{fig:xs}}
\end{figure}

First, although we have demonstrated how extended mediator-induced decay cascades might potentially 
emerge from our model, we must also ensure that the overall {\it cross-sections}\/ for producing 
these cascades are sufficiently large that the resulting multi-jet signal could actually be detected 
over background.  While these cross-sections are certainly affected by the cascade probabilities 
discussed above, their overall magnitudes are set by the simpler cross-sections associated with the 
sub-processes for the production of the initial states that trigger these cascades.
For the diagrams sketched in Figs.~\ref{fig:ChiChiDiagram}--\ref{fig:PhiPhiDiagram},
these production cross-sections are respectively given by 
\beqn
 \sigma_{\chi\chi} ~&\equiv&~ \sum_{m,n=0}^{N-1} \sigma(pp\to \chi_m \overline{\chi}_n)~\nonumber\\ 
 \sigma_{\phi\chi} ~&\equiv&~ \sum_{m=0}^{N-1} \sigma(pp\to \phi \chi_m)~\nonumber\\ 
 \sigma_{\phi\phi} ~&\equiv&~ \sigma(pp\to \phi^\dagger \phi)~.
\label{eq:productionprocesses}
\eeqn
Calculating these cross-sections is relatively straightforward, and in Fig.~\ref{fig:xs} we 
display our results as functions of $m_\phi$ for a center-of-mass (CM) energy of 
$\sqrt{s}=13$~TeV.~  In particular, the solid curves correspond to the parameter choices 
$m_0= 500$~GeV, $\Delta m= 50$~GeV, $c_0 = 0.1$, and $\delta=1$ with $\gamma = 1$, while the 
dashed curves correspond to the same values of $m_0$, $\Delta m$, $c_0$, 
and $\delta$, but with $\gamma=3$.  We note that since $\sigma_{\phi\phi}$ has no dependence 
at leading order on the mass spectrum of the ensemble constituents (and therefore on the values 
of the parameters $m_0$, $\Delta m$, and $\gamma$), the corresponding curves for both of 
these parameter choices are identical.  We also note that the wiggles which appear in the 
curves for $\sigma_{\chi\chi}$ and $\sigma_{\chi\phi}$, especially at small $m_\phi$, are 
the consequence of threshold effects which arise due to the discrete changes in $N$ that 
occur as $m_\phi$ changes, in accordance with Eq.~(\ref{eq:Ndeff}).   
   
We observe from Fig.~\ref{fig:xs} that the cross-section for $\phi\phi$ pair-production 
dominates for small $m_\phi$, but falls rapidly from $\sigma_{\phi\phi}\sim 500$~fb 
to $\sigma_{\phi\phi} \sim 10^{-3}$~fb as the mass of the mediator increases from 
$m_\phi=500$~GeV to $m_\phi=2500$~GeV.~  By contrast, the cross-sections for the other two 
production processes either grow with $m_\phi$ or fall less sharply over the range of $m_\phi$ 
shown.  This is primarily a consequence of the corresponding increase in $N$, which in turn 
results in more individual production processes involving different $\chi_n$.  Since 
increasing $\gamma$ in turn increases the individual production cross-sections for the 
heavier $\chi_n$, both $\sigma_{\chi\chi}$ and $\sigma_{\phi\chi}$ are noticeably larger 
for $\gamma = 3$ than for $\gamma=1$.  We also note that across the entire range of $m_\phi$ 
shown, $\sigma_{\chi\chi}$ and $\sigma_{\phi\chi}$ are both larger than $0.01$~fb, indicating
that these processes could potentially lead to observable signals at the LHC.

We now turn to examine how general considerations involving the coupling 
structure of our model serve to constrain the coupling parameter $c_0$. 
Since all of the couplings $c_n$ in our model are proportional to $c_0$, we see that 
$c_0$ serves as an overall proportionality factor for both $\Gamma_\phi$ and 
$\Gamma_n$.  In particular, our results in Eqs.~(\ref{eq:Gammaphi}) and (\ref{eq:Gammanmeq}) 
imply that $\Gamma_\phi \propto c_0^2$ and $\Gamma_n \propto c_0^4$.
Fortunately, the value of $c_0$ is constrained by a number of theoretical consistency conditions 
and phenomenological constraints.  For example, given the perturbative treatment leading to
the results in Eqs.~(\ref{eq:Gammaphi}) and (\ref{eq:Gammanmeq}), self-consistency 
requires that we must impose the
perturbativity requirement that $c_n \lesssim 4\pi$ for all $0\leq n \leq N-1$.  
Given the general expression in Eq.~(\ref{eq:CouplingScaling}), we see that the value of $c_n$ 
generally increases as a function of $n$ for $\gamma > 0$ and decreases for $\gamma <0$.
For any combination of model parameters we must therefore demand that
\begin{equation}
  c_0 ~\lesssim~ 
  \begin{cases}
       4\pi \left[1 + \frac{\Delta m}{m_0}(N-1)^\delta \right]^{-\gamma} & {\rm for}~~ \gamma \geq 0 \cr
       4 \pi & {\rm for}~~ \gamma \leq 0~.
 \end{cases}
\label{eq:PerturbativityConstraint}
\end{equation} 
In addition, for cases in which the decay chains are initiated through the direct production of the 
mediator $\phi$, we are assuming that $\phi$ behaves like a physical particle rather than a broad 
resonance.  We must therefore also demand that $c_0$ be sufficiently small that 
$\Gamma_\phi \ll m_\phi$, which in turn requires
\begin{equation}
  c_0 ~\ll~ 4\sqrt{\pi} \left[\sum_{n=0}^{N-1} \left(\frac{m_n}{m_0}\right)^\gamma
    \left(1 - \frac{m_n^2}{m_\phi^2}\right)^2  \right]^{-1/2}
    \label{eq:c0phiwidth}
\end{equation}
This latter constraint can occasionally surpass the one in Eq.~(\ref{eq:PerturbativityConstraint}).
For example, for $\gamma=0$ we learn from Eq.~(\ref{eq:PerturbativityConstraint}) that 
$c_0\lsim 4\pi$, yet even in such cases $\Gamma_\phi$ can occasionally exceed $m_\phi$, even with 
only a few ensemble constituents.  

In addition to these criteria for theoretical consistency, 
there are also a number of further constraints which we shall take into account in defining our 
region of interest within the full parameter space of our model.  We emphasize that these 
are not necessarily inviolable constraints on the model, but rather conditions which we shall 
impose either for sake of clarity in simplifying our analysis or in order to restrict our focus 
within the model parameter space to regions in which long decay chains arise.

For example, in order for a decaying particle ensemble to give rise to observable signatures of 
mediator-induced decay cascades at the LHC, many of the $\chi_n$ constituents must of course
decay promptly within the detector. 
In general, the decay length $L_n$ of $\chi_n$ in the detector frame is given by 
$L_n \equiv \beta\gamma c \tau_n$, where $\tau_n = \Gamma_n^{-1}$ is the proper lifetime of $\chi_n$ 
and where $\beta = v/c$ and $\gamma = (1-\beta^2)^{-1/2}$ are the usual relativistic factors.  
Since we shall generally be interested in decay chains with many steps --- chains in which the 
dominant individual decays produce daughters that are not overwhelmingly lighter than their parents --- 
none of the ensemble constituents will be excessively boosted upon production.  We can therefore 
treat the relativistic factor $\beta \gamma$ as a mere ${\cal O}(1)$ numerical coefficient in order 
to obtain an order-of-magnitude estimate of the bound.  This is particularly convenient since these 
factors generally depend on the detailed structure of the decay chain and therefore differ from one 
event to the next.  We will therefore estimate the characteristic length scale at which a given 
ensemble constituent decays as $c \tau_n$.  Broadly speaking, if $c\tau_n \gg 1$~cm, a particle of 
species $\chi_n$ will typically appear as either a displaced vertex or as $\met$ at the LHC.~  
By contrast, if $c\tau_n \lesssim \mathcal{O}(1\mathrm{~cm})$, such a particle will tend to decay 
promptly within the detector.  It is these latter decays which are our focus.

\begin{figure}[b!]
\centering
  \includegraphics[width=7.7cm]{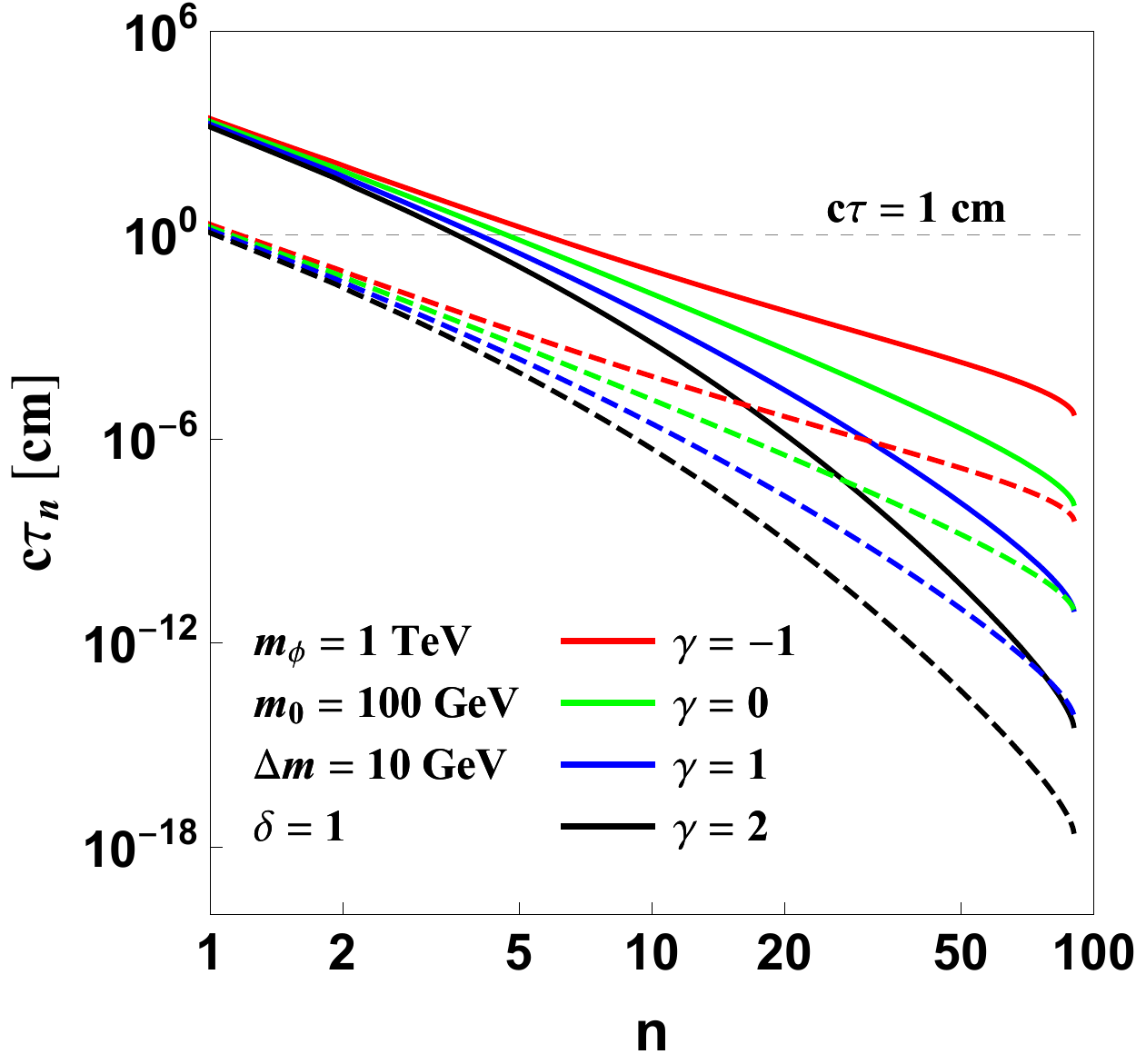}
\caption{The decay lengths $c\tau_n$ of the ensemble constituents $\chi_n$, plotted as functions 
  of $n$ for several different choices of model parameters.  For all curves shown in this plot we 
  have taken $m_\phi = 1$~TeV, $m_0 = 100$~GeV, $\Delta m = 10$~GeV, and $\delta = 1$.  The red, 
  green, blue, and black curves correspond to the parameter choices $\gamma = \{-1,0,1,2\}$, 
  respectively.  Likewise, the solid curves correspond to the choice $c_0 = 0.02$, while the dashed 
  curves correspond to the choice $c_0 = 0.1$.  In general we see that increasing $c_0$ in this way 
  has the effect of decreasing the decay lengths of our ensemble states and ultimately ensuring that 
  all of the ensemble constituents decay within the detector.
  \label{fig:varParamsGamma}}
\end{figure}

In Fig.~\ref{fig:varParamsGamma}, we plot the length scales $c\tau_n$ as functions of $n$ for 
several different choices of model parameters.  The red, green, blue, and black curves correspond to 
the parameter choices $\gamma = \{-1,0,1,2\}$, respectively.  The solid curves correspond to the 
choice $c_0 = 0.02$, while the dashed curves correspond to the choice $c_0 = 0.1$.  The values 
of the remaining model parameters are taken to be $m_\phi = 1$~TeV, $m_0 = 100$~GeV, 
$\Delta m = 10$~GeV, and $\delta = 1$ for all curves shown.  We emphasize that the perturbativity 
criterion in Eq.~(\ref{eq:PerturbativityConstraint}) is satisfied for all curves shown.  Note that 
for the parameters shown, the decay lengths tend to decrease as functions of $n$.  This remains 
true even if $\gamma = -1$, indicating that the total phase space available for the decays of 
$\chi_n$ increases with $n$ more rapidly than the associated couplings $c_n$ might decrease.   
For $c_0 = 0.02$, we see from Fig.~\ref{fig:varParamsGamma} that a significant number of the 
ensemble constituents have $c \tau_n \gg {\cal O}(1~\mathrm{cm})$ and therefore do not decay 
promptly within the detector.   Indeed, depending on the amount by which $c\tau_n$ exceeds 
${\cal O}(1~{\rm cm})$, these $\chi_n$ would either decay a measurable distance away from the 
primary vertex (thereby giving rise to a displaced vertex), or else appear in the detector as 
$\met$.  By contrast, for $c_0 = 0.1$, we see that all $\chi_n$ with $n>0$ in the ensemble 
have $c \tau_n \lesssim \mathcal{O}(1~\mathrm{cm})$.  

In general, long decay chains can certainly arise even in cases for which the lighter $\chi_n$ 
have values of $c \tau_n$ exceeding  ${\cal O}(1~\mathrm{cm})$.  In such cases the decays of 
relevance for our purposes would simply be the decays of the heavier constituents, with the decays 
of the lighter constituents subsequently occurring either with displaced vertices or completely 
outside the detector.  Indeed, such situations could potentially give rise to many interesting 
signatures which will be discussed further in Sect.~\ref{sec:conclusions}.~  However, for 
simplicity in what follows, we shall henceforth restrict our attention to the region of
parameter space within which 
\beq
   c \tau_n ~\lesssim~ \mathcal{O}(1\mathrm{~cm})~~~{\rm for~all}~~ n>0~.
  \label{insidedetector}
\eeq
In such cases, all possible decays of our ensemble constituents will
occur within the detector, thereby allowing us to regard our decay chains
as terminating only when the collider-stable ensemble ``ground state'' $\chi_0$ is reached.

\begin{figure}[tbh]
\centering
  \includegraphics[width=3.0 in]{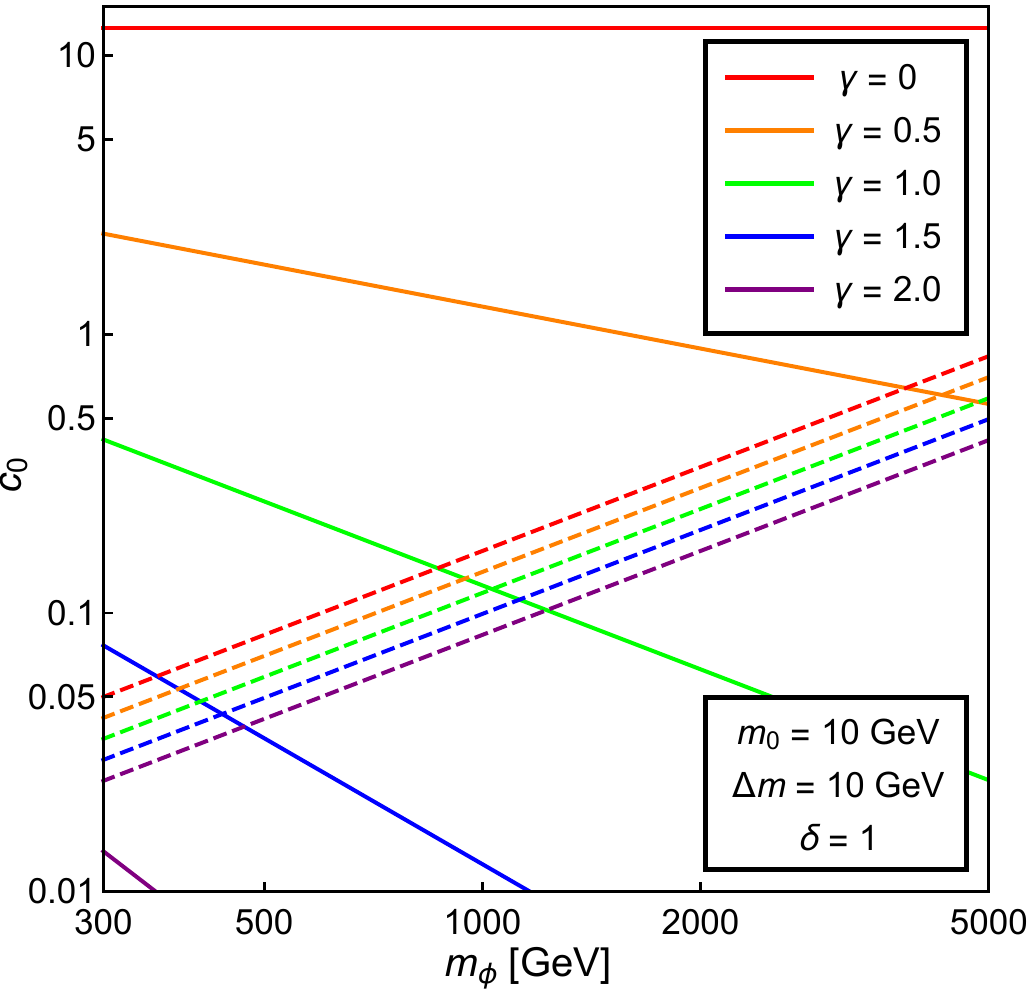}
  ~~~~
  \includegraphics[width=3.0 in]{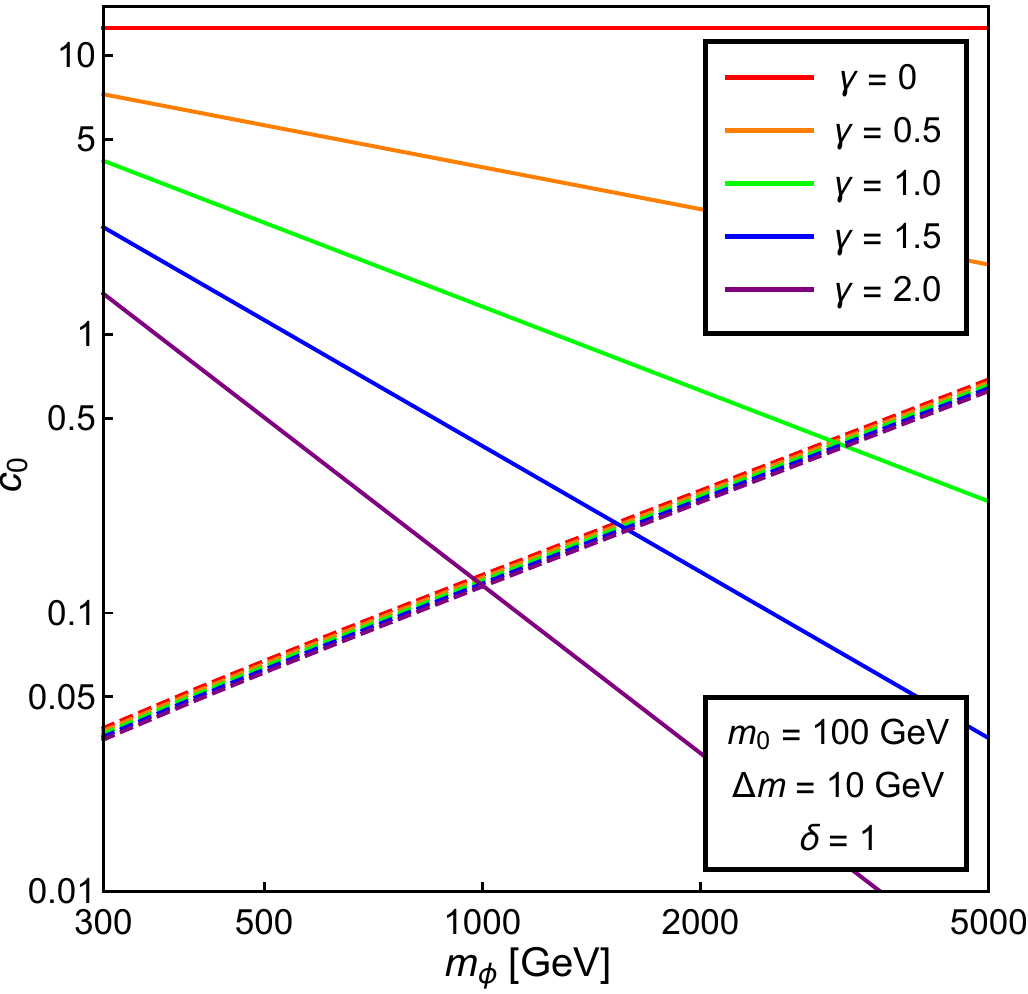}
\caption{Allowed values of $c_0$ in our model, plotted as functions of $m_\phi$ for different 
   values of $\gamma$.  The results in both panels assume $\Delta m=10$~GeV and $\delta=1$, while 
   $m_0= 10$~GeV (top  panel) or $m_0= 100$~GeV (bottom panel).  In each panel, the solid curves 
  indicate the upper bounds on $c_0$ arising from the perturbativity constraint in 
  Eq.~(\protect\ref{eq:PerturbativityConstraint}), while the dashed curves indicate the lower bounds 
  on $c_0$ arising from the prompt-decay constraint in Eq.~(\protect\ref{insidedetector}).
  In general we see that there exists an ample allowed range for $c_0$ within which both constraints 
  can be satisfied simultaneously, but this range becomes increasingly narrow as $m_\phi$ or 
  $\gamma$ becomes large or as $m_0$ becomes small.
  \label{fig:c0bounds}}
\end{figure}

Since $\tau_n \propto c_0^{-4}$, requiring that our ensemble constituents satisfy the criterion 
in Eq.~(\ref{insidedetector}) is tantamount to imposing a {\it lower}\/ bound on $c_0$ for any 
particular assignment of the remaining parameters which characterize our example model.  Indeed, 
as illustrated in Fig~\ref{fig:varParamsGamma}, reducing $c_0$ below this bound only inhibits the 
decay rates of our ensemble constituents to a point beyond which some of the lighter ensemble 
constituents will begin to exhibit displaced vertices or decay outside the detector.  However, since 
$c_0$ is also bounded from above by the perturbativity constraint in 
Eq.~(\ref{eq:PerturbativityConstraint}) and/or by our requirement that $\Gamma_\phi\ll m_\phi$, 
we see that there is a tension between these two groups of constraints.

In Fig.~\ref{fig:c0bounds}, we illustrate how the competition between the perturbativity constraint 
and the prompt-decay constraint play out within the parameter space of our model.  The solid curves
appearing within each panel of this figure represent the {\it upper}\/ bounds on $c_0$ arising from 
the constraint in Eqs.~(\ref{eq:PerturbativityConstraint}) and~(\ref{eq:c0phiwidth}), plotted as 
functions of $m_\phi$ for a variety of different values of $\gamma$.  By contrast, the dashed lines 
represent the {\it lower}\/ bounds on $c_0$ arising from our prompt-decay criterion in
Eq.~(\ref{insidedetector}).  While the contours in both panels in Fig.~\ref{fig:c0bounds}
correspond to $\Delta m= 10$~GeV and $\delta=1$, those in the top panel correspond to the choice 
$m_0 = 10$~GeV while those in the bottom  panel correspond to the choice $m_0 = 100$~GeV.~ 

As is evident from Fig.~\ref{fig:c0bounds}, there are indeed regions of parameter space within which 
both the perturbativity constraint and the prompt-decay condition can be simultaneously satisfied.  
Nevertheless, it is also evident from this figure that as $\gamma$ increases, a significant tension 
rapidly develops between these two bounds.  As we have already seen, the regions of parameter space 
within which $\gamma \gtrsim 1$ turn out to be the regions in which extended mediator-induced 
decay cascades develop.  As a result, this tension will ultimately have important consequences for 
our model.  

It is also relatively straightforward to understand the differences between the top and bottom panels 
of Fig.~\ref{fig:c0bounds}.  In general, for $\gamma \geq 0$ the perturbativity constraint in
Eq.~(\ref{eq:PerturbativityConstraint}) depends on the properties of $\chi_{N-1}$.  By contrast,
the prompt-decay condition in Eq.~(\ref{insidedetector}) depends on the properties of $\chi_1$.  
Given the functional form for $c_n$ in Eq.~(\ref{eq:CouplingScaling}), we see that $c_1$ is 
essentially insensitive to $\gamma$ in the $\Delta m \ll m_0$ regime, as indicated in the bottom 
panel of Fig.~\ref{fig:c0bounds}.  Likewise, the perturbativity bound becomes increasingly
sensitive to $\gamma$ as the ratio $m_{N-1}/m_0$ increases. 

For all of these reasons, we shall limit our attention in this paper to regions of parameter space
in which $\gamma\gsim 1$, $c_0=0.1$.  Indeed, as we have seen, these are the regions in which 
the processes illustrated in Figs.~\ref{fig:ChiChiDiagram}--\ref{fig:PhiPhiDiagram} can give rise 
to observable signatures involving relatively large numbers of jets at the parton level.


\FloatBarrier
\section{From Parton Level to Detector Level:~ When You're a Jet, Are You a Jet All the Way?\label{sec:expobs}}


While it is certainly instructive to examine the collider phenomenology of our model 
at the parton level, what ultimately matters, of course, are the signatures that can actually 
be observed at the 
detector level.  Indeed, not all of the parton-level ``jets'' produced from mediator-induced 
decay cascades at the parton level ultimately translate to individual reconstructed jets at 
the detector level.  Moreover effects associated with initial-state radiation (ISR), final-state
radiation (FSR), and parton-showering can give rise to additional jets at the detector level.
Thus, it is critical that we investigate how the parton-level results we have 
derived in Sect.~\ref{sec:pheno} are modified by these considerations at the detector level.

Toward this end, our analysis shall proceed as follows.  For any given choice of model 
parameters, we generate signal events for the initial 
pair-production processes $pp\rightarrow \phi\chi_m$, $pp\rightarrow \chi_m\bar{\chi}_n$, 
and $pp\rightarrow \phi^\dagger\phi$ at the $\sqrt{s}=13$~TeV LHC using the 
\texttt{MG5\_aMC@NLO}~\cite{Alwall:2014hca} code package.  We then evaluate the 
cross-sections for these processes using this same code package.
Due to the complexity of the decay chains which arise in our model, we treat 
the final-state particles produced during each step of the chain as being strictly 
on shell and simulate the decay kinematics using our own Monte-Carlo code.
We have confirmed that the kinematic distributions obtained using our decay code
agree well with those obtained from a full implementation of our model 
in \texttt{MG5\_aMC@NLO} in cases in which the decay chains are short and such a 
comparison is feasible.  The resulting set of three-momenta for the final-state 
particles in each event was then passed to \texttt{Pythia~8}~\cite{Sjostrand:2007gs} 
for parton-showering and hadronization.  Detector effects were simulated using 
\texttt{Delphes~3}~\cite{deFavereau:2013fsa}.  Jets were reconstructed 
in \texttt{FastJet}~\cite{Cacciari:2011ma} using the anti-$k_T$ 
clustering algorithm~\cite{Cacciari:2008gp} with a jet-radius parameter $R = 0.4$.

This procedure has the practical benefit of allowing us to examine the kinematics of long 
decay chains.  However, it is important to note that this procedure
neglects certain considerations which
can slightly modify the kinematics of the decay cascades and have $\mathcal{O}(1)$ 
effects on the cross-sections for the relevant final states.  
First, our procedure neglects the interference between two distinct contributions to the 
overall amplitude for the process $pp\rightarrow \chi_m\bar{\chi}_n + j$, the first coming from
$pp\rightarrow \phi\chi_m$ production followed by the decay $\phi\rightarrow\bar\chi_n j$ of 
the on-shell $\phi$ particle, and the second coming from processes similar to 
$pp\rightarrow \chi_m\bar{\chi}_n$, but in which an additional quark or gluon is produced 
as initial-state radiation or radiated off the internal $\phi$ line.
However, since we find that the former contribution vastly dominates the 
latter, the effect of neglecting these interference effects is not expected to be
significant.  Second, our procedure does not employ any jet-matching scheme\footnote{The phrase 
``jet-matching'' here refers to the set of computational techniques involved in accurately 
interfacing between matrix-element generators and showering algorithms in collider simulations.    
This is not to be confused with the parton-jet matching which is performed during jet 
reconstruction at the detector level.} in order to 
correct for double-counting in regions of phase space populated both by  
matrix-element-generation and parton-showering algorithms.  Since the event-selection 
criteria we impose in our detector-level analysis involve significant threshold cuts   
on the $p_T$ values of the relevant jets, this effect is not expected to have a significant 
impact on our results.    
Third, our procedure also ignores the possibility 
that any  $\chi_n$ which appear in decay chains or any of the mediators produced 
by the processes $pp\rightarrow \phi\chi_m$ or $pp\rightarrow \phi^\dagger\phi$ could be off 
shell.  Once again, the impact on our results is not expected to be significant.

We begin by examining several experimental observables which are potentially useful 
for discriminating between signal and SM backgrounds.  Clearly, the most distinctive
feature of these extended mediator-induced decay cascades is the sheer multiplicity of 
``jets'' at the parton level.  Thus, given limited statistics, observables which 
characterize the overall properties of the event as a whole are likely to provide more 
distinguishing power than  the observables which involve particular combinations of 
the momenta of individual jets in the event, due to the combinatorial issues 
associated with the latter.  We therefore focus primarily on the former class of
observables in what follows.  These observables include $\Njet$ and $\met$, the
distributions of the magnitude $p_{T_j}$ of the transverse 
momentum of all jets in the event, and the scalar sum
\begin{equation}
  H_T ~=~ \sum_{j=1}^{\Njet} p_{T_j}~.
  \label{eq:HT}
\end{equation}

In order to assess the extent to which showering, hadronization, and detector effects modify 
the distributions of $p_{T_j}$, $\Njet$, $\met$, and $H_T$, it is useful to compare the 
parton-level distributions of these observables  to the corresponding detector-level distributions.     
In constructing the parton-level distributions of all of these collider observables, we consider 
each quark and anti-quark  in the final state to be a ``jet'', regardless of its proximity in 
$(\eta_j,\phi_j)$-space to any other such ``jets'' in the event, where $\eta_j$ and $\phi_j$ 
respectively denote the pseudorapidity and azimuthal angle of a given jet.  Moreover, we 
impose no cuts on either $p_{T_j}$ or $\eta_j$.  By contrast, in constructing the 
detector-level distribution of $p_{T_j}$, we require that every jet in a given event satisfy 
$p_{T_j} > 20$~GeV and $|\eta_j| < 5$.  Furthermore, in order to be counted as a jet at
the detector level, a would-be jet must be separated from every other, more energetic jet in the 
event by a distance $\Delta R_{jj} \equiv \sqrt{(\Delta\eta_{jj})^2 + (\Delta\phi_{jj})^2} > 0.4$ 
in $(\eta_j,\phi_j)$-space.

\begin{figure*}[b]
\centering
  \includegraphics[width=5.5cm]{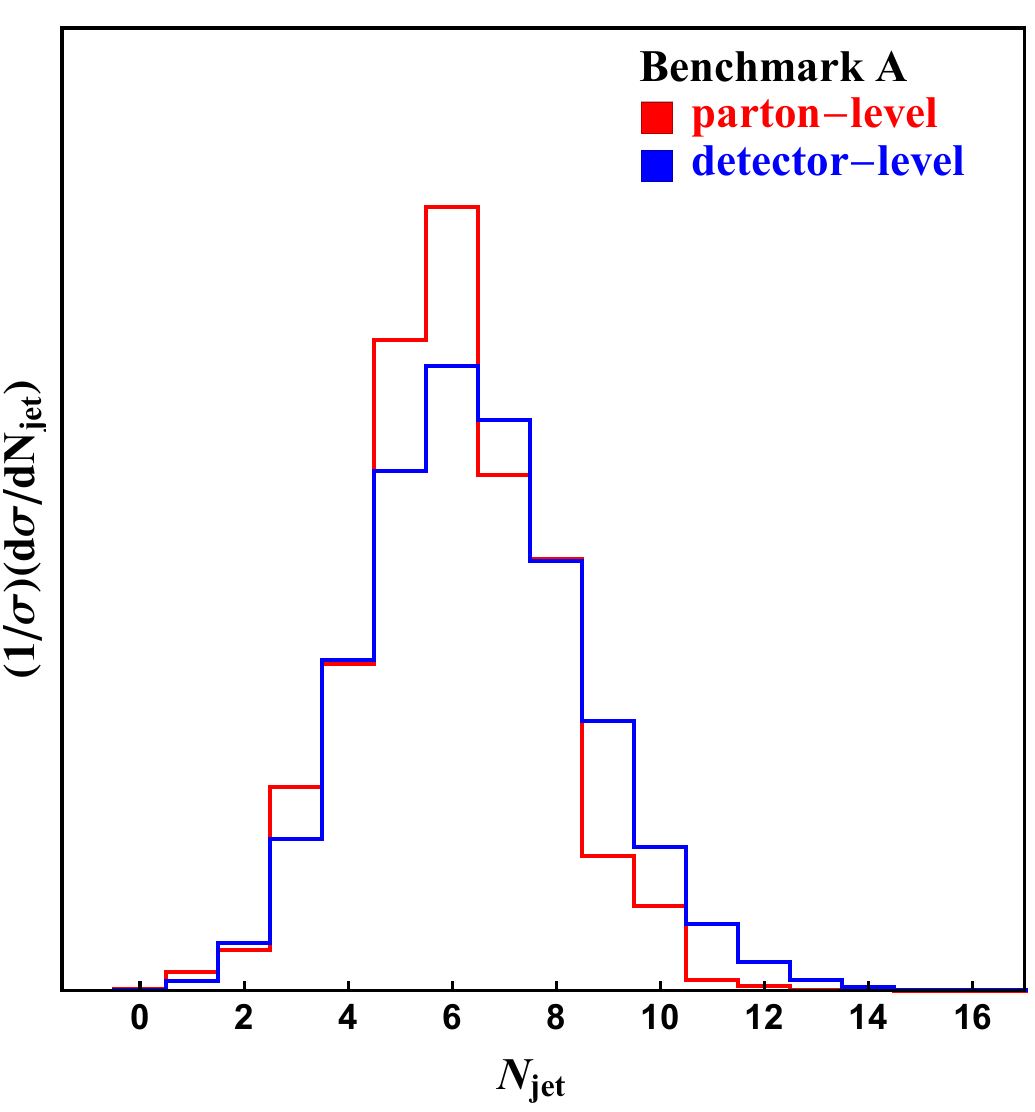}~~
  \includegraphics[width=5.5cm]{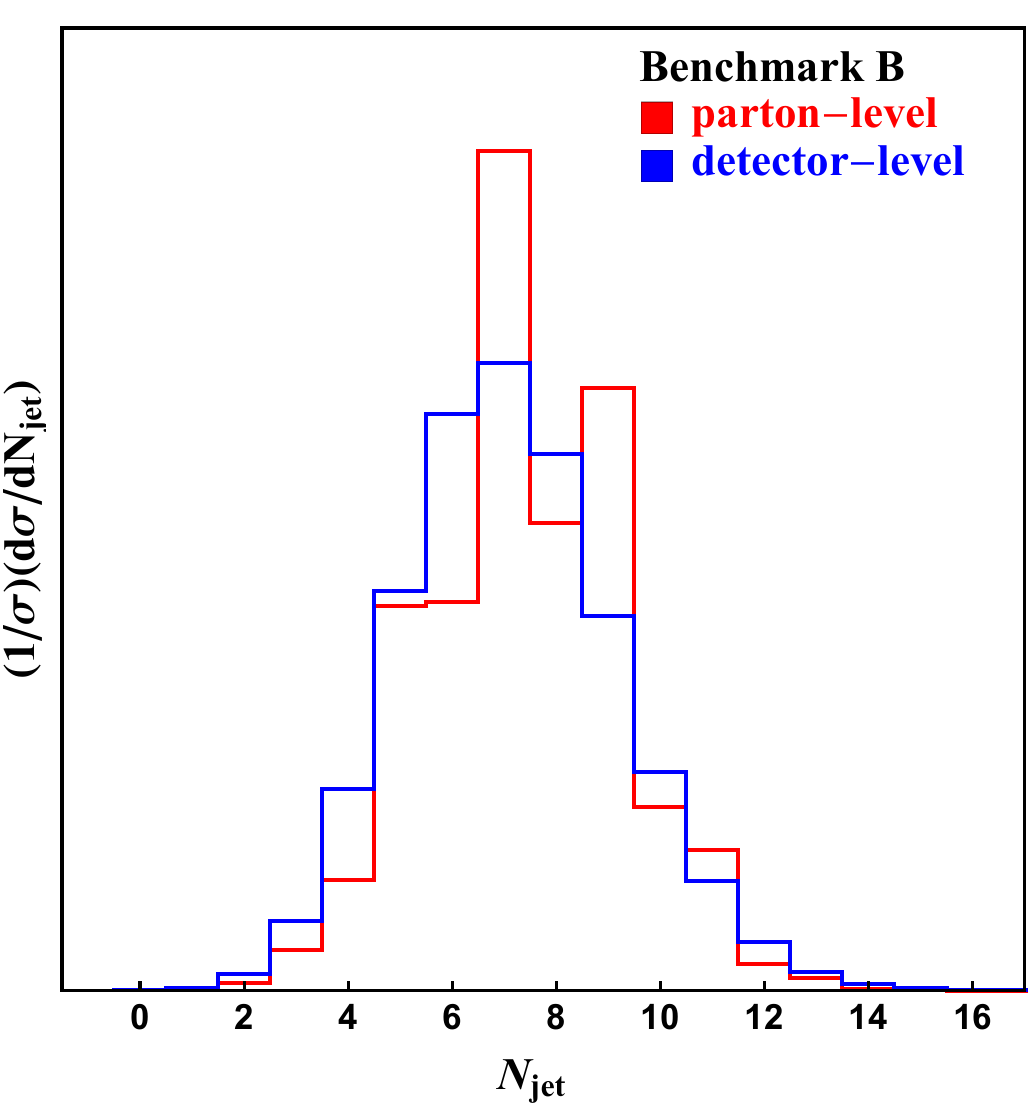}~~
  \includegraphics[width=5.5cm]{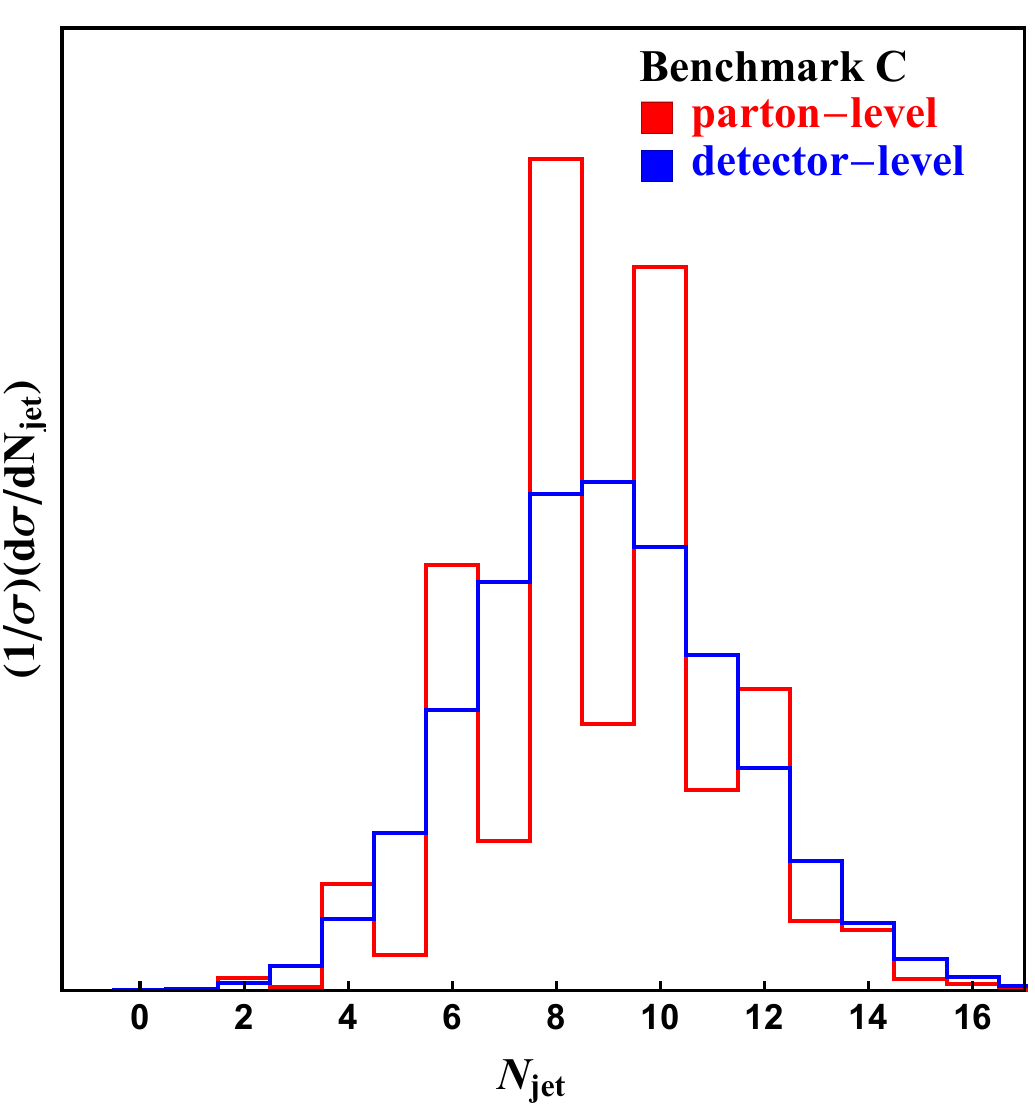}
\caption{Normalized $\Njet$ distributions for Benchmarks~A (left panel), B (middle panel), 
  and~C (right panel).  The red histogram in each panel shows the distribution obtained at the 
  parton level (with quarks, anti-quarks, and gluons considered to be ``jets''), while the 
  blue histogram shows the corresponding distribution at the detector level.
  \label{fig:num} }
\end{figure*}
 
For purposes of illustration, we identify three representative benchmark points within the 
parameter space of this model for which these criteria discussed in Sec.~\ref{secondpart} are 
satisfied, but for which different classes of production processes dominate the event rate in the 
multi-jet channel at large $\Njet$.  The parameter choices associated with these benchmarks are 
provided in Table~\ref{tab:BenchmarkPts}.~   Benchmark~A is representative of the regime in which 
both $pp\rightarrow \phi^\dagger\phi$ and $pp \rightarrow \phi\chi_m$ provide significant 
contributions to the event rate in the multi-jet channel at large $\Njet$, with these two 
processes contributing at roughly the same order.  Benchmark~B is representative of the regime in 
which $pp \rightarrow \phi\chi_m$ dominates the event rate, while Benchmark~C is representative of 
the regime in which $pp\rightarrow\chi_m\bar{\chi}_n$ dominates.

In Fig.~\ref{fig:num}, we show the normalized distributions of $\Njet$ obtained for Benchmarks~A 
(left panel), B (middle panel), and~C (right panel).  The distributions shown include the 
individual contributions from $pp\rightarrow\phi\chi_m$, $pp\rightarrow\chi_m\bar{\chi}_n$, 
and $pp\rightarrow\phi^\dagger\phi$, each weighted by the cross-section for the 
corresponding process.  The red histogram in each panel shows the distribution obtained at the 
parton level (with quarks, anti-quarks, and gluons considered to be ``jets''), while the blue 
histogram shows the corresponding distribution at the detector level.       

\begin{figure*}[t]
\centering
  \includegraphics[width=5.5cm]{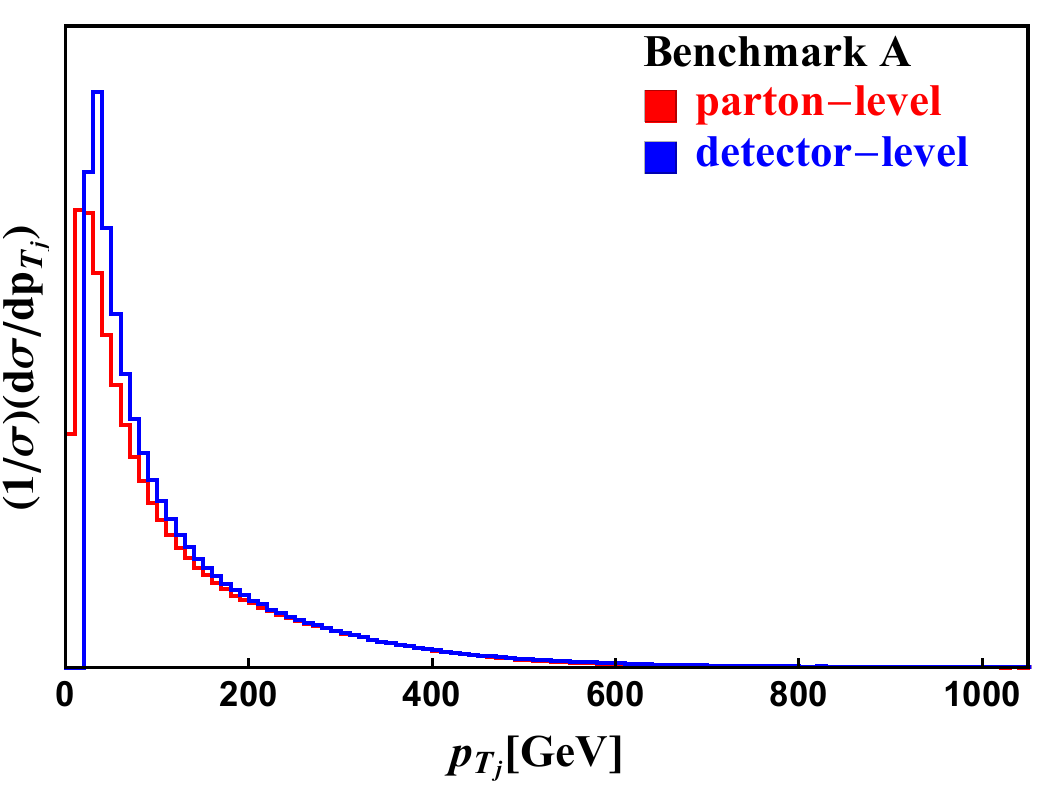}~~
  \includegraphics[width=5.5cm]{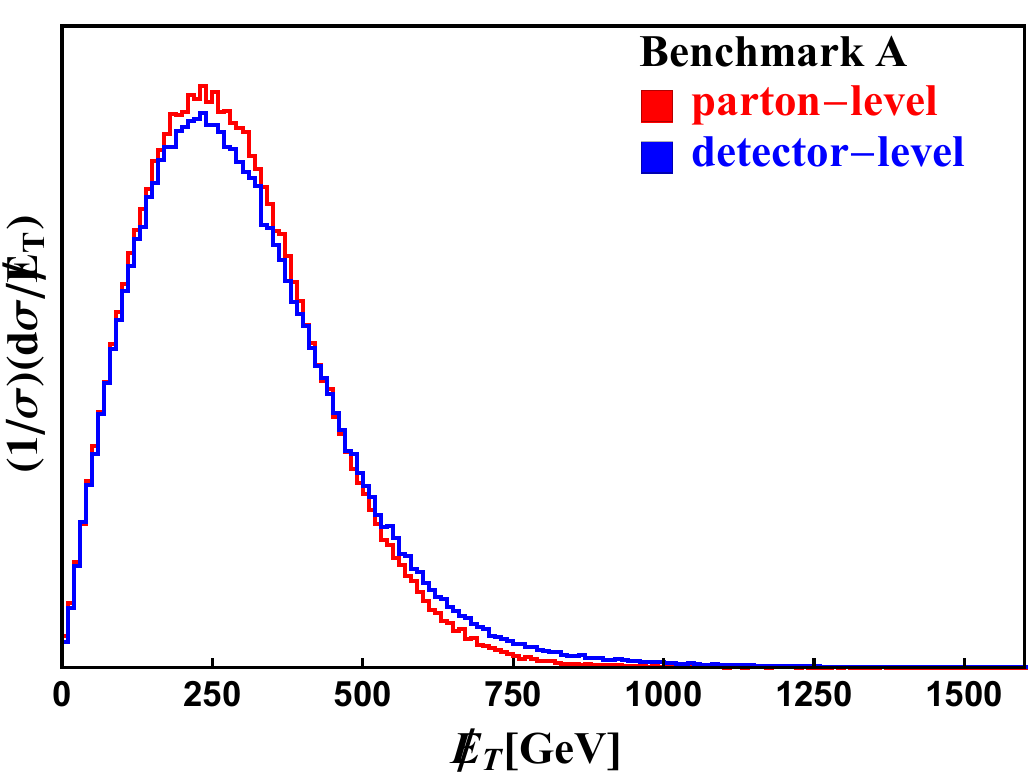}~~
  \includegraphics[width=5.5cm]{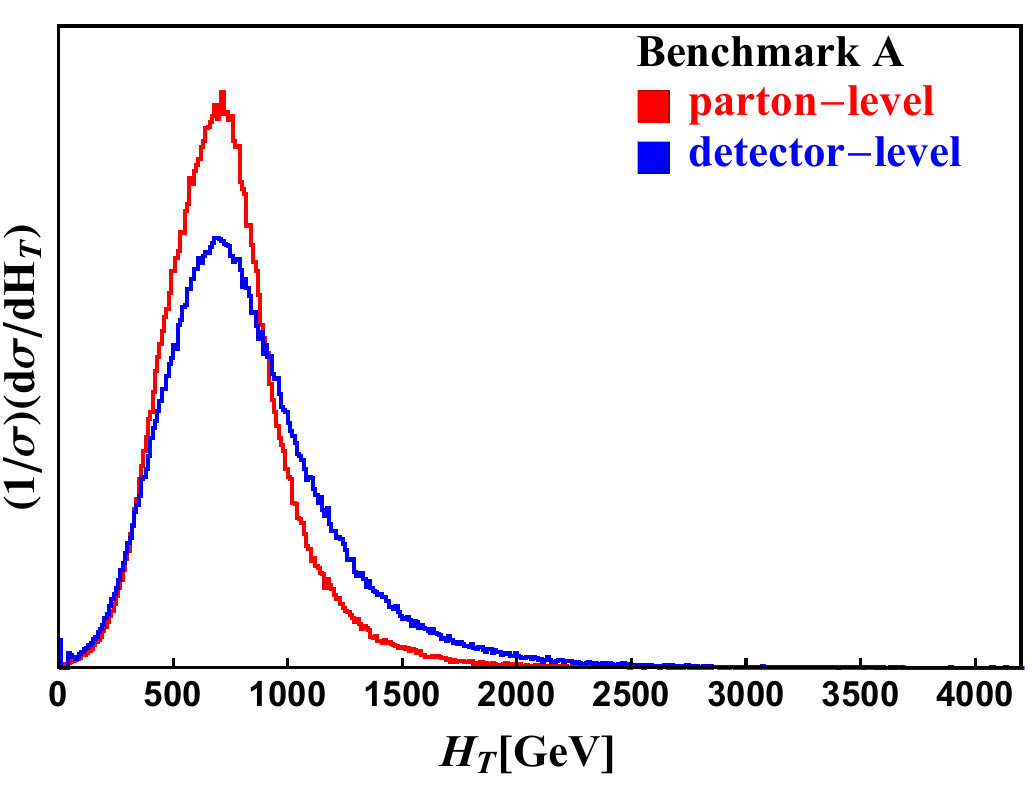} \\
  \includegraphics[width=5.5cm]{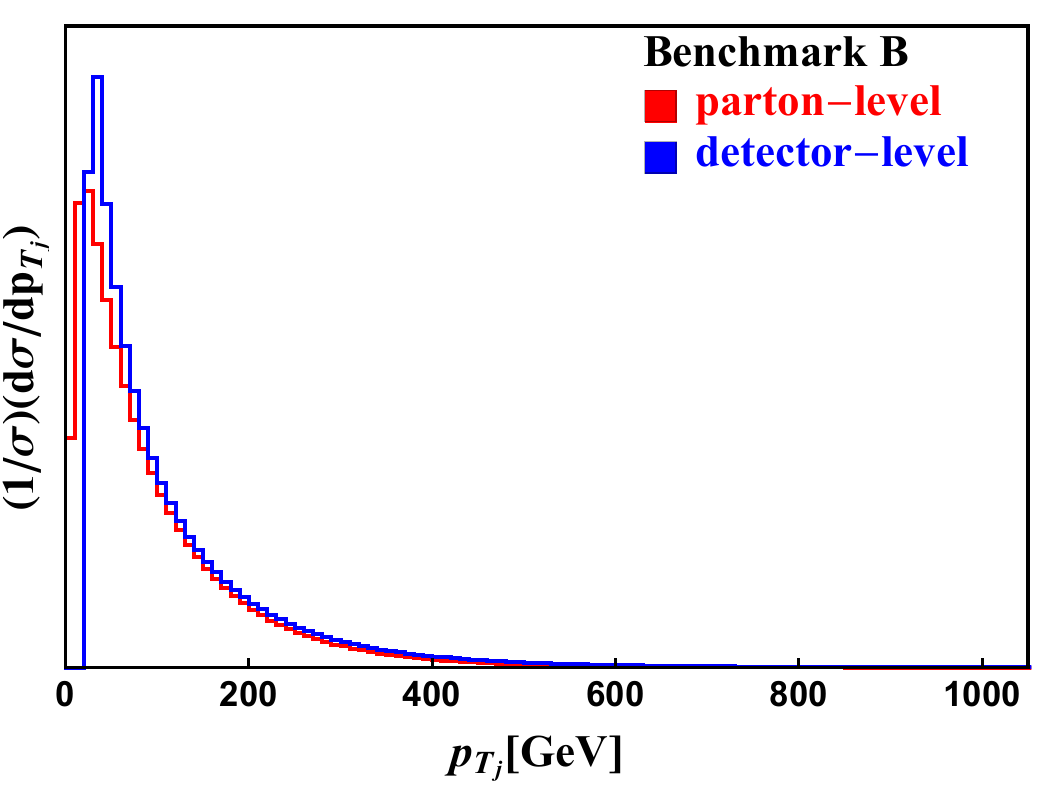}~~
  \includegraphics[width=5.5cm]{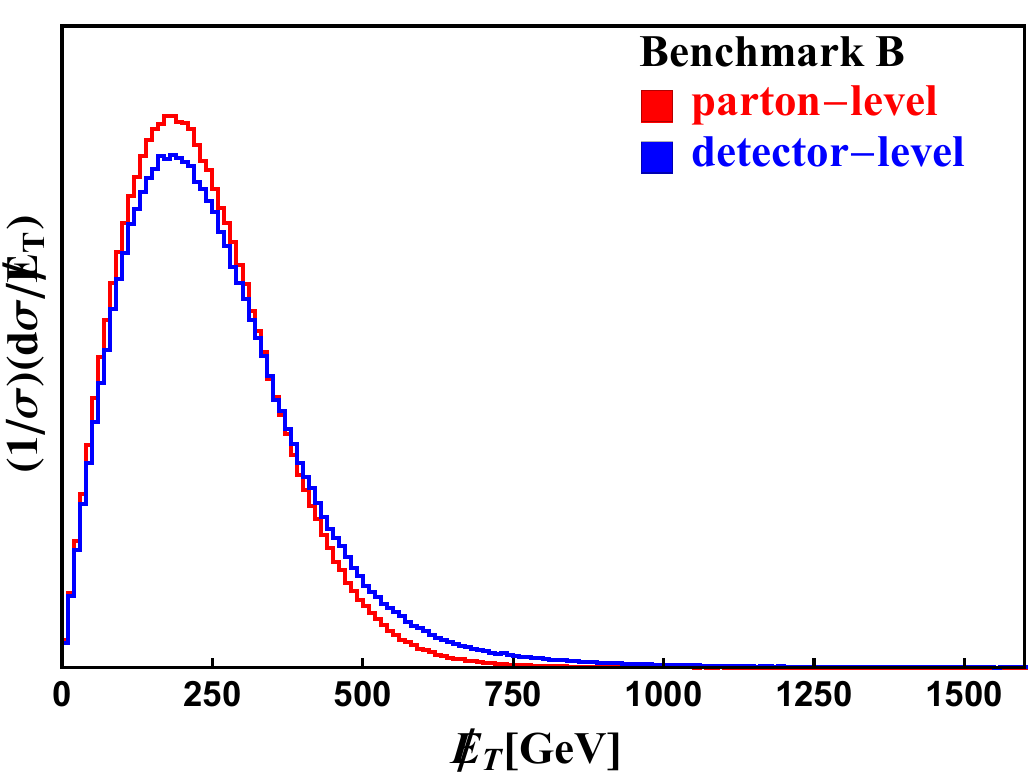}~~
  \includegraphics[width=5.5cm]{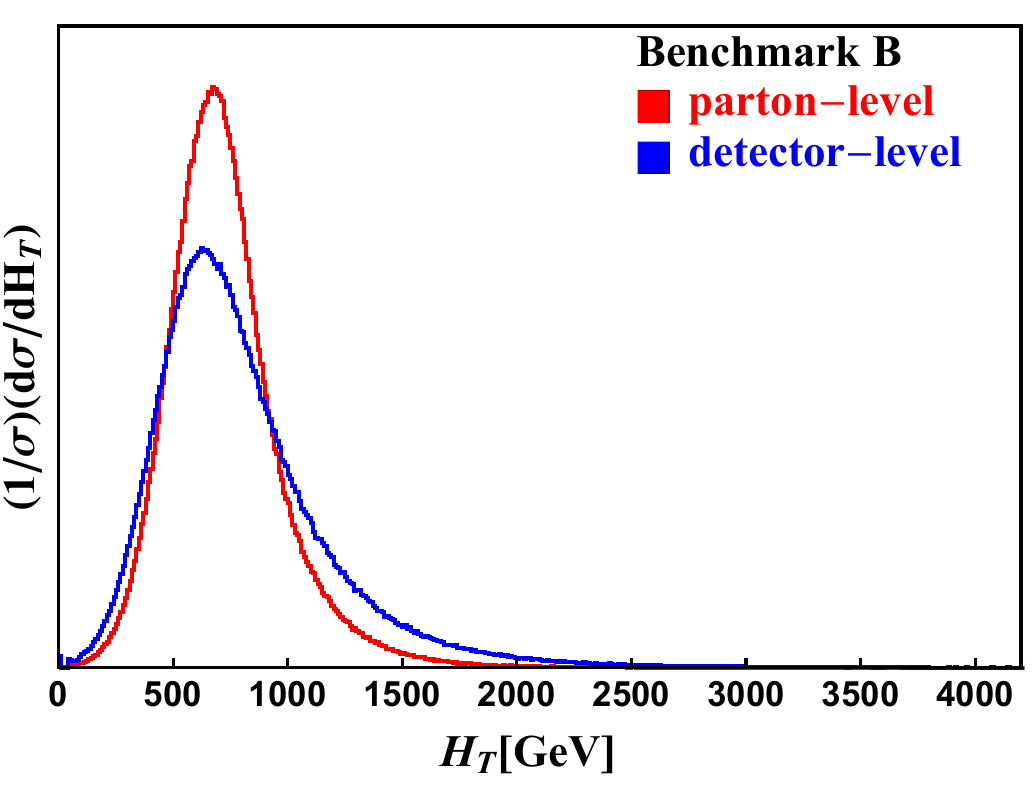} \\
  \includegraphics[width=5.5cm]{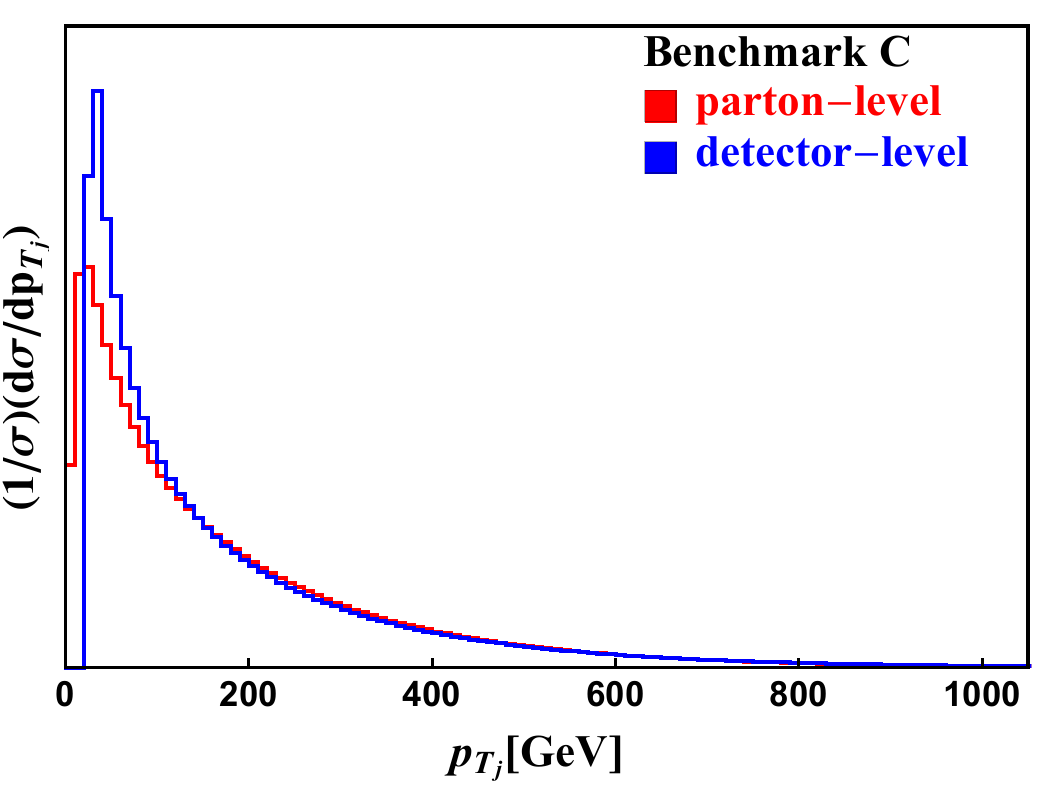}~~
  \includegraphics[width=5.5cm]{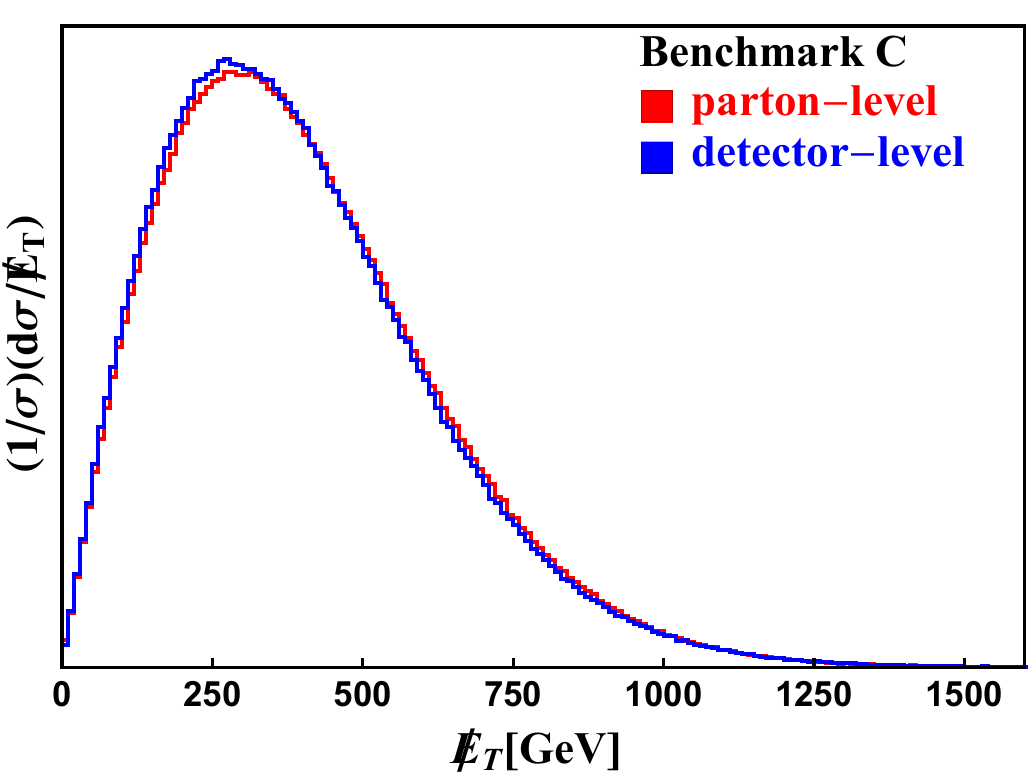}~~
  \includegraphics[width=5.5cm]{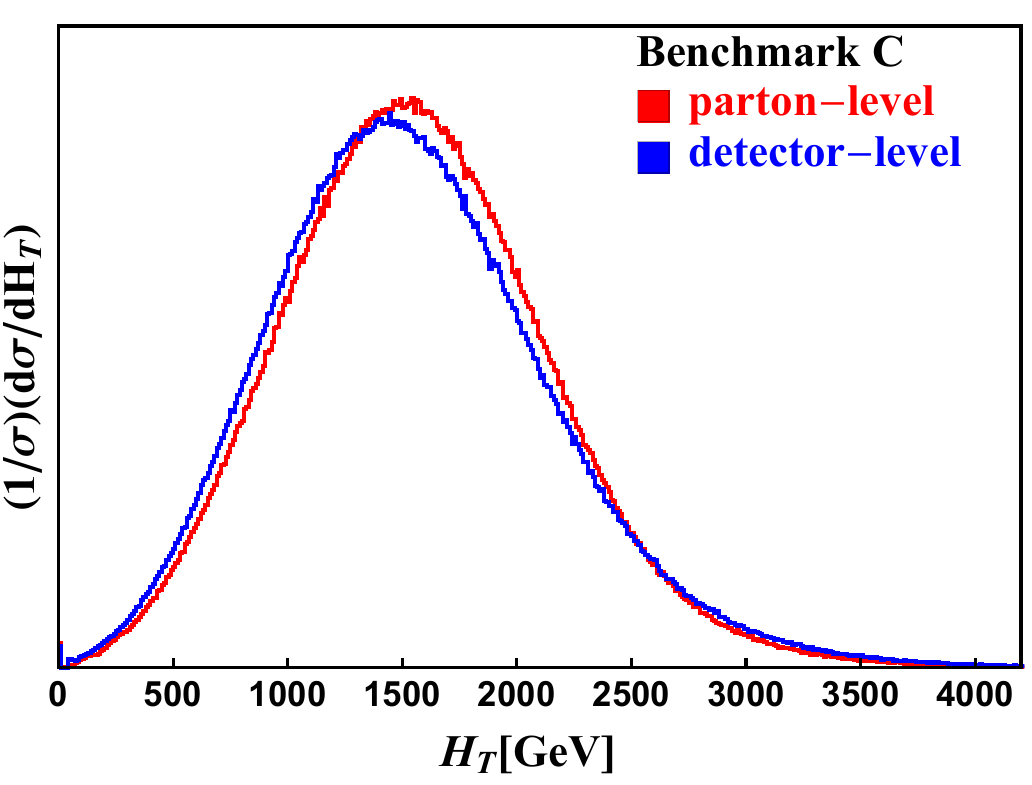} \\
  \caption{Normalized distributions of  $p_{T_j}$ (left column), $\met$ (middle column), 
   and $H_T$ (right column) for the three parameter-space benchmarks defined in 
   Table~\protect\ref{tab:BenchmarkPts}.~  The distributions in the top, middle, and bottom rows
   of the figure correspond to Benchmarks~A, B, and~C, respectively.  The red histogram in 
   each panel shows the distribution obtained at the parton level (with quarks, anti-quarks, 
   and gluons considered to be ``jets''), while the blue histogram shows the corresponding 
   distribution at the detector level.   
  \label{fig:ptetametht} }
\end{figure*}

For Benchmark~A, we see from Fig.~\ref{fig:num} that the parton-level and detector-level 
$\Njet$ distributions look quite similar and that both of these distributions peak 
at around $\Njet = 6$.  For Benchmark~B, 
by contrast, the parton-level distribution exhibits local maxima at both $\Njet = 7$ and at
$\Njet = 9$.  This behavior follows from the fact that processes of the form 
$pp\rightarrow \phi\chi_n$, which yield an odd number of parton-level jets, dominate the 
production cross-section for this benchmark.  Moreover, we observe that in going from the 
parton level to the detector level, the $\Njet$ distribution shifts to slightly lower values.
Several effects contribute to this reduction in $\Njet$.  First, jets associated with soft, 
isolated quarks or anti-quarks may fall below the $p_{T_j} > 20$~GeV detector-level 
threshold for jet identification.  Moreover, due to the large multiplicity of jets in these 
events, the hadrons associated with one or more of these jets frequently end up in such 
close proximity in $(\eta_j,\phi_j)$ space that they will be clustered together as a single 
jet at the detector level.  For Benchmark~C, the parton-level $\Njet$ distribution peaks around 
$\Njet = 10$, with most of the final states containing even numbers of jets.  The distribution is 
smoothed out at the detector level, but otherwise retains the same overall shape.

\begin{table}[t!] 
\begin{center}
\begin{tabular}{||c||c|c|c|c|c|c||}
 \hline\hline
 Benchmark  & ~$m_\phi$~ & ~$m_0$~ & ~$\Delta m$~ & ~$\delta$~ & 
   ~$\gamma$~ & ~$c_0$~ \\ \hline \hline
 \rule[-6pt]{0pt}{16pt} 
 A & \,1~TeV\,  & \,500 GeV\, & \,50~GeV\, & 1 & 1 & \,0.1\,  \\ \hline 
 \rule[-6pt]{0pt}{16pt}
 B & \,1~TeV\,  & \,500 GeV\, & \,50~GeV\, & 1 & 3 & \,0.1\,  \\ \hline
 \rule[-6pt]{0pt}{16pt}
 C & \,2~TeV\,  & \,500 GeV\, & \,50~GeV\, & 1 & 1.5 & \,0.1\,  \\ \hline\hline
\end{tabular}
\end{center}
\caption{Parameter choices which define our three representative benchmark points.  Benchmark~A 
  is representative of the regime in which $pp\rightarrow \phi^\dagger\phi$ and 
  $pp\rightarrow \phi{\chi}_m$ both contribute significantly (and at roughly the same order) to the 
  event rate.  By contrast, Benchmark~B is representative of the regime in which 
  $pp\rightarrow \phi{\chi}_m$ dominates the event rate.  Benchmark~C is representative of the 
  regime in which $pp\rightarrow \chi_m\bar{\chi}_n$ dominates.
  \label{tab:BenchmarkPts} }
\end{table}

One of the primary messages of Fig.~\ref{fig:num} is that our benchmarks all give rise to a 
significant population of events with large jet multiplicities even at the detector level.
Indeed, for Benchmarks~A, B, and~C, we find that the fraction of events for which $\Njet \geq 9$
at the detector level is  $16.3\%$, $24.3\%$, and  $54.8\%$, respectively.   
 
In Fig.~\ref{fig:ptetametht}, we show the normalized distributions for the other collider 
observables we consider in our analysis for our three parameter-space benchmarks.  
From left to right, the panels in each row of the figure correspond to the observables 
$p_{T_j}$, $\met$, and $H_T$.  The distributions in the top, middle, and bottom rows
of the figure correspond to Benchmarks~A, B, and~C, respectively.  The red histogram in each 
panel once again shows the distribution obtained at the parton level, while the blue histogram 
shows the corresponding distribution at the detector level.

In interpreting the results displayed in Fig.~\ref{fig:ptetametht}, we begin by 
noting that the parton-level $p_{T_j}$ distributions for all of our benchmarks are sharply peaked 
toward small values of $p_{T_j}$.  In other words, as one might expect, given the length of the 
decay chains in these decay-cascade scenarios, a significant fraction of the quarks and anti-quarks
produced in these decay chains tend to be extremely soft.  However, we also note that the 
distributions for Benchmarks~A and~B are more sharply peaked than the distribution for 
Benchmark~C.~  This is ultimately a result of $m_\phi$ being larger for this latter benchmark 
than for the other two.  A larger value of $m_\phi$ implies a larger value of $N$, and the fact
that $\gamma >0$ for Benchmark~C implies that production processes involving the heavier $\chi_n$ 
present in the ensemble will dominate.  The average CM energy associated with any of the production
processes in Figs.~\ref{fig:ChiChiDiagram}--\ref{fig:PhiPhiDiagram} is consequently larger for 
Benchmark~C than it is for Benchmark~A or~B, which results in a higher 
average $p_{T_j}$.  We also observe that since a $p_{T_j} > 20$~GeV threshold 
is required for jet identification at the detector level, many of the soft ``jets'' present
at the parton level for each of our benchmarks do not translate into jets at the detector level.  

In comparison with the $p_{T_j}$ distributions shown in Fig.~\ref{fig:ptetametht}, the 
corresponding $\met$ and $H_T$ distributions vary more dramatically from one benchmark to
the next.  Perhaps not unsurprisingly, the parton-level $H_T$ distribution for Benchmark~C peaks 
at a higher value $H_T$ than do the distributions of this same variable for Benchmarks~A and~B, 
again owing to the fact that $m_\phi$ is larger for this benchmark.  More interestingly, however, 
we also see that the parton-level and detector-level $H_T$ distributions for Benchmark~C are 
almost identical, while the detector-level $H_T$ distributions for Benchmarks~A and~B differ 
drastically from the corresponding distributions at parton level.  The discrepancy between the 
parton-level and detector-level $H_T$ distributions for these two benchmarks is ultimately a 
result of the $p_{T_j} > 20$~GeV threshold for jet-identification at the detector level.
As discussed above, the jets produced through mediator-induced decay cascades have a
higher average $p_{T_j}$ for Benchmark~C than they do for Benchmarks~A or~B, and consequently
the $H_T$ distribution for this benchmark is affected less by the cuts.  A similar effect,
albeit less pronounced, is also observed in the $\met$ distributions for our benchmarks.  
We also note that in general, the detector-level $\met$ and $H_T$ distributions for
all three of these benchmarks exhibit slightly longer tails than do the corresponding 
parton-level distributions.        

The results displayed in Fig.~\ref{fig:ptetametht} indicate that the shapes of the 
parton-level $p_{T_j}$, $\met$, and $H_T$ distributions resulting from mediator-induced 
decay cascades vary across the parameter space of our model.  Moreover, we see that the 
extent to which the parton-level and detector-level distributions of the same variable differ 
also depends non-trivially on the location within that parameter space. 
 
 
\FloatBarrier
\section{Detection Channels \label{sec:LHCseaches}}


A  variety of different search strategies sensitive to particular kinds of physics beyond 
the SM which give rise to large numbers of jets have been implemented by both the ATLAS 
and CMS Collaborations~\cite{Sirunyan:2017cwe,Aaboud:2017hdf,
Aad:2015mzg,Sirunyan:2018xwt,ATLASDisplacedJet,CMSDisplacedJet}.  Some of these turn 
out to be more suitable for detecting and constraining the large-jet-multiplicity events 
produced by the mediator-induced decay cascades in our example model than others.  

One such class of search strategies are those primarily tailored to the 
detection of microscopic black holes and sphalerons.  The leading constraints on
such exotic objects are currently those from a CMS analysis~\cite{Sirunyan:2018xwt}
performed with $35.9\mathrm{~fb}^{-1}$ of integrated luminosity at $\sqrt{s} = 13$~TeV.
The constraints obtained from a similar ATLAS study~\cite{Aad:2015mzg} performed with
$3.6\mathrm{~fb}^{-1}$ at the same CM energy are less competitive.  These searches turn out 
to be less effective for our model due to the high $H_T$ threshold for signal-event 
selection: $H_T > 900$~GeV in the CMS search and $H_T > 800$~GeV in the ATLAS 
search.  These cuts are imposed in order to reduce the SM multi-jet background.  By contrast,
for our signal events, either the $H_T$ distribution is peaked below 800 GeV or the signal 
cross-section is too small to be significant.  With only $35.9\mathrm{~fb}^{-1}$ of integrated 
luminosity, no meaningful constraints can be derived on our   model parameter space from the 
analysis in Ref.~\cite{Sirunyan:2018xwt}.
 
Another class of search strategies commonly adopted in new-physics searches in
channels involving large jet multiplicities are those tailored to the detection of 
scenarios involving long-lived hidden-sector states~\cite{ATLASDisplacedJet,CMSDisplacedJet}.
In searches of this sort, events are selected on the basis of one or more displaced 
vertices being present.  Such searches can indeed be relevant for the detection of extended
decay cascades in our example model, but only within the regime in which one or more of the 
$\chi_n$ are sufficiently long-lived that they give rise to such vertices.  Since we have 
focused in this paper on the region of parameter space within which region all of the $\chi_n$ 
with $n > 0$ decay promptly within the ATLAS or CMS detector, such searches also have no 
bearing on our analysis.   

\begin{figure*}[p!]   
\begin{center}
\centering
  \includegraphics[width=0.33\textwidth]{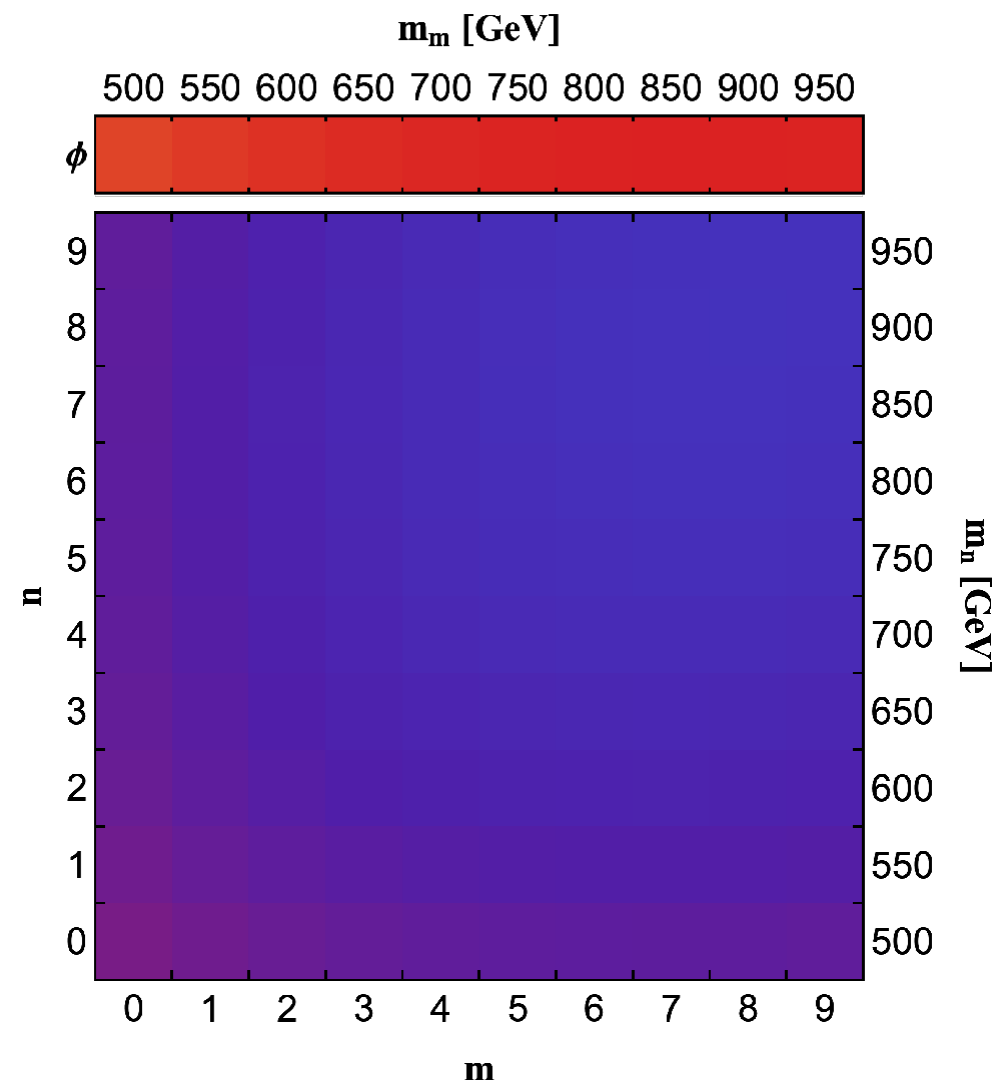}~
  \includegraphics[width=0.33\textwidth]{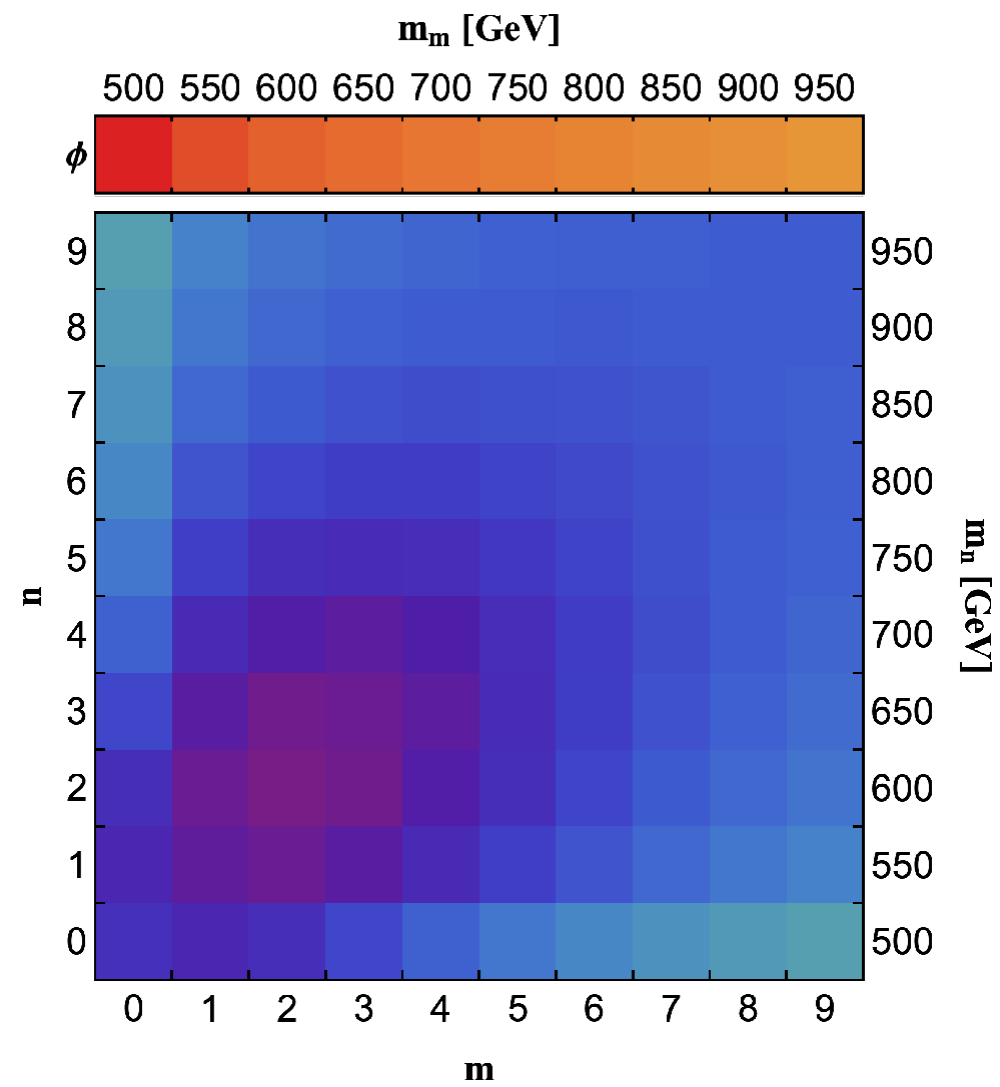}~
  \includegraphics[width=0.33\textwidth]{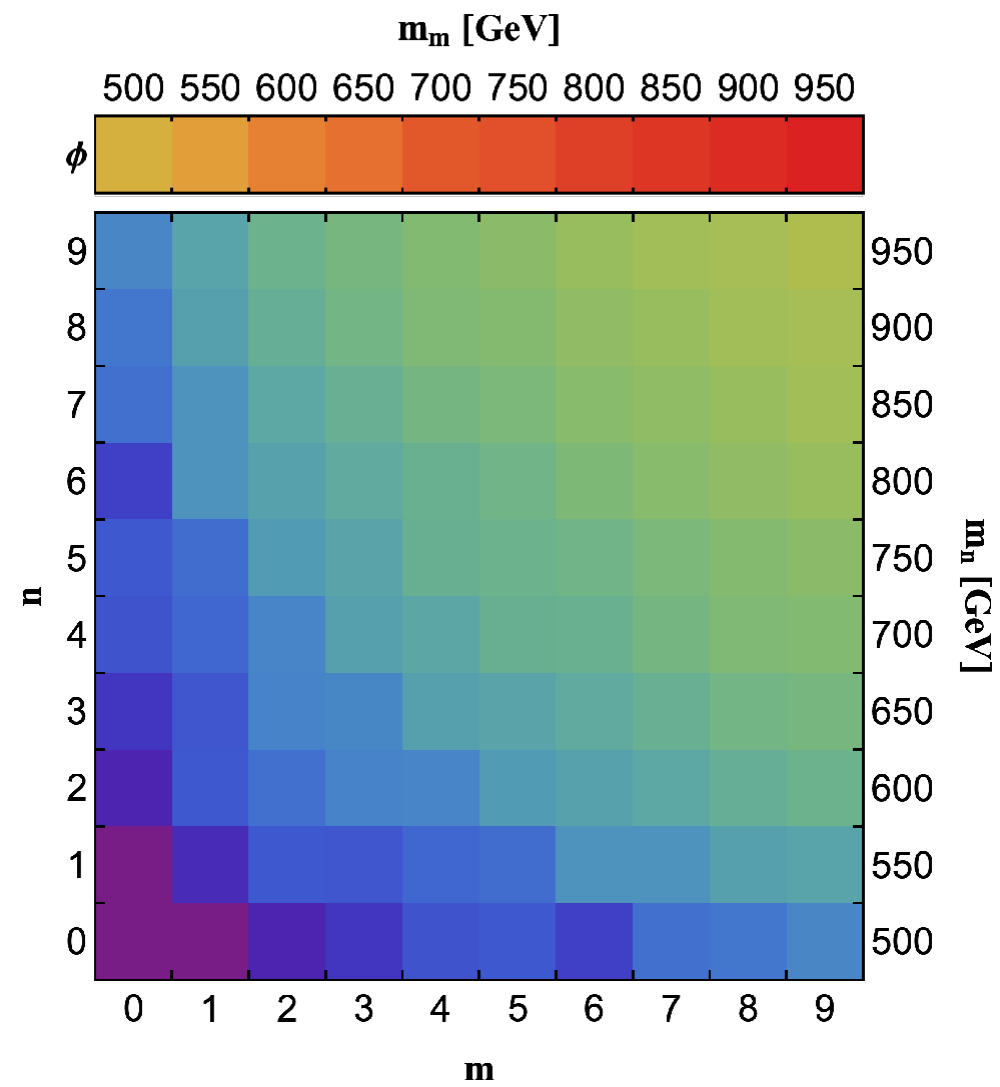} \\
  \includegraphics[width=0.33\textwidth]{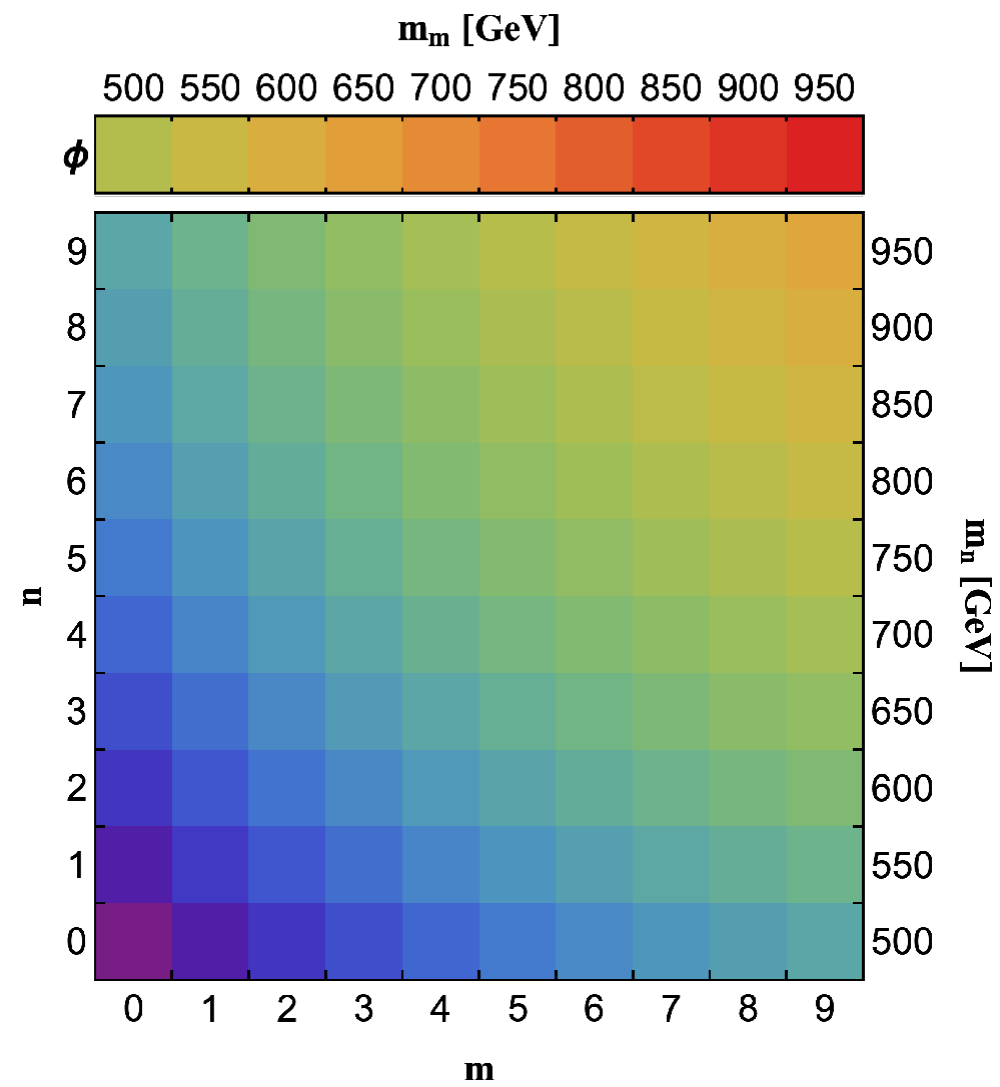}~
  \includegraphics[width=0.33\textwidth]{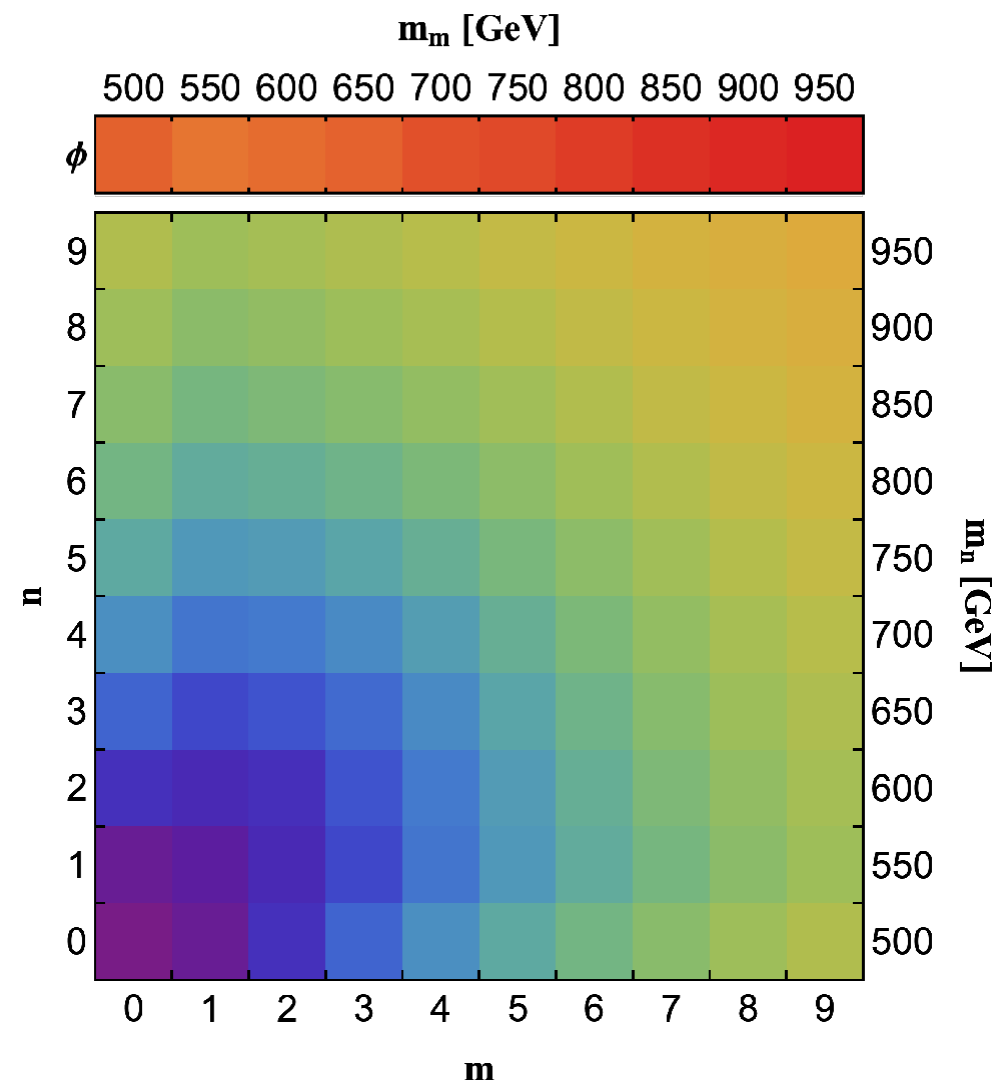}~
  \includegraphics[width=0.33\textwidth]{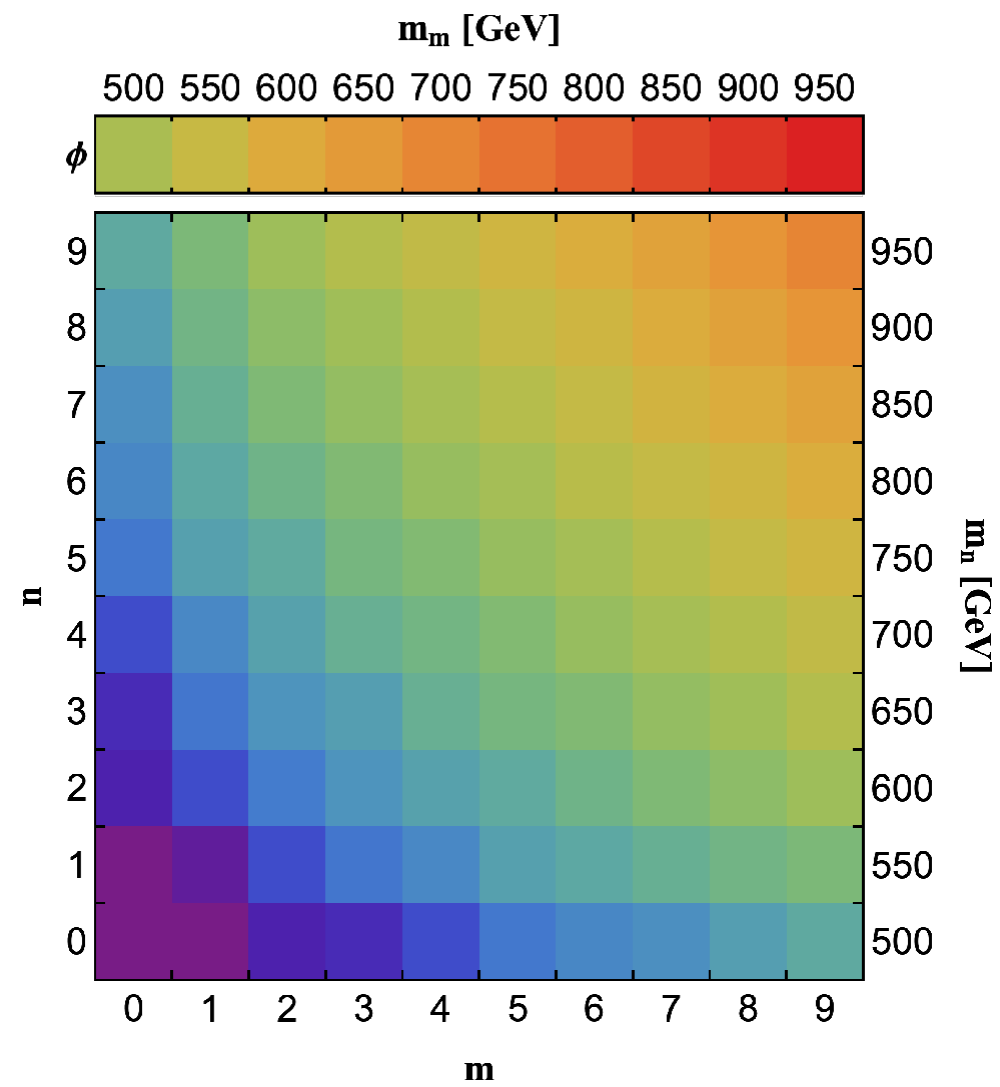} \\
  \includegraphics[width=0.33\textwidth]{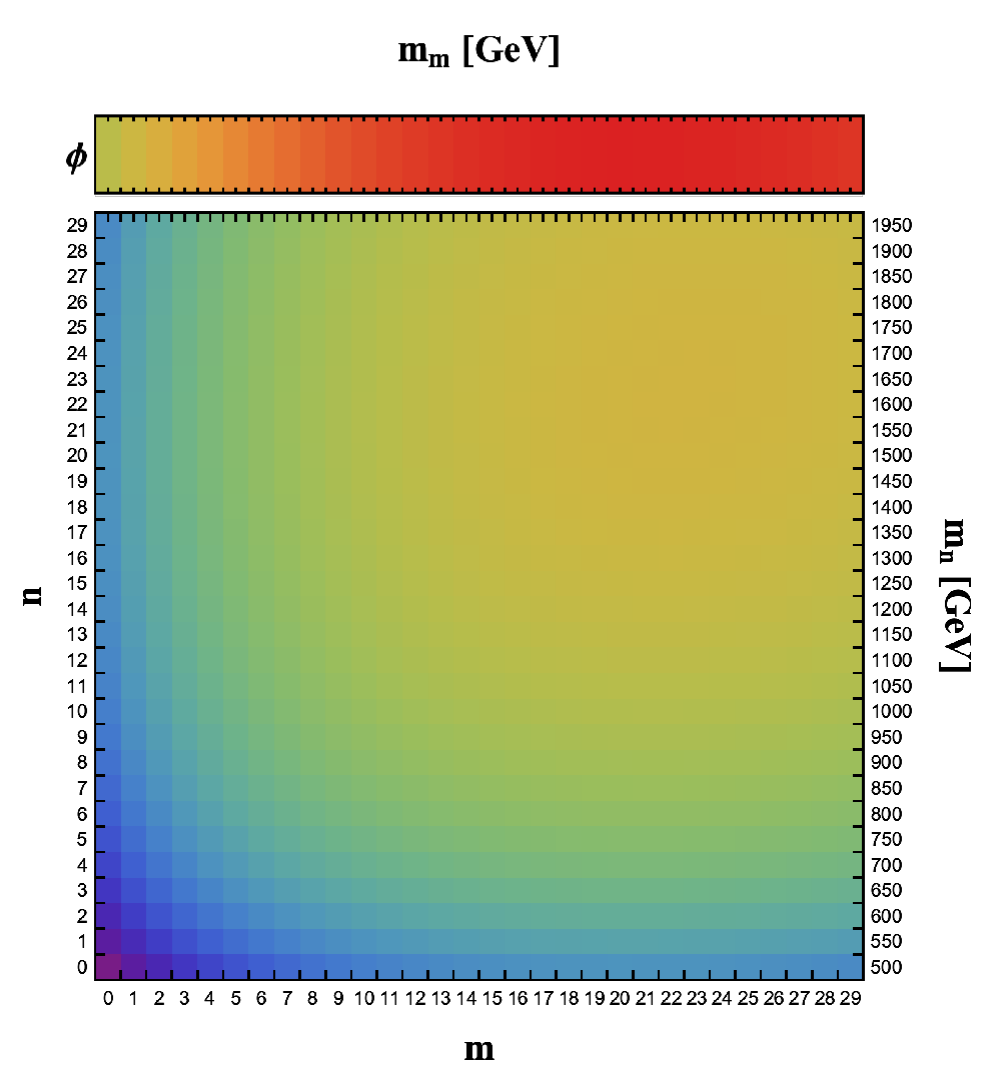}~
  \includegraphics[width=0.33\textwidth]{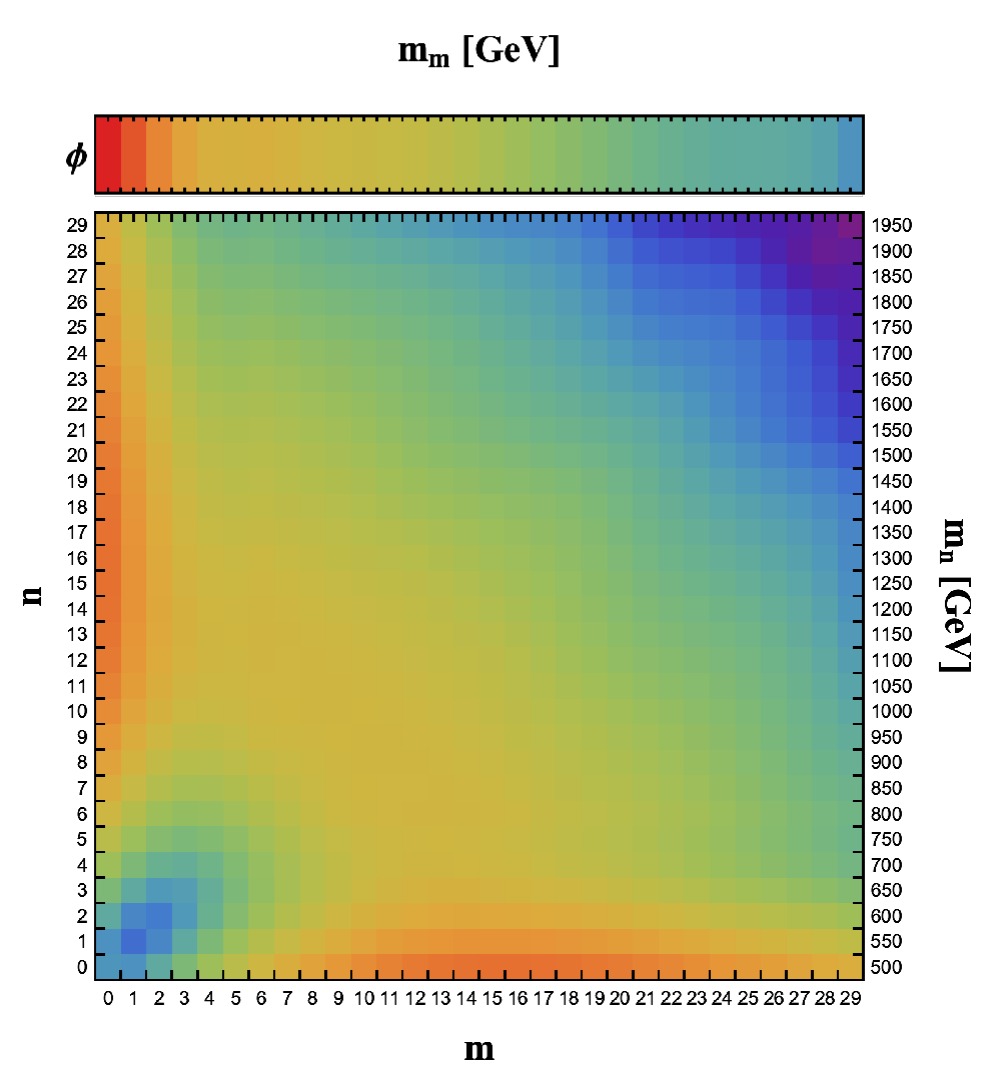}~
  \includegraphics[width=0.33\textwidth]{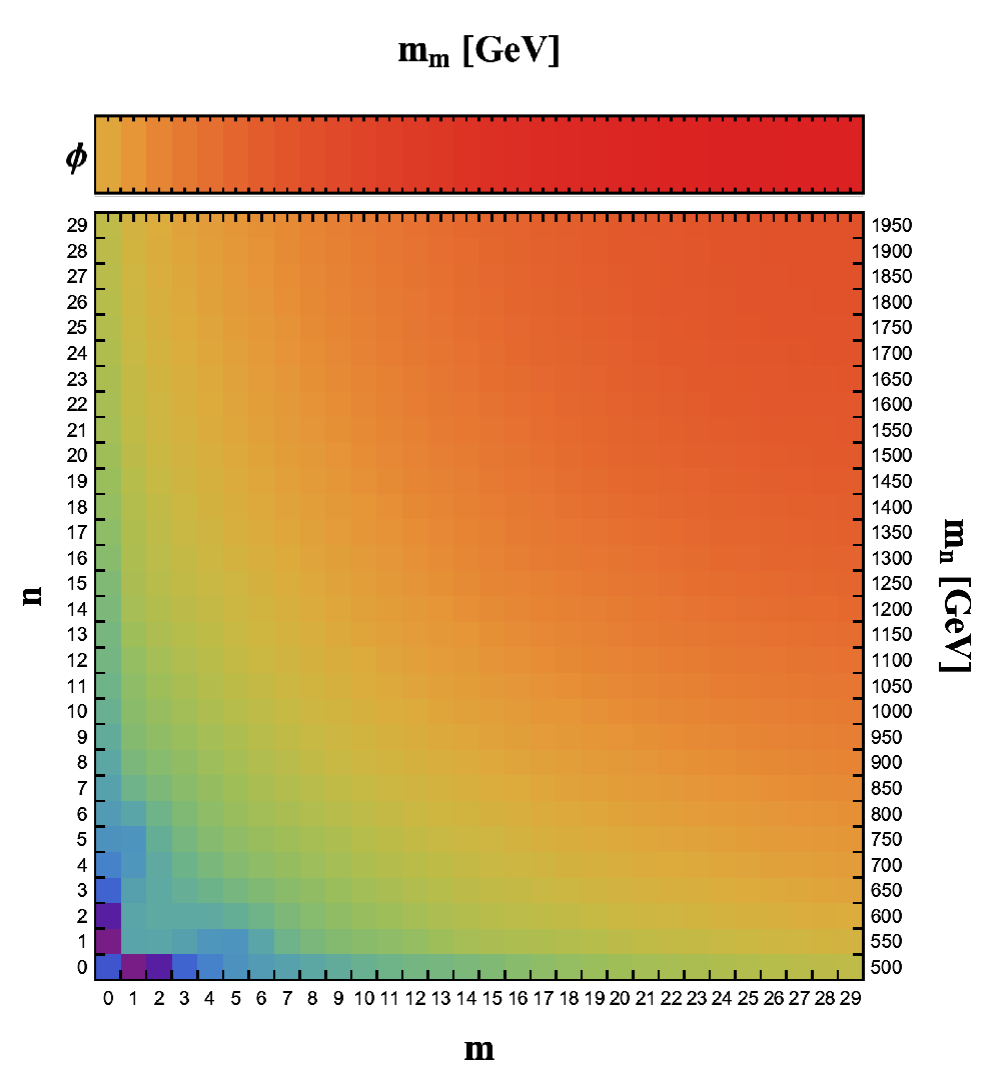} \\
  \includegraphics[width=0.33\textwidth]{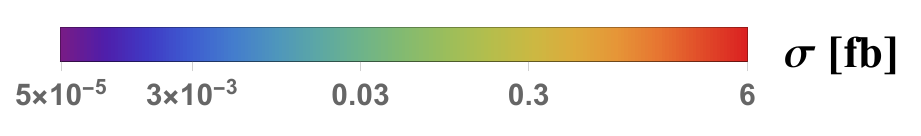}~
  \includegraphics[width=0.33\textwidth]{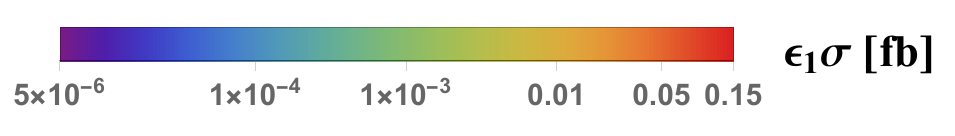}~
  \includegraphics[width=0.33\textwidth]{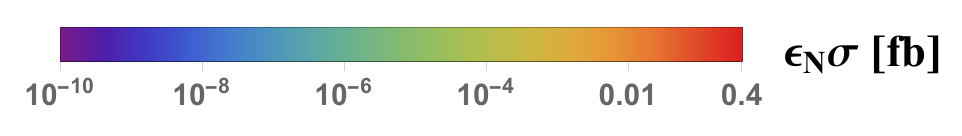} \\  
\end{center}
\caption{Production cross-sections before and after cuts for the processes 
  $pp\rightarrow \phi\chi_m$ and $pp\rightarrow \chi_m\bar{\chi}_n$, calculated for the three
  benchmarks defined in Table~\protect\ref{tab:BenchmarkPts} at the $\sqrt{s} = 13$~TeV LHC.~   
  The left column shows the cross-sections for these processes before any cuts
  are applied, while the center and right columns show the corresponding 
  cross-sections after the application of the event-selection criteria associated with 
  the monojet and multi-jet analyses described in the text, respectively.
  The results displayed in the top, middle, and bottom rows of the figure 
  correspond to Benchmarks~A, B, and~C, respectively.  The bar at the top of each panel 
  shows the individual cross-sections $\sigma(pp\rightarrow \phi\chi_m)$ for different values 
  of the index $m$, while the density plot below it shows the cross-sections 
  $\sigma(pp\rightarrow \chi_m\bar{\chi}_n)$ for different values of the indices $m$ 
  and $n$.    We emphasize that a different color scheme is used in each column, owing to the 
  significant difference in the overall scale of the cross-sections before and after cuts are 
  applied. 
  \label{fig:BenchmarkABC} }
\end{figure*} 

By contrast, it turns out that the search strategies which are particularly relevant for 
probing the parameter space of our model are those commonly adopted in searches for 
supersymmetry in the multi-jet $+~\met$ channel.  In searches of this sort, signal events are 
selected primarily on the basis of $\Njet$ and $\met$.  The leading constraints on our model 
from such searches are currently those from LHC $\sqrt{s} = 13$~TeV  searches by the ATLAS 
Collaboration~\cite{Aaboud:2017hdf} with $36.1\mathrm{~fb}^{-1}$ of integrated luminosity and 
those by the  CMS Collaboration~\cite{Sirunyan:2017cwe} with 
$35.9~\mathrm{~fb}^{-1}$ of integrated luminosity.  The ATLAS search turns out to be the 
more relevant of the two for constraining our example model, primarily because the CMS analysis 
includes a sizable $\met$ cut.  This leads to a significant reduction in statistics for our 
signal process.   

For this reason, we assess the constraints on our model from the multi-jet channel by
modeling our triggering requirements and event-selection criteria after those employed in 
Ref.~\cite{Aaboud:2017hdf}.  In particular, we adopt the same triggering criteria that 
we used in constructing the detector-level $\Njet$, $\met$, and $H_T$ distributions in 
Sect.~\ref{sec:expobs}.  In addition, primarily in order to reduce the SM multi-jet 
background, we impose the $\met$ cut 
\begin{equation}
   \frac{\met}{\sqrt{H_T}} ~>~ 5~\mathrm{GeV}^{1/2}~.
\end{equation} 
Following Ref.~\cite{Aaboud:2017hdf}, we include only the three-momenta of 
jets with pseudorapidities in the range $|\eta_j| < 4.5$ when calculating 
$\met$ for a given event; likewise, we include only those jets with 
$p_{T_j} > 40$~GeV and $|\eta_j| < 2.8$ within the scalar sum in Eq.~(\ref{eq:HT}) 
when calculating $H_T$.  Finally, we impose a cut on the total number of jets
in the event which exceed a given $p_{T_j}$ threshold.  More specifically, we 
define $\Njet^{50}$ to be the number of jets with $p_{T_j} > 50$~GeV in a given
event and $\Njet^{80}$ to be the number of jets with $p_{T_j} > 80$~GeV.~  We
then perform an inclusive search involving a number of different signal 
regions defined by different combinations of the threshold cuts
$\Njet^{50} \geq \{8, 9, 10, 11\}$ and $\Njet^{80} \geq \{7, 8, 9\}$.
For each channel, we impose the corresponding constraint on the parameter space of 
our example model by comparing the number of signal events $N_s$ after cuts 
with the 95\%~C.L. upper limit on $N_s$ in Ref.~\cite{Aaboud:2017hdf}.
We emphasize that these signal regions are equivalent to those adopted in
Ref.~\cite{Aaboud:2017hdf} for searches in the ``heavy-flavor channel'' with 
$N_{b-\mathrm{tag}} \geq 0$ --- \ie, with no additional $b$-tagging requirement
imposed.  By contrast, searches in the ``jet-mass channel,''  which are particularly suited
for probing new-physics scenarios involving highly-boosted massive particles
which give rise to large-radius jets, are less constraining within our parameter-space 
region of interest.  Highly-boosted $\phi$ or $\chi_n$ particles are not
produced at any significant rate within this region, and the requirement that 
large-radius jets with jet masses above a few hundred GeV be present leads to 
a significant reduction in signal events.

While the most striking signals to which our example model gives rise would 
be detected in the multi-jet channel, this model can also give rise to observable signals
in other channels relevant for new-physics searches.  We must therefore ensure 
that our model is consistent with the results of existing searches in these 
channels within our parameter-space region of interest.  For example,
diagrams of the sort depicted in Fig.~\ref{fig:ChiPhiDiagram} contribute to the event 
rate in the monojet $+~\met$ channel, as do diagrams similar to that
shown in Fig.~\ref{fig:ChiChiDiagram} in which an additional quark or gluon is produced
as initial-state radiation or radiated off the internal $\phi$ line.  Such diagrams 
clearly contribute to the event rate in the monojet $+~\met$ channel whenever the 
ensemble constituents $\chi_m$ and $\overline{\chi}_n$ in the final state are both 
stable on collider timescales and therefore appear as $\met$ within a collider detector.
Searches in this channel play an important role in constraining single-particle 
dark-sector models with a similar mediator coupling structure~\cite{TChannelMonojet},
and thus can be anticipated to play an an important role in constraining the parameter
space of our model as well.

Moreover, diagrams of this sort in which $\chi_m$ and/or $\overline{\chi}_n$ decay within 
the detector can also potentially contribute to the nominal signal-event rate in the 
monojet $+~\met$ channel.  This is because the event-selection criteria adopted in 
searches in this channel typically permit a small number of additional hadronic jets to be 
present in the final state.  Thus, in assessing the monojet constraints on our example model, 
we must account for events in which the number of jets collectively produced by the decays 
of $\chi_m$ and/or $\overline{\chi}_n$ is sufficiently small that these event-selection 
criteria are satisfied.

The most stringent constraints on our model from searches in the monojet $+~\met$ 
channel are those obtained by the ATLAS Collaboration with $36.1\ \mathrm{fb}^{-1}$ of 
integrated luminosity at the $\sqrt{s} = 13$~TeV LHC~\cite{ATLASMonojet}.
In assessing the constraints on our example model from searches in the monojet $+\met$ 
channel, we model our triggering requirements and event-selection criteria after those 
employed in Ref.~\cite{ATLASMonojet}.  In particular, we select events in which
$\met > 250$~GeV and in which the leading jet has $p_{T_j} > 250$~GeV and 
$|\eta_j| < 2.4$.  In addition, we require that there exist no more than four 
jets in the event with $p_{T_j} > 30$~GeV and $|\eta_j| < 2.8$.  We also impose 
the criterion $\Delta \phi(\mpt,\vec{p}_j)> 0.4$, where $\Delta \phi(\mpt,\vec{p}_j)$ 
is the difference in azimuthal angle between the missing-transverse-momentum vector $\mpt$ and 
the three-momentum vector $\vec{p}_j$ of any reconstructed jet in the event.

Finally, we note that while the most striking multi-jet signatures which arise in 
our model are those involving large jet multiplicities, channels involving a more
modest number of jets and $\met$ can also potentially be relevant for constraining
the parameter space of our model.  Indeed, Fig.~\ref{fig:num} indicates that 
a significant number of events with 5--6 jets can be produced even within regions of 
parameter space where the peak on the $\Njet$ distribution is much higher.   
The leading constraints of this sort turn out to be those from an ATLAS 
search~\cite{Aaboud:2017vwy} for squarks and gluinos in events involving 2--6 hadronic 
jets and substantial $\met$.  However, as we shall see, constraints from such 
moderate-jet-multiplicity searches turn out to be subleading compared to those from the 
monojet $+~\met$ and multi-jet $+~\met$ searches discussed above.

\begin{table*}[t!]
\begin{center}
\begin{tabular}{||c||c|c|c||c|c|c||c|c|c||}
 \hline\hline &
 \multicolumn{3}{c||}{Before Cuts} & \multicolumn{3}{c||}{After Monojet Cuts} &
 \multicolumn{3}{c||}{After Multi-Jet Cuts} \\ \hline
 ~Benchmark~  \rule[-6pt]{0pt}{16pt} & 
   ~$\sigma_{\chi\chi}$ (fb)~ & ~$\sigma_{\phi\chi}$ (fb)~ & ~$\sigma_{\phi\phi}$ (fb)~ &
   ~$\epsilon_1\sigma_{\chi\chi}$ (fb)~ & ~$\epsilon_1\sigma_{\phi\chi}$ (fb)~ & 
   ~$\epsilon_1\sigma_{\phi\phi}$ (fb)~ &
   ~$\epsilon_N\sigma_{\chi\chi}$ (fb)~ & ~$\epsilon_N\sigma_{\phi\chi}$ (fb)~ & 
   ~$\epsilon_N\sigma_{\phi\phi}$ (fb)~ 
   \\ \hline \hline
 \rule[-6pt]{0pt}{16pt} 
 A & 0.28 & 4.19 & 4.29 & 0.015 & 0.41 & 0.32 & $7.6\times 10^{-4}$ & 0.058 & 0.12 \\ \hline 
 \rule[-6pt]{0pt}{16pt}
 B & 9.72 & 23.9 & 4.29 & 0.32 & 0.77 & 0.10 & 0.10 & 0.87 & 0.24 \\ \hline
  \rule[-6pt]{0pt}{16pt}
 C & 3.06 &  0.92 & $9.1\times 10^{-3}$ & 0.065 & $6.0\times 10^{-3}$ & $1.4\times 10^{-5}$ & 0.62 & 
   0.34 & $4.6\times 10^{-3}$ \\ \hline
  \rule[-6pt]{0pt}{16pt}
 LHC Limit & \multicolumn{3}{c||}{ } & \multicolumn{3}{c||}{531} & \multicolumn{3}{c||}{7.2} \\ 
 \hline\hline
\end{tabular}
\end{center}
\caption{The inclusive cross-sections $\sigma_{\chi\chi}$, $\sigma_{\phi\chi}$, and 
  $\sigma_{\phi\phi}$ defined in Eq.~(\protect\ref{eq:productionprocesses}) at the 
  $\sqrt{s} = 13$~TeV LHC, as well as the corresponding cross-sections after the   
  application of the event-selection associated with the monojet search 
  and multi-jet searches described in the text.  Also shown are the corresponding 
  experimental upper limits on the overall production cross-section after cuts for both 
  of these monojet and multi-jet searches. 
  \label{tab:BenchmarkXSecs} }
\end{table*}

In Fig.~\ref{fig:BenchmarkABC}, we present our results for the individual cross-sections
$\sigma(pp\rightarrow \phi\chi_m)$ for different values of the index $m$ and the individual 
cross-sections $\sigma(pp\rightarrow \chi_m\bar{\chi}_n)$ for different combinations of the 
indices $m$ and $n$ for the three benchmarks defined in Table~\ref{tab:BenchmarkPts} at the 
$\sqrt{s} = 13$~TeV LHC.~  The results in the top, middle, and bottom rows of the figure
correspond to Benchmarks~A, B, and~C, respectively.  The left panel in each row of the figure 
shows these cross-sections before any cuts are applied, while the center and right panels in
the same row show the corresponding cross-sections after the application of the 
event-selection criteria associated with searches in the monojet and multi-jet channels, 
respectively.  More specifically, the monojet results shown here correspond the event-selection 
criteria associated with Signal Region IM1 of Ref.~\cite{ATLASMonojet} with $\met>250$~GeV, 
while the multi-jet results correspond to the Signal Region $\Njet^{50}\geq 8$ of 
Ref.~\cite{Aaboud:2017hdf} with $N_{b-\mathrm{tag}}\geq 0$.

In interpreting the results shown in Fig.~\ref{fig:BenchmarkABC}, we begin by observing that 
for Benchmark~A, the individual cross-sections $\sigma(pp\rightarrow \phi\chi_m)$ before cuts 
are larger for heavier $\chi_m$, due primarily to the fact that $\gamma$ is positive.  This remains 
true even after the application of the multi-jet cuts, as shown in the top right panel of the figure.  
By contrast, after the monojet cuts are applied, $\sigma(pp\rightarrow\phi\chi_0)$ is by far the 
largest of the $\sigma(pp\rightarrow\phi\chi_m)$ for this benchmark.  This is primarily a consequence 
of the upper limit on $\Njet$ included among these cuts.  Similar behavior is also apparent 
for this benchmark within the $\chi\chi$ channel.
 The results obtained for Benchmark~B are qualitatively similar 
to those obtained for Benchmark~A, except that the individual contributions 
$\sigma(pp\rightarrow \phi \chi_m)$ and $\sigma(pp\rightarrow \chi_m \bar{\chi}_n)$ involving 
heavier $\chi_m$ contribute more significantly even after the monojet cuts.  This is primarily a 
reflection of the fact that $\gamma$ is larger for Benchmark~B than it is for Benchmark~A.~ 
For Benchmark~C, the larger value of $m_\phi$ implies that the number of states in the
ensemble is significantly larger than it is for the other two benchmarks.  This larger
value of $N$ notwithstanding, the results for this benchmark are also qualitatively similar 
to those obtained for Benchmark~A.~  The most salient difference between the results 
obtained for these two benchmarks is the significant decrease in  
$\sigma(pp\rightarrow\chi_m\bar{\chi}_n)$ when both $m$ and $n$ become large.  This 
is simply a reflection of the fact that both of the ensemble constituents are quite
heavy in this regime.

The total production cross-sections $\sigma_{\chi\chi}$, $\sigma_{\phi\chi}$, and 
$\sigma_{\phi\phi}$ obtained by summing the contributions from all relevant 
individual production processes are provided in Table~\ref{tab:BenchmarkXSecs}.
The cross-sections before the application of any cuts are provided, as well as the 
corresponding cross-sections obtained after the application of our monojet
and multi-jet cuts.  Once again, the monojet results correspond the event-selection 
criteria associated with Signal Region IM1 of Ref.~\cite{ATLASMonojet} with $\met>250$~GeV, 
while the multi-jet results correspond to the Signal Region $\Njet^{50}\geq 8$ of 
Ref.~\cite{Aaboud:2017hdf} with $N_{b-\mathrm{tag}}\geq 0$.
Current limits on the overall production cross-section from 
LHC monojet and multi-jet searches are also included in the bottom row of the 
figure for purposes of comparison.  For Benchmark~A, we observe that $\sigma_{\phi\chi}$ and 
$\sigma_{\phi\phi}$ are approximately equal and both much larger than $\sigma_{\chi\chi}$ before 
cuts.  However, $\epsilon_1\sigma_{\phi\chi}$ is slightly larger than $\epsilon_1\sigma_{\phi\phi}$ 
after the monojet cuts are applied, and $\epsilon_N\sigma_{\phi\phi}$ dominates the overall 
production rate after the application of the multi-jet cuts.  For Benchmark~B, 
$\sigma_{\phi\chi}$ dominates the total production cross-section both before and after each set 
of cuts is applied.  Likewise, for Benchmark~C, $\sigma_{\chi\chi}$ 
dominates both before and after cuts, though the contribution from $\epsilon_N\sigma_{\phi\chi}$ 
after the application of the multi-jet cuts, while subleading in comparison with 
$\epsilon_N\sigma_{\chi\chi}$, is non-negligible.  

More importantly, however, we observe that all three of these benchmark points are consistent 
with LHC limits from both monojet and multi-jet searches, despite the fact that a different 
production process provides the leading contribution to the overall event rate in the 
multi-jet channel in each case.  Thus, we see that a variety of qualitatively different
scenarios which give rise to mediator-induced decay cascades can be consistent with current
constraints and therefore potentially within the discovery reach of future collider searches.


\FloatBarrier
\section{Surveying the Parameter Space \label{sec:scan}}


Having gained from our benchmark studies a sense of the range of phenomenological possibilities 
which can arise within our model, we now expand our analysis by performing a more systematic 
survey of the phenomenological possibilities that arise across the full parameter space of 
this model.  The purpose of this survey is not only to assess the impact of current 
experimental constraints, but also to determine which of the production processes discussed in 
Sect.~\ref{sec:model} dominates the event rate within different regions.  
In performing this survey, we shall vary the mediator mass $m_\phi$ and the scaling exponent 
$\gamma$ which determines how the mediator interacts with the fields of the dark sector 
while holding fixed the parameters $m_0 = 500$~GeV, $\Delta m = 50$~GeV, and $\delta = 1$ 
which characterize the internal structure of the dark sector itself.
For simplicity, and in order to maintain consistency with the constraints outlined in 
Sect.~\ref{sec:pheno} across the $(m_\phi,\gamma)$-plane, we fix $c_0 = 0.1$.      
More specifically, we sample $m_\phi$ and $\gamma$ at a variety of discrete values within the 
ranges $0.6\mbox{~TeV}\leq m_\phi \leq 2.5\mbox{~TeV}$ and $0 \leq \gamma \leq 3.5$. 
For each such combination of $m_\phi$ and $\gamma$, we then evaluate the aggregate cross-sections 
$\sigma_{\phi\phi}$, $\sigma_{\phi\chi}$, and $\sigma_{\chi\chi}$ according to the 
event-generation and event-selection procedures outlined in Sect.~\ref{sec:expobs}.~
In addition, in order to provide a measure of the fraction of events associated with any 
particular combination of these parameters have truly large jet multiplicities,
we also define the parameter $\Njet^{10\%}$, which represents the maximum value of 
$\Njet$ for which at least 10\% of the events in a given data sample have 
$\Njet \geq \Njet^{10\%}$.     

\begin{center}
\begin{figure*}[t!] 
\begin{center}
  \includegraphics[width=0.9\textwidth]{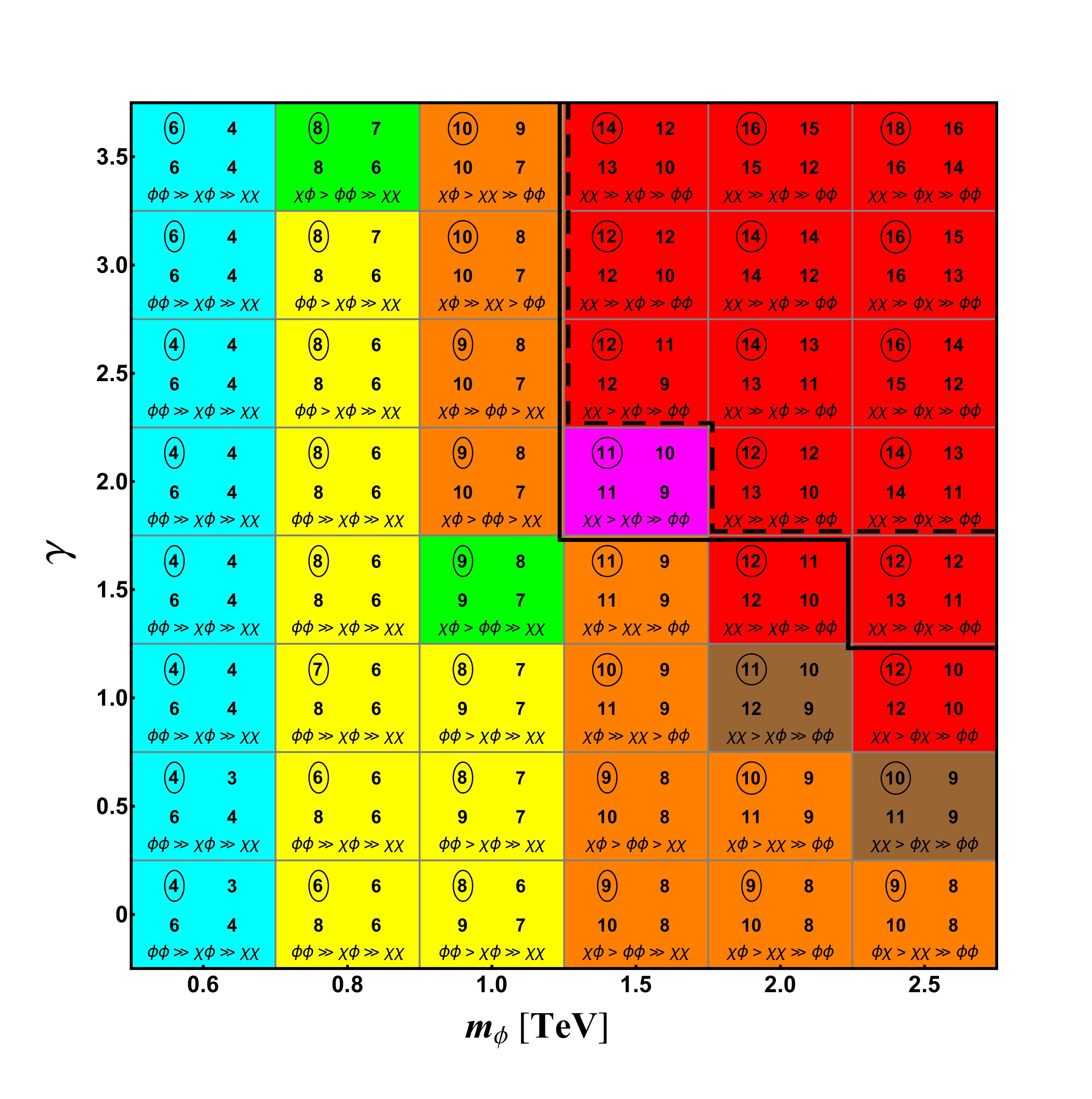}~~~~~~ \\
  \vspace{-5pt}
  \includegraphics[width=0.375\textwidth]{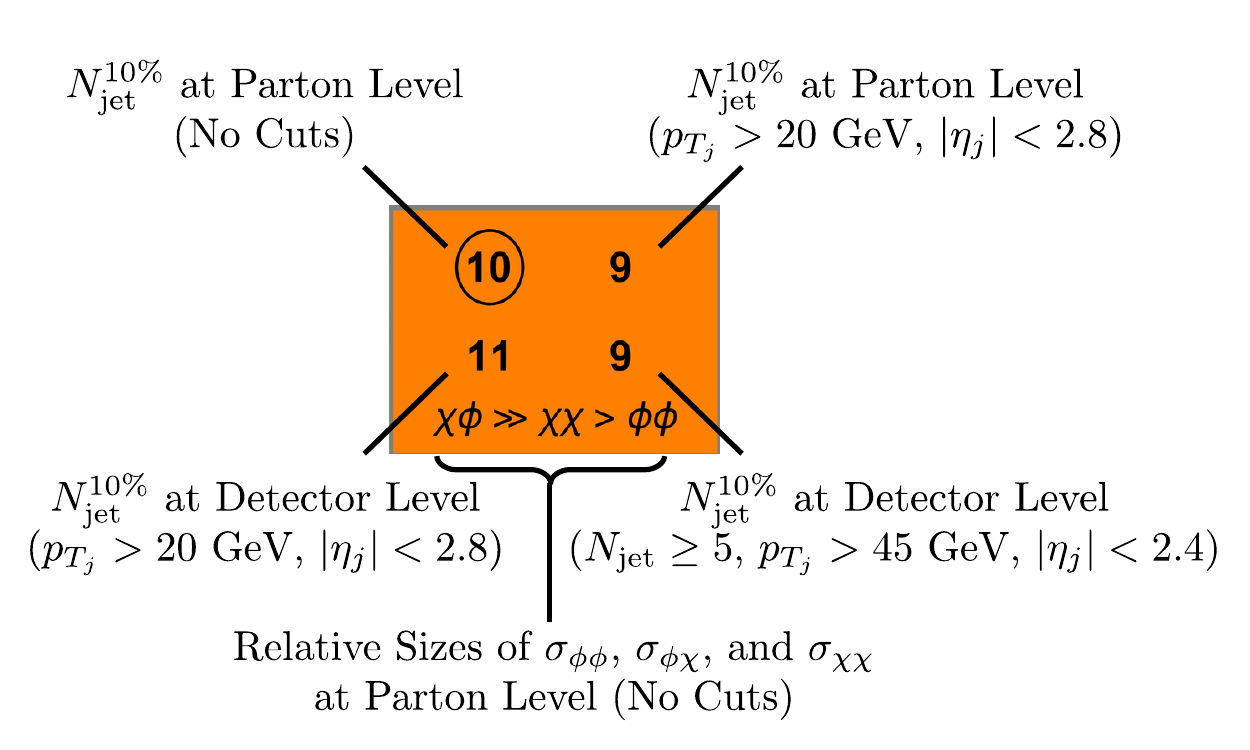} ~~
  \includegraphics[width=0.575\textwidth]{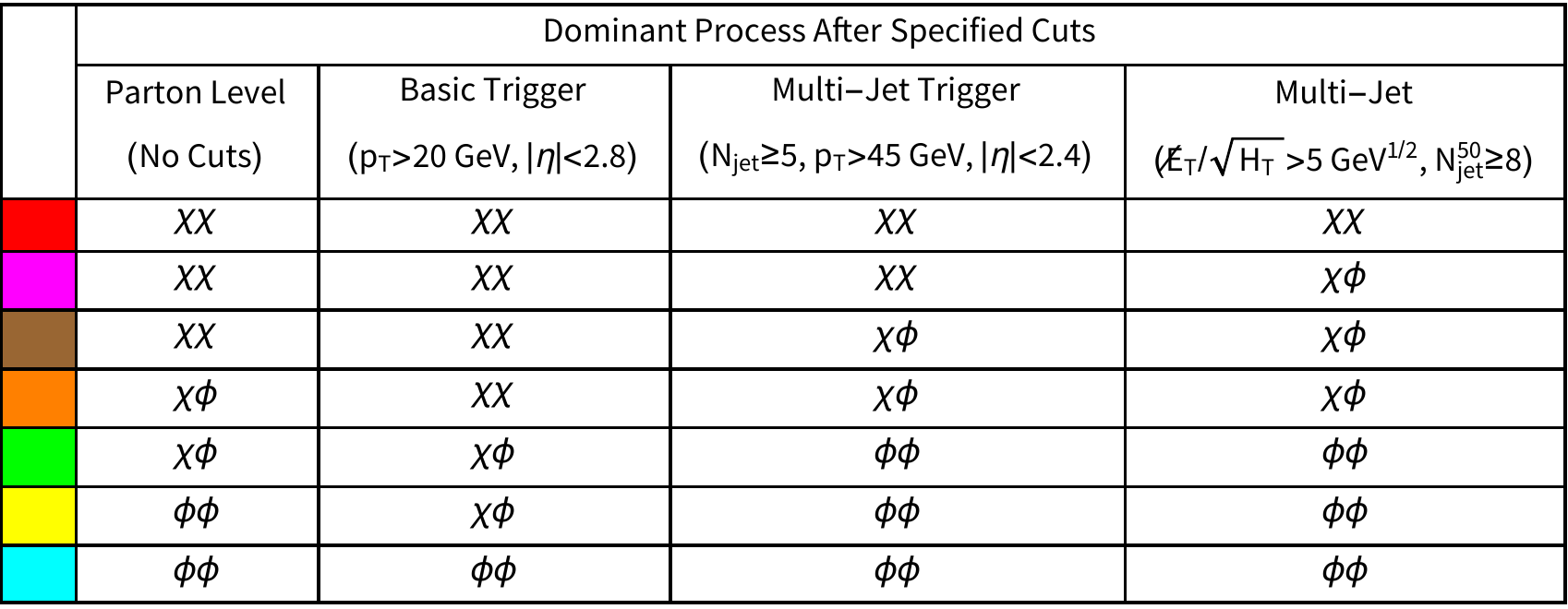}
\end{center}
\caption{Summary of the collider-phenomenology consequences of the mediator-induced 
decay cascades within our model, plotted for discrete points within 
the $(m_\phi,\gamma)$ plane with $m_0 = 500$~GeV, $\Delta m = 50$~GeV, $\delta = 1$, 
and $c_0 = 0.1$ held fixed.  Each box within the figure corresponds to a particular combination 
of $m_\phi$ and $\gamma$.  The four numbers displayed in each box indicate the values of 
$\Njet^{10\%}$ obtained after the application of the cuts specified in the key at the
bottom left.  The text at the bottom of each box indicates the relative sizes of the 
cross-sections $\sigma_{\phi\phi}$, $\sigma_{\phi\chi}$, and $\sigma_{\chi\chi}$ at the 
parton level, before cuts.  The color of each box indicates which production process dominates 
the overall cross-section for decay-cascade events after the application of the different sets 
of event-selection criteria described in the legend at the bottom right.  As discussed in 
the text, the thick, black solid contour represents the bound from multi-jet searches, while 
the thick, black dashed contour represents the corresponding bound from moderate-jet-multiplicity
searches.  The regions above and to the right of these contours are excluded.  
  \label{fig:scan} }
\end{figure*}
\end{center}

The results of this parameter-space survey are shown in Fig.~\ref{fig:scan}.  Each 
individual box within the figure corresponds to a particular combination of $m_\phi$ and 
$\gamma$.  The four numbers displayed within each box indicate the value of 
$\Njet^{10\%}$ at four different stages of our analysis, as indicated in the key
at the bottom left of the figure.  The number enclosed within
a black circle in the upper left of each box indicates the value of $\Njet^{10\%}$
at the parton level with no additional cuts, while the number in the upper right indicates
the corresponding value obtained at the parton level with the basic trigger cuts 
$p_{T_j} > 20$~GeV and $|\eta_j|<2.8$ applied.  Similarly, the number in 
the lower left indicates the value of $\Njet^{10\%}$ obtained at detector level with the 
same basic trigger applied, while the number in the lower right indicates the 
value of $\Njet^{10\%}$ obtained after the application of the multi-jet trigger cuts 
$\Njet\geq 5$, $p_{T_j}>45$~GeV, and $|\eta_j|<2.4$.  The 
text at the bottom of each box indicates the relative size of the cross-sections
$\sigma_{\phi\phi}$, $\sigma_{\phi\chi}$, and $\sigma_{\chi\chi}$ at the parton level,
before the application of any cuts.  The color of each box indicates which production 
process dominates the overall cross-section for mediator-induced decay-cascade events 
after the application of the different sets of event-selection criteria described in the 
legend at the bottom right of the figure.  We note that the event-selection criteria 
associated with the results shown in the ``Multi-Jet'' column of the legend include not 
only the cuts explicitly listed in the heading of that column, but also the cuts associated 
with the multi-jet trigger.

Comparing the $\Njet^{10\%}$ values appearing in the upper left and upper right corners 
of a given box provides a sense of how rudimentary cuts associated with jet-energy thresholds 
and detector geometry affect the $\Njet$ distribution, while comparing the values shown in
the upper left and lower left corners provides information about the effects of 
ISR, FSR, and parton-showering.  We observe that 
throughout the region of the $(m_\phi,\gamma)$-plane shown in the figure, geometric and 
jet-energy-threshold effects do not have a significant impact on $\Njet^{10\%}$.          
We also observe that while the effects of ISR, FSR, and parton-showering are less uniform
across the $(m_\phi,\gamma)$-plane, leading to an increase in $\Njet^{10\%}$ in some 
regions and a reduction in others, the overall impact on these effects is not particularly 
dramatic within any region of the plane.  The reduction in $\Njet^{10\%}$ which results
from the application of the multi-jet cuts is typically more pronounced.  However, the overall
message is that whenever mediator-induced decay chains tend to generate a significant number of 
``jets'' at the parton level, this typically translates into a significant population of events 
with large jet multiplicities at the detector level as well.    

In addition to information about jet multiplicities, Fig.~\ref{fig:scan} also provides 
information about how the bounds discussed in Sect.~\ref{sec:LHCseaches} constrain the parameter 
space of our model.  In particular, the solid black jagged line separates the points within 
out parameter-space scan which satisfy the bound from the multi-jet search limits derived in 
Ref.~\cite{Aaboud:2017hdf} from the points which do not.  Similarly, the dashed black jagged line
separates the points within our parameter-space scan which satisfy the bound from the 
moderate-jet-multiplicity search limits derived in Ref.~\cite{Aaboud:2017vwy} from the points 
which do not.  The regions above and to the right of each contour are excluded by the 
corresponding constraint.  By contrast, we find that the constraints from the monojet search 
limits derived in Ref.~\cite{ATLASMonojet} do not exclude any of the parameter space shown.    

We see from Fig.~\ref{fig:scan} that the region of parameter space in which $m_\phi$ and $\gamma$ 
are both large --- and in which processes of the form $pp\rightarrow \chi_m\bar{\chi}_n$ dominate 
the event rate --- is the region most severely impacted by the constraints from 
multi-jet searches (which supersede the moderate-jet-multiplicity searches throughout the region 
shown).  Nevertheless, we observe that regions of parameter space remain within which 
such processes dominate the event rate both before and after cuts are applied, while at the 
same satisfying these constraints.  While the values of $\Njet^{10\%}$ are largest 
within this excluded region at all stages of our analysis, we note that there exists a substantial 
region of the allowed parameter space wherein $\Njet^{10\%} \geq 8$ even after the 
application of the multi-jet cuts.  This is the region within which $m_\phi$ is large, 
$\gamma$ is small, and processes of the form $pp\rightarrow \phi\chi_n$ dominate the overall 
event rate.  By contrast, within regions of parameter space where $m_\phi$ is small, $N$ is 
likewise small and the number of individual processes of the form $pp\rightarrow \phi\chi_n$ 
or $pp\rightarrow \chi_m\bar{\chi}_n$ which contribute to the overall event rate is 
comparatively small.  As a result, $pp\rightarrow \phi^\dagger\phi$ tends to dominate
the event rate in this region and $\Njet^{10\%}$ tends not to be terribly high in 
comparison with the results obtained for larger values of $m_\phi$.  That said, we note
that reasonably large jet multiplicities can still arise within this region, especially
for cases in which $\gamma$ is large.            

The results shown in Fig.~\ref{fig:scan} demonstrate that while existing LHC 
searches impose non-trivial constraints on the parameter space of our model, there 
nevertheless exists a substantial region of that parameter space within which 
extended mediator-induced decay cascades arise without violating these constraints.
The prospects for probing these regions of parameter space at future colliders ---
or through use of alternative search strategies at the LHC --- will be discussed in
Sect.~\ref{sec:conclusions}. 

\FloatBarrier
 

\section{Implications Beyond Colliders: A Consistency Check \label{sec:noncolliderpheno}}


Our primary aim in this paper has been to investigate the general
properties of mediator-induced decay chains and the multi-jet signatures to which
such decay chains can give rise at colliders.  Indeed the model which we introduced 
in Sect.~\ref{sec:model} was chosen simply for purposes of illustration, and 
similar signatures can arise across a broad class of new-physics 
scenarios in which multiple dark-sector states interact with the fields of the SM
through a common mediator.  For this reason, we have thus far focused exclusively on 
collider considerations in placing constraints on this model.

That said, one might wonder whether other constraints --- such as those from flavor 
physics or cosmology --- generically exclude new-physics scenarios which are capable of 
yielding observable jet-cascade signatures of this sort at the LHC or at future 
colliders.  Thus, in this section, we demonstrate that our illustrative model 
is indeed consistent with these additional
constraints.  In so doing, we furnish a proof of concept that new-physics scenarios
which would give rise to multi-jet collider signals of mediator-induced decay chains at
the LHC or at future colliders can indeed be compatible with
all applicable constraints. 

\subsection{Flavor-Physics Considerations} 
 
The first set of constraints we consider are those from flavor physics.
As noted in Sect.~\ref{sec:model}, our illustrative model can potentially give rise to FCNCs, 
which are tightly constrained by data.  However, there exist standard methods  
through which constraints on FCNCs can easily be satisfied in models of this type.  

Most such methods are rooted in the principle of minimal flavor violation~\cite{MFV} --- \ie, 
the principle that there exists a unique source for the breaking of the 
$U(3)_Q\times U(3)_u \times U(3)_d$ flavor symmetry that would exist in the quark sector of 
the SM in the absence of Yukawa couplings.  Indeed, an approach along these lines was employed 
in Ref.~\cite{TChannelHaipeng} within the context of a model involving a coupling structure 
identical to that in Eq.~(\ref{eq:Lint}), but involving only one dark-sector particle 
species.  In this approach, one posits that $\phi$ 
transforms as a triplet under the $U(3)_u$ symmetry which acts on the right-handed up-type 
quarks.  The simplest way to ensure that the mass term for $\phi$ accords with the 
principle of minimal flavor violation is to posit that the mass matrix for the component
fields in $\phi$ is proportional to the identity matrix in flavor space.  In this case,
this mass matrix remains invariant under arbitrary $U(3)_u$ transformations and is 
therefore diagonal in the basis in which the quark-mass matrix is likewise diagonal.  The three 
physical fields $\phi_u$, $\phi_c$, and $\phi_t$ within the $\phi$ multiplet are degenerate
in mass, with $m_{\phi_u} = m_{\phi_d} = m_{\phi_t}$, and each couples at tree level 
exclusively to a single quark species.  As a result, the leading contributions to FCNC 
processes that would otherwise arise as a consequence of Eq.~(\ref{eq:Lint}) vanish, 
and all higher-order corrections are suppressed by powers of the small, off-diagonal elements
in the Cabibbo-Kobayashi-Maskawa matrix.  Thus, even for reasonably large values of $N$,
such modifications render our model compatible with experimental constraints on FCNCs. 
Moreover, we emphasize that more complicated structures in which $m_{\phi_u}$, $m_{\phi_c}$, 
and $m_{\phi_t}$ are {\it not}\/ degenerate can also be consistent with the principle of 
minimal flavor violation and a similar suppression to FCNCs.    

The results we have explicitly derived in this paper, which pertain the case in which
there exists a single mediator $\phi$ with a coupling structure such that $c_{0u} \neq 0$
and $c_{0q} = 0$ for $q = \{c,t\}$, can also be viewed as pertaining to the regime in which a 
hierarchy $m_{\phi_u} \ll m_{\phi_c},m_{\phi_t}$ exists among the masses of these three 
mediator generations.  However, it is also interesting to consider how these results are 
modified in the regime in which $m_{\phi_u}$, $m_{\phi_c}$, and $m_{\phi_t}$ are degenerate.  
We begin by considering the effect of introducing additional generations of mediator
particles on the decay phenomenology of our model.
In the regime in which $m_{\phi_u} = m_{\phi_d} = m_{\phi_t}$, the branching 
fractions $\mathrm{BR}_{\phi n}$ and $\mathrm{BR}_{n \ell}$ are the same as in the regime in which
$m_{\phi_u} \ll m_{\phi_c},m_{\phi_t}$.  Moreover, given that $m_c \ll \Delta m$ within
our parameter-space region of interest, the total widths of $\phi_u$ and $\phi_c$ are 
approximately equal in the degenerate-mass regime and both approximately equal to the 
result for $\Gamma_\phi$ obtained in the hierarchical-mass regime.  By contrast, the total width 
of each $\chi_n$ is effectively doubled, since contributions to each partial width $\Gamma_{n\ell}$ 
arise both from the process $\chi_n \rightarrow \bar{u} u \chi_\ell$, which involves a virtual 
$\phi_u$, and from the process $\chi_n \rightarrow \bar{c} c \chi_\ell$, which involves a virtual 
$\phi_c$.  (Contributions to $\Gamma_{n\ell}$ from $\chi_n \rightarrow \bar{t} t\chi_\ell$ are 
suppressed in comparison with these other processes, due to the mass of the top quark.)  
This modification has the effect of relaxing the lower limit on $c_0$ shown in 
Fig.~\ref{fig:c0bounds}, thereby opening up additional regions of model-parameter space.    

Introducing additional generations of mediator particles also has an effect on the 
cross-sections for the various production processes which give rise to mediator-induced
decay chains in our model.  The cross-sections for the production processes 
$pp \rightarrow \chi_m \overline{\chi}_n$ and $pp \rightarrow \phi \chi_n$ will not change 
significantly, given that parton-distribution function for $u$ within the 
proton is much larger than that for $c$ at all relevant momentum fractions and energy scales.  
By contrast, in the degenerate-mass regime, the combined cross-section for 
$pp \rightarrow \phi_u\phi_u$ and $pp \rightarrow \phi_c\phi_c$ --- processes which proceed 
primarily via gluon fusion and give rise to effectively indistinguishable decay cascades --- is 
approximately double the cross-section for $pp \rightarrow \phi_u\phi_u$ in the 
hierarchical-mass regime.  (The process $pp \rightarrow \phi_t\phi_t$ leads to entirely
different event topologies necessarily involving at least one top-quark pair, and therefore 
must be considered separately.)  However, even when this doubling is taken into account, 
we find that the boundary between the allowed and excluded regions of parameter space in 
Fig.~\ref{fig:scan} does not change.  Thus, in going from the hierarchical-mass regime to the 
degenerate-mass regime, the main conclusions of Sect.~\ref{sec:scan} are unchanged. 
 
\subsection{Cosmological Considerations}
 
The second set of considerations we consider are those from early-universe cosmology.
Since we are assuming that all $\chi_n$ with $n > 0$ decay promptly within a collider 
detector, these particles decay far too early within the history of the universe to
have any appreciable effect on cosmological observables.  However, no contribution 
to the width of $\chi_0$ arises as a consequence of Eq.~(\ref{eq:Lint}), and we have 
thus far made no assumptions about the lifetime $\tau_0$ of this particle, other than that
it is stable on collider timescales.  Indeed, it is possible that $\chi_0$ decays as a 
result of additional, highly suppressed interactions on far longer timescales.   
If $\tau_0$ is sufficiently long-lived, a cosmological population of this lightest ensemble 
constituent could have observable consequences for early-universe cosmology or particle 
astrophysics.  There are essentially two relevant regimes we must consider, based on the 
relationship between $\tau_0$ and the present age of the universe.  We shall address these 
two possibilities in turn. 

The first possibility is that $\tau_0$ far exceeds collider timescales, but is 
nevertheless significantly shorter than the observable age of the universe.  In this case, we 
simply stipulate that $\tau_0$ is sufficiently short that constraints from Big-Bang 
nucleosynthesis~\cite{CyburtBBN,JedamzikBBN1,KawasakiBBN1,JedamzikBBN2,KawasakiBBN2}, 
spectral distortions to the cosmic microwave background~\cite{HuSilk1,HuSilk2,Chluba1,Chluba2}, 
modifications of the ionization history
of the universe~\cite{SlatyerDarkAges,SlatyerCMB,PoulinIonization}, \etc, are 
satisfied.  Indeed, this is the simplest way in which the illustrative model we introduced
in Sect.~\ref{sec:model} can be rendered consistent with cosmological constraints.   
Moreover, we note that for certain values of $\tau_0$ within this regime, an unstable 
$\chi_0$ could potentially also give rise to signals at a dedicated surface detector like 
MATHUSLA~\cite{MATHUSLA,MATHUSLAWhitePaper}.   

The second possibility is that $\chi_0$ is cosmologically --- and perhaps even 
absolutely --- stable.  In this case, we must ensure that cosmological and astrophysical 
constraints on whatever relic population of $\chi_0$ particles is generated during the 
early universe are satisfied within the region of model-parameter space wherein multi-jet 
signatures of mediator-induced decay chains would be accessible at the LHC or at future colliders.  
Moreover, it is also interesting to consider whether a relic population of $\chi_0$ particles 
could account for the present-day abundance of dark matter in our universe within this 
same parameter-space regime.  In order for this possibility to be realized, $\chi_0$ 
must not only acquire a present-day cosmological abundance $\Omega_0$ similar to the dark-matter 
abundance $\OmegaDM \approx 0.26$ inferred from Planck data~\cite{Planck}, but also satisfy 
constraints from direct- and indirect-detection experiments.  

In principle, there are two contributions to $\Omega_0$ which one must consider within the
context of our model.
First, since the scattering processes that follow from Eq.~(\ref{eq:Lint}) maintain 
thermal equilibrium between $\chi_0$ and the visible-sector particles in the radiation bath at 
high temperatures $T \gg m_0$, this lightest ensemble constituent generically acquires a 
contribution to its abundance from thermal freeze-out.
Second, the late decays of long-lived $\chi_n$ with $n > 0$ can give rise to an additional, 
non-thermal contribution to this abundance.  However, assuming that $\chi_0$ freezes 
out during a radiation-dominated epoch, the time $t_{\rm fr}$ at which freeze-out effectively
occurs is given by 
\begin{equation}
  t_{\rm fr} ~\approx~ \sqrt{\frac{45}{2\pi^2}}\,g^{-1/2}_\ast(T_{\rm fr})
    \frac{M_P}{T_{\rm fr}^2}~,
\end{equation}
where $M_P$ is the reduced Planck mass, where $T_{\rm fr} \approx m_0/20$ is the freeze-out
temperature, and $g_\ast(T_{\rm fr})$ is the number of effectively massless degrees of freedom 
in the radiation bath at this temperature.  Comparing $t_{\rm fr}$ to the  
the lifetimes $\tau_n$ for all of the $\chi_n$ with $n > 0$ within our 
parameter-space region of interest, we typically find that $\tau_n \ll t_{\rm fr}$ for all of
these heavier $\chi_n$.  Any $\chi_0$ particles produced in the early universe via the decays 
of these heavier $\chi_n$ therefore 
attain thermal equilibrium with the radiation bath prior to freeze-out.  Moreover, since 
all of the $\chi_n$ with $n > 0$ effectively decay away well before $\chi_0$ freezes out,
the effect of coannihilation processes of the form $\overline{\chi}_n\chi_m \rightarrow \bar{q}q$
on $\Omega_0$ is also negligible.  Thus, within our parameter-space region of interest, the 
only relevant contribution to $\Omega_0$ is that from thermal freeze-out, where 
the annililation rate of $\chi_0$ is governed solely by $t$-channel processes of the form 
$\overline{\chi}_0\chi_0 \rightarrow \bar{q}q$ involving a virtual $\phi$.

Taking these considerations into account, we evaluate $\Omega_0$ numerically using the 
{\tt MadDM}~\cite{MadDM} code package for the same parameter choices $m_0 = 500$~GeV and 
$c_0 = 0.1$ as in Fig.~\ref{fig:scan} and a sampling of mediator masses within the range 
$0.6\mathrm{~TeV} \leq m_\phi \leq 2.5\mathrm{~TeV}$.  For all values of $m_\phi$ within this 
range, we find that $\Omega_0$ exceeds $\OmegaDM$ --- at least within the context of the 
standard cosmology.  However, within the context of a modified cosmology --- for example, 
one in which the abundance of $\chi_0$ is diluted at late times due to an injection of 
entropy from late-decaying particles after freeze-out has occurred~\cite{GelminiGondolo} --- the 
abundance of such a particle can easily be reduced to an acceptable level.  Within such a
modified cosmology, then, a cosmologically stable $\chi_0$ can acquire an abundance    
$\Omega_0 \approx \OmegaDM$ and constitute the majority of the dark matter.
 
Of course, in order for $\chi_0$ to account for a non-negligible fraction of $\OmegaDM$, 
constraints from direct- and indirect-detection experiments must also be satisfied.  We begin by 
discussing the constraints from direct detection.  The leading contribution to the event rate 
for elastic scattering between $\chi_0$ particles in the local dark-matter halo and the atomic 
nuclei in a dark-matter detector arises due to fundamental processes of the form 
$q \overline{\chi}_0 \rightarrow q \overline{\chi}_0$ involving an $s$-channel mediator.  
Such processes yield both a spin-dependent and a spin-independent contribution to the event 
rate~\cite{TChannelHaipeng}.  For the case we are considering here, in which $\chi_0$ is a Dirac 
fermion, the dominant contribution is from spin-independent scattering.  We shall therefore focus 
on this spin-independent contribution in what follows.  In doing so, we note that  
inelastic-scattering processes of the form $q \overline{\chi}_0 \rightarrow q \overline{\chi}_n$ 
with $n > 0$ are irrelevant for direct-detection phenomenology, given that 
$\Delta m \gg \mathcal{O}({\rm MeV})$ throughout our parameter-space region of interest.

The most stringent constraint on $\sigma^{\rm (SI)}$ for a dark-matter particle with this mass is 
the bound $\sigma^{\rm (SI)} < 6.44 \times 10^{-46}$~cm$^2$ set by the XENON1T 
experiment~\cite{XENON1T}.  In order to assess whether this constraint can be satisfied within 
our parameter-space region of interest, we evaluate the spin-independent cross-section per 
nucleon $\sigma^{\rm (SI)}$ for the parameter choices used in 
Fig.~\ref{fig:scan} using the {\tt MadDM}~\cite{MadDM} code package.  We find that 
the spin-independent cross-section per nucleon decreases with increasing $m_\phi$ from 
$\sigma^{\rm (SI)} \approx 4.88 \times 10^{-44}$~cm$^2$ for $m_\phi = 0.6$~TeV to 
$\sigma^{\rm (SI)} \approx 4.86 \times 10^{-47}$~cm$^2$ for $m_\phi = 2.5$~TeV, and that the bound from 
XENON1T is satisfied for $m_\phi \gtrsim 1.5$~TeV.~  Moreover, we note that for certain choices 
of $c_0$ and $m_0$ which differ only slightly from these benchmark values, even lower values of
$m_\phi$ can be accommodated.

The leading indirect-detection constraints on a cosmologically stable population of $\chi_0$
particles are those derived from Fermi-LAT observations of dwarf spheroidal galaxies and other 
Milky-Way satellites~\cite{Fermi2016}.  In order to assess the implications of these constraints
within our parameter-space region of interest, we compute the expected gamma-ray flux from
$\overline{\chi}_0\chi_0\rightarrow \bar{q}q$ annihilation within the dark-matter halos of 
these Milky-Way satellites using the {\tt MadDM}~\cite{MadDM} code package, which incorporates 
the $J$-factors derived in Ref.~\cite{DwarfProfiles}.  For the same $m_0$ and $c_0$ as 
in Fig.~\ref{fig:scan} and a mediator mass of 
$m_\phi = 600$~GeV, we find that the velocity-averaged annihilation cross-section is 
$\langle \sigma v\rangle \approx 2.25 \times 10^{-29}$~cm$^3$/s, which is well below
the corresponding bound $\langle \sigma v\rangle \leq 2.30 \times 10^{-25}$~cm$^3$/s from 
Fermi-LAT data.  Since increasing $m_\phi$ for the same $m_0$ further suppresses 
$\langle \sigma v\rangle$, we find that a cosmological population of $\chi_0$ with 
$\Omega_0 \approx \OmegaDM$ is consistent with the Fermi-LAT data.

We therefore conclude that a cosmologically stable $\chi_0$ is a viable
dark-matter candidate within our parameter-space region of interest.  Indeed, 
the phenomenological consequences of such a particle do not conflict with bounds 
from direct- and indirect-detection data.  Moreover, in the presence of a source of late
entropy injection subsequent to thermal freeze-out, this lightest ensemble constituent 
can also acquire an appropriate abundance.


\section{Conclusions and Outlook \label{sec:conclusions}}


In this paper, we have investigated the collider phenomenology of scenarios in which 
multiple dark-sector particles with similar quantum numbers couple to the fields of the visible 
sector via a common massive mediator.  In such scenarios, the mediator not only plays an 
important role in providing a portal through which the dark and visible sectors interact, but 
also necessarily gives rise to decay processes wherein heavier dark-sector particles decay to final 
states which include both lighter dark-sector particles and visible-sector fields.
In cases in which these visible-sector fields are quarks or gluons, successive decays of this 
sort give rise to extended decay cascades involving large numbers of hadronic jets at hadron 
colliders.  We have investigated the structure of these mediator-induced decay cascades and 
examined how existing LHC searches constrain the parameter spaces associated with such 
scenarios.  We have also shown that there exist large regions of parameter space within which 
all applicable constraints from these searches are satisfied, but within which extended decay 
cascades of this sort develop and within which jet multiplicities are characteristically large.  
Thus, striking signatures of this sort could potentially manifest themselves at forthcoming LHC 
runs or at future colliders.  Such signatures could therefore provide a way of probing the 
properties of the dark sector and the mediator through which it couples to the SM. 

Many possible extensions of our analysis can be envisioned.  For example,
in this study, we have chosen to focus on the region of parameter space in which the number of 
jets with $p_T$ sufficient to satisfy the applicable jet-identification criteria is effectively
maximized.  Thus, we have chosen our model parameters such that $m_{N - 1} < m_\phi$ and such 
that the lifetimes of all $\chi_n$ with $n > 0$ are sufficiently short that these particles 
typically decay promptly within a collider detector.
However, it would be interesting to examine the discovery prospects for our model within 
other regions of parameter space as well --- regions within which extended mediator-induced
decay cascades still arise, but within which the collider phenomenology nevertheless differs in 
salient ways.

One such alternative possibility arises in the regime in which $m_{N-1} \ll m_\phi$  In such 
cases, any ensemble constituent $\chi_n$ initially produced by the decay of an on-shell mediator 
$\phi$ is highly boosted.  In this regime, particles produced by the subsequent decay of this 
$\chi_n$ will be collimated in the direction of its three-momentum vector.  A similar situation 
can also in principle arise in situations in which significant mass gaps occur within
the mass spectrum of $\chi_n$.  Such possibilities are under investigation~\cite{DDMboosted}.
     
In this connection, we also note that while the results of existing LHC searches are effective 
in probing and constraining scenarios involving mediator-induced decay cascades, alternative 
search strategies may be even more efficient in resolving the particular kinds of multi-jet 
signals which arise in these scenarios from SM backgrounds.  For example, a variety of jet-shape 
variables and other jet-substructure techniques could potentially provide a way of improving the 
discovery reach for signatures of these cascades when such
highly boosted particles arise.  Such techniques can also be advantageous in probing regions 
in which parton-level ``jet'' multiplicities $\Njet$ are typically so large that multiple jets 
within the same event inevitably overlap in $(\eta,\phi)$-space.  In such cases, a moderate 
jet multiplicity at the detector level might therefore belie a much higher value of $\Njet$. 
The application of jet-substructure techniques will be especially relevant at future hadron 
colliders with CM energies significantly higher than that of the LHC.~  However, since the 
substructure of jets arising from mediator-induced decay cascades differs from the 
substructure of the jets produced by the decays of the heavy SM particles $W^\pm$, $Z$, and $t$, 
alternative jet-shape variables and clustering algorithms may be required~\cite{DDMboosted}.

Several additional phenomenological possibilities also arise in the regime in which 
many of the $\chi_n$ are sufficiently long-lived that they do not tend to decay promptly within 
a collider detector.  For example, any $\chi_n$ with a characteristic decay length in the range 
$\mathcal{O}(1\mathrm{~cm}) \lesssim L_n \lesssim \mathcal{O}(10\mathrm{~m})$ 
will give rise to events in which the jet cascades associated with the decays of 
the more massive, promptly-decaying constituents in the ensemble are accompanied 
by one or more macroscopically displaced vertices.  
Moreover, depending on the choice of model parameters, it is possible that the 
final-state ensemble constituent $\chi_m$ produced at one of these displaced 
vertices might itself decay within the detector a macroscopic distance away, and 
so forth.  This could lead to spectacular and completely novel signatures involving 
multiple displaced vertices arising from the same decay chain.    
An investigation into the prospects for realizing and detecting such signatures at the
LHC and at future colliders is currently underway~\cite{DDMDisplaced}.

Furthermore, any $\chi_n$ with decay lengths $L_n \gtrsim \mathcal{O}(10\mathrm{~m})$ 
will manifest themselves as $\met$ within a collider detector.  Whenever additional 
$\chi_n$ with $n>0$ have decay lengths in this range, the decay chains precipitated by any 
of the production processes depicted in Figs.~\ref{fig:ChiChiDiagram}--\ref{fig:PhiPhiDiagram} 
effectively terminate not merely when a $\chi_0$ particle is produced, but whenever {\it any}\/ of 
these ensemble constituents is produced.  This has a salient impact on the resulting 
multi-jet phenomenology.  For example, for the case of $pp\rightarrow \phi^\dagger\phi$ 
production discussed in detail in Sect.~\ref{sec:pheno},
the probability $\mathcal{P}_\phi(S)$ for such a decay chain
to involve a particular number of steps $S$ would differ from the result in
Eq.~(\ref{eq:prob}).  We also note that long-lived particles within these ensembles 
could also give rise to observable signals at a dedicated surface detector such as 
MATHUSLA~\cite{MATHUSLAWhitePaper,DDMMATHUSLA} --- signals which could then be correlated
with large-jet-multiplicity signatures in the main LHC detectors.
          
In all cases, however, our main message is clear.  If the dark sector contains multiple components 
with similar quantum numbers, and if this sector communicates with the visible sector through 
a mediator, then this mediator has the potential to induce extended decay cascades yielding large
multiplicities of SM particles.   Moreover, as we have demonstrated in this paper, scenarios of 
this sort can be consistent with existing constraints.  Thus, the detection of the corresponding
collider signatures of these scenarios remains a viable future possibility.  Such signatures 
might therefore provide an important route for uncovering and probing not only the dark 
sector but also the mediator through which it couples to the SM.

\acknowledgements

The research activities of KRD, DK, HS, SS, and DY are supported in part
by the Department of Energy under Grant DE-FG02-13ER41976 (de-sc0009913).  
The research activities of KRD are also supported in part by the National 
Science Foundation through its employee IR/D program.  The research activities 
of DK are also supported in part by the Department of Energy under Grant de-sc0010813. 
The research activities of BT are supported in part by the National Science Foundation under 
Grant PHY-1720430.  Portions of this work were performed at the Aspen Center for Physics, 
which is supported in part by the National Science Foundation under Grant PHY-1607611.
The opinions and conclusions expressed herein are those of the authors, and do not represent 
any funding agencies.


\end{document}